\documentclass{aa}
\usepackage{graphicx}
\usepackage{txfonts}
\usepackage[colorlinks=true, linkcolor=blue, citecolor=blue, urlcolor=blue]{hyperref}
\usepackage{subfiles}
\usepackage{pdflscape}
\usepackage{booktabs}
\usepackage{placeins}
\usepackage{subcaption}
\begin{document}

   \title{Velocity evolution of broad-lined type-Ic supernovae with and without gamma-ray bursts}
   \author{
   G. Finneran\,\href{https://orcid.org/0000-0001-7590-2920}{\includegraphics[height=10pt]{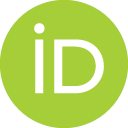}}\inst{1,\thanks{gabriel.finneran@ucdconnect.ie}}\and L. Cotter\,\href{https://orcid.org/0000-0002-7910-6646}{\includegraphics[height=10pt]{figures/ORCIDLogomark128x128.png}}\inst{1}\and 
  A. Martin-Carrillo\,\href{https://orcid.org/0000-0001-5108-0627}{\includegraphics[height=10pt]{figures/ORCIDLogomark128x128.png}}\inst{1}
   }
   \institute{School of Physics and Centre for Space Research, University College Dublin, Belfield, Dublin 4, Ireland}
  
  \date{}
 
  \abstract
   {More than 60 broad-lined type Ic (Ic-BL) supernovae (SNe) are associated with a long gamma-ray burst (GRB). However, many type Ic-BL SNe exhibit no sign of an associated GRB. On average, the expansion velocities of GRB-associated type Ic-BL SNe (GRB-SNe) are greater than those of type Ic-BL SNe without an associated GRB. It has been proposed that this is the result of energy transfer between the ultra-relativistic GRB jet and the SN ejecta. However, this cannot fully explain the discrepancy, as some type Ic-BL SNe without a GRB detection (ordinary type Ic-BL SNe) may also harbour GRB jets.}
   {This work presents the largest spectroscopic sample of type Ic-BL SNe with and without GRBs to date, consisting of 61 ordinary type Ic-BL SNe and 13 GRB-SNe, comprising a total of 875 spectra. The goal of this work is to compare the evolution of SN expansion velocities in cases where an ultra-relativistic jet has been launched (GRB-SNe) and cases where no GRB jet is inferred from observations (ordinary type Ic-BL SNe). This will help us to understand whether the presence of the jet affects the evolution of the expansion velocity, possibly allowing us to infer the existence of jets in cases where GRB emission is not detected.}
   {We measured the expansion velocities of the Fe II [5169\,\AA and Si II [6355\,\AA] features observed in the spectra of type Ic-BL SNe using a spline fitting method. We fit the expansion velocity evolution with single and broken power-laws. In each analysis we compared two populations: ordinary type Ic-BL SNe and GRB-SNe.}
   {The expansion velocities of the Fe II and Si II features reveal considerable overlap between the two populations. Although some GRB-SNe expand more rapidly than ordinary type Ic-BL SNe, the difference between the population medians is not statistically significant. Our analysis confirms that type Ic-BL SNe and GRB-SNe generally expand more rapidly than type Ic SNe. The marginalised Fe II and Si II power-law indices indicate that GRB-SNe decline at similar rates to ordinary type Ic-BL supernovae. Broken power-law evolution appears to be more common for the Si II feature, which always follows a shallow-steep decay. In contrast the broken power-law Fe II decays are predominantly steep-shallow. The Si II velocity evolution of PTF12gzk and SN2016coi (engine-driven SNe) are similar to GRB060218-SN2006aj, with both showing broken power-law decay. This observation may hint at a two-component ejecta model, such as a GRB jet or a cocoon.}
   {Neither the velocities nor their evolution can be used to distinguish between ordinary type Ic-BL SNe and GRB-SNe. Velocities consistent with broken power-law evolution may indicate the presence of a GRB jet in some of these ordinary type Ic-BL SNe, but this is likely not as robust as late-time radio surveys. These results suggest that GRB-SNe and ordinary type Ic-BL SNe are drawn from the same underlying population of events.}

   \maketitle
%
\nolinenumbers
\section{Introduction}
   Broad-lined type Ic (type Ic-BL) supernovae (SNe) are an observationally rare class of supernovae. Photometrically these events are among the brightest supernovae, reaching peak absolute magnitudes of around -18.5 mag after 10-15 days \citep{Taddia.2019}. Spectroscopically they are distinguished by the absence of hydrogen and helium absorption features and presence of broad absorption features of Fe, Si and Ca in their photospheric spectra (for reviews of SN classifications see \cite{HandbookSNeGalYam.2017,Filippenko.1997}). These broad features are the result of the rapid expansion of the outer layers of the SN ejecta, with typical ejecta velocities being in the $\sim$15000-20000\,km/s range at peak light \citep{Srinivasaragavan.2024,Taddia.2019,Modjaz.2016}. Type Ic-BL SNe are believed to result from the death of a Wolf-Rayet star, which has been stripped of its hydrogen and helium layers, either through episodic/eruptive mass loss, mass loss due to winds, or interaction in a binary system \citep[e.g.][]{Smith.2014, Crowther.2007, Mokiem.2007}.

    Gamma-ray bursts (GRBs) are brief, intense pulses of gamma-rays and have been the subject of extensive theoretical and observational interest from the astronomy community for over five decades. The duration of the gamma-ray prompt emission has lead to the separation of GRBs into long and short bursts \citep{Kouveliotou.1993}. Long GRBs (those with prompt emission longer than 2 seconds) were linked to the death of massive stars by the `collapsar' model \citep{Woosley.1993, MacFadyen.1999}. This model proposes that long-GRBs are the result of a jet launched by a compact object created when a massive star undergoes core-collapse. The deceleration of this jet in the ambient medium outside of the former star is responsible for the GRB afterglow \citep{Mesz+Rees1997AG}, whilst dissipation of energy within the jet powers the prompt emission \citep{Daigne.1998,Rees.1994}. 
    
    The most notable prediction from the collapsar model is that a supernova (SN) should accompany nearly every long GRB, with the star's outer layers being accelerated by deposition of energy from the central engine within the stellar material. This type of SN is known as an engine-driven SN. In 1998, the detection of GRB980425 and its associated broad-lined type Ic supernova (type Ic-BL SN) SN1998bw put the collapsar model at the forefront of long GRB science \citep{Galama.1998}. In the years since the discovery of GRB980425-SN1998bw, more than 60 `GRB-SN associations' have been detected (e.g. \cite{Finneran.2024A,Cano.201723}). Nearly all of the supernovae associated with GRBs are type Ic-BL SNe\footnote{SN2011kl, a super-luminous, magnetar-powered SN associated with GRB111209A is the sole exception to this pattern \citep{Greiner.2015}.}.

   \begin{figure*}[!ht]
      \centering
      \includegraphics[width=\linewidth]{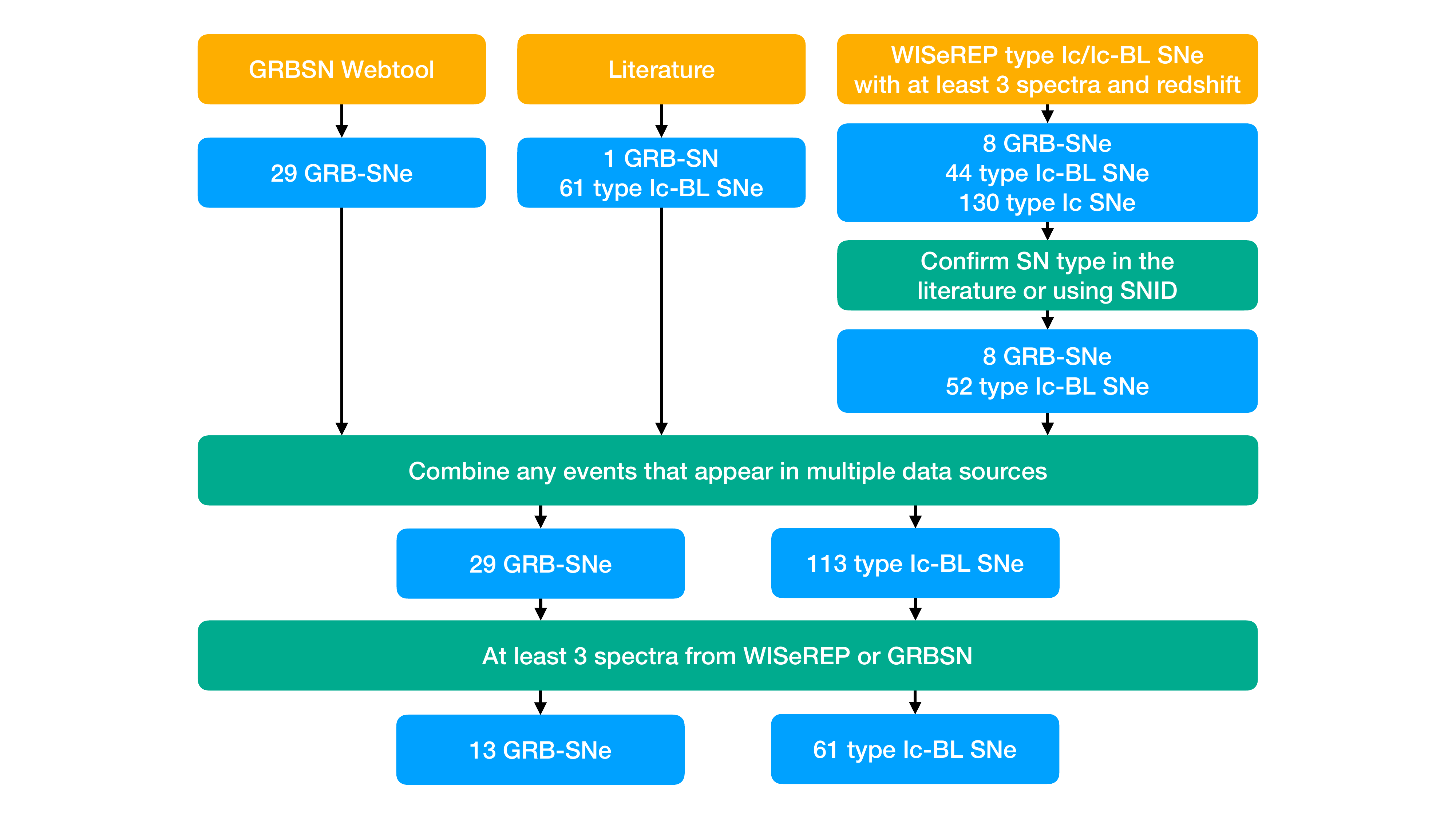}
      \caption{Flow of sample selection showing the data sources in orange, number of SNe at each stage of the process in blue and filtering criteria in green. The number of each SN type at each stage of the process is shown. The search on WISeREP produced 8 GRB-SNe, 44 type Ic-BL SNe and 130 type Ic SNe. Since the classifications reported on WISeREP may be unreliable, we performed a literature search for these SNe to confirm their types. For those SNe that could not be found in the literature, we classified them using SNID \citep{Blondin.2007}. This step produced 52 confirmed type Ic-BL SNe and 8 GRB-SNe. Whilst searching the literature to confirm SN types we identified an additional 1 GRB-SN and 61 type Ic-BL SNe. A total of 29 GRB-SNe were gathered from the GRBSN webtool, some of which had already been found in the literature and WISeREP searches. Combining data for events found in multiple data sources and imposing a minimum of three spectra from GRBSN or WISeREP resulted in a final sample of 61 type Ic-BL SNe and 13 GRB-SNe. Note that we also obtained spectra for GRB180427A-SN2018fip from the ESO archive, bringing the number of spectra above the threshold of three and allowing for its inclusion in the final dataset.}
      \label{fig:samplenumnbers}
   \end{figure*}

    The association between GRBs and type Ic-BL SNe has been extensively investigated over the past 25 years (for a review of GRBs associated with SNe (hereafter GRB-SNe), see \cite{Cano.201723}; a review of GRB jets in SNe may be found in \cite{Corsi.2021}). Observations of long GRBs at low redshifts indicate that the majority of these events are associated with a supernova, with simulations showing that the same central engine can power both the GRB and the type Ic-BL SN \citep{Barnes.2018}. In cases where an extensive follow-up campaign was conducted, supernova-like emission is ruled out for just five events at low redshift: GRB 060505 \citep{Fynbo.2006}, GRB 060614 \citep{Fynbo.2006, Gal-Yam.2006,Valle.2006i}, GRB 111005A \citep{MichalowskI.2018, Tanga.2018}, GRB 211211A \citep{Rastinejad.2022, Troja.2022}, and GRB 230307A \citep{Levan.2024,Yang.2024}. Evidence for kilonova emission has been found for GRB230307A \citep{Rastinejad.2022, Troja.2022} and GRB211211A \citep{Levan.2024,Yang.2024}, and this has accelerated the shift away from the typical long/short classification of GRBs.

    In contrast with the abundance of type Ic-BL SNe associated with nearby GRBs, fewer than one in four type Ic-BL supernovae have an associated GRB detection\footnote{This figure excludes supernovae associated with GRBs for which no spectroscopic confirmation of the supernova type was available.}. We term these SNe `ordinary type Ic-BL SNe' in this work. Despite this, type Ic-BL SNe appear to have very similar parameters regardless of whether or not they are associated with a GRB. Type Ic-BL SNe with GRBs synthesise $\sim$0.4$\pm$0.2 M$_\odot$ of nickel, with ejecta of $\sim$6$\pm$4 $M_\odot$ \citep[e.g][]{Cano.201723}, while ordinary type Ic-BL SNe synthesise $\sim$0.3$\pm$0.2 M$_\odot$ of nickel, with ejecta of $\sim$4$\pm$3 M$_\odot$ \citep[e.g.][]{Taddia.2019,Prentice.2016,Lyman.2016,Taddia.2015, Drout.2011}. These two populations also have similar supernova kinetic energies ($\sim$10$^{52}$ ergs) \citep[c.f][]{Taddia.2019, Cano.201723}. 

   These kinetic energies are extremely high relative to other SNe \citep{Corsi.2021}. This has been seen as evidence for an additional energy source within these events, as these energies cannot be achieved through gravitational collapse alone \citep{Corsi.2021,Barnes.2018}. It has been inferred that the source of this additional energy may be a rapidly rotating accreting black hole or magnetar (an engine) that launches a GRB jet within the star \citep{Corsi.2021}. High velocity material generated by the propagation of this jet through the stellar material may be responsible for the high kinetic energies of these SNe. For this reason they are sometimes referred to as engine-driven SNe. However, because GRB jet-like emission has not been observed for almost three quarters of type Ic-BL SNe, it is necessary to consider mechanisms by which a jet may be created by a type Ic-BL supernova but go undetected.

    The ultra-relativistic nature of GRB jets gives rise to the phenomenon of `orphan' afterglows. These are GRB afterglows observed without a corresponding prompt gamma-ray component \citep{Rhoads.1997, Meszaros.1998}. Observers whose line of sight lies outside the beaming cone of the GRB radiation will not see the prompt emission, and will only observe the afterglow radiation once the jet begins to spread laterally as it decelerates. In contrast, the radiation emitted by a type Ic-BL SN is visible in all directions. It is therefore possible that many type Ic-BL supernovae could have a contemporaneous GRB, invisible because the observer is too far off-axis. Late-time radio observations suggest that there may be hidden GRBs in some type Ic-BL SNe \citep{Corsi.2023, Corsi.2016,Soderberg.2006}. At X-ray wavelengths, simulations using the Python package \texttt{afterglowpy} \citep{Ryan.2020} indicate that typical GRBs associated with type Ic-BL SNe may be invisible at observing angles above 25-30 degrees. To date, SN2020bvc is the only candidate off-axis afterglow associated with a type Ic-BL SN \citep{Izzo.2020}. 
    
    Another justification for the lack of GRB emission in three quarters of type Ic-BL SN events may be the phenomenon of jet choking. As the GRB jet propagates through the outer layers of the progenitor, it may become stalled, in a scenario known as a choked jet. If this happens, little to no GRB prompt emission will be produced \citep{Nakar.2017}. This process may transfer energy to the progenitor, producing a high velocity, hot cocoon around the jet \citep{Nakar.2017, Ramirez.2002}. Such a cocoon has been inferred from spectroscopic observations of some GRB-SNe \citep{Izzo.2019}. Such energy transfer may be responsible for the high-velocity features of type Ic-BL SNe.

   To look for evidence of energy injection from a GRB jet in a type Ic-BL SN, \cite{Modjaz.2016} measured the Fe II expansion velocities of a small sample of type Ic-BL SNe with and without an associated GRB. Their results showed that the mean expansion velocities of GRB-SNe are $\sim$6000\,km/s higher than those of ordinary type Ic-BL SNe. However, we note that the average velocities of both populations found by \cite{Modjaz.2016} are consistent within one sigma, due to large uncertainties on the average velocities of GRB-SNe. Subsequent theoretical work by \cite{Barnes.2018} suggested that the deposition of jet energy within the SN ejecta does not significantly impact the supernova spectra or feature velocities. For this reason, it is not clear why there should be velocity differences between ordinary type Ic-BL SNe and GRB-SNe. It is therefore possible that all type Ic-BL SNe harbour a GRB jet.
    
    Previous studies of type Ic-BL SN velocities have relied on relatively small sample sizes \citep[e.g. 11 GRB-SNe and 10 ordinary type Ic-BL SNe in][]{Modjaz.2016}. With the rapid increase in the detection rate of type Ic-BL supernovae enabled by all-sky survey telescopes such as the \textit{Zwicky Transient Facility} (ZTF) \citep{Dekany.2020,ZTF2019}, it has become possible to study a much larger dataset than ever before (e.g. 26 type Ic-BL SNe in \citet{Srinivasaragavan.2024}; 34 type Ic-BL SNe in \citet{Taddia.2019}). The abundance of spectral data in online repositories allows for a quantitative investigation of velocity evolution in these supernovae.
    
    This paper investigates the spectral velocity evolution of the largest sample of type Ic-BL SNe and GRB-SNe collected to date, consisting of almost 900 spectra. A spline fitting method is adopted to measure the velocities of the Fe II and Si II absorption features within the type Ic-BL SN spectrum. The focus of the investigation is to determine whether the velocities of these events can help us to distinguish between Ic-BL SNe with and without GRBs. Section \ref{sec:sample} describes the collection of sample data for this research, highlighting some of the challenges involved in the collation of a large heterogeneous dataset. Section \ref{sec:methods} details the  methodology and its application to the dataset. The results of this analysis are described in Sect. \ref{sec:results}, with the discussion and conclusions presented in Sections \ref{sec:discussion} and \ref{sec:conclusions} respectively.
       

\section{Sample selection and data collection}

   \subsection{Sample selection}\label{sec:sample}

    The supernovae studied in this analysis were identified via an extensive search of the literature and online repositories. In particular, we made extensive use of the Weizmann Interactive Supernova Data Repository\footnote{\url{https://www.wiserep.org/}} (WISeREP) \citep{WISEREP} and the GRBSN webtool\footnote{\url{https://grbsn.watchertelescope.ie}} \citep{Finneran.2024A}. Our sample selection process is illustrated in Fig. \ref{fig:samplenumnbers}.

    We began by searching WISeREP for supernovae of types Ic and Ic-BL with at least 3 spectral epochs and measured redshift. This allowed us to include supernovae that may have been labelled as type Ic SNe prior to the common usage of the Ic-BL SN type in the early 2010s. This approach also ensures that supernovae which have an ambiguous type are included in the initial sample. For example, \cite{Gangopadhyay.2020} mention that SN2016P has spectral features whose widths are in between type Ic and type Ic-BL SNe, while \citet{Prentice.2019} say that its velocities are more akin to type Ic SNe. We also cross-checked the list of type Ic-BL SNe provided by WISeREP with those on the Transient Name Server\footnote{\url{tns-wis.org}} (TNS). This search is believed to be complete up to early April 2024.
    
    The initial WISeREP search produced 8 GRB-SNe, 130 type Ic and 44 type Ic-BL supernovae. However, transient classifications reported on WISeREP may not be accurate, as they tend to be based on very early observations, often from only a single spectrum. Later observations may alter the classification of the SN, but this information may not be updated on WISeREP. This could lead to contamination of our sample with events which do not meet the criteria of a type Ic/Ic-BL supernova. To confirm the classification provided by WISeREP, we followed two approaches. First, a literature search was performed for each SN. This method was used to confirm the SN type of 8 GRB-SNe and 43 type Ic-BL supernovae.
    
    For those SNe where no reliable classification could be found in the literature, we classified the SNe using the Supernova Identification (SNID) tool \citep{Blondin.2007}. To improve the classification accuracy of SNID for Ic-BL SNe, we included the stripped-envelope supernova (SESN) templates provided by the NYU supernova group\footnote{\url{https://github.com/nyusngroup/SESNspectraLib}} \citep{Modjaz.2016, Liu.2016}.

   During classification, we set the SN redshift to the value published on WISeREP. The age range of the SNID template spectra used for classification was restricted to $t_{peak}$-10 and $t_{peak}$+45 days. Epochs larger than 45 days can show emission features that are relatively similar among different SN types and can lead to uncertain classification. The -10 day restriction limits the chance that a continuum dominated template spectrum is used during the fitting. This analysis identified 9 type Ic-BL SNe which (as of April 2024) have not yet been published.

   The literature search and SNID classification of our WISeREP list resulted in a confirmed sample of 8 GRB-SNe and 52 type Ic-BL SNe. A separate literature search for type Ic-BL SNe identified an additional GRB-SN and 61 type Ic-BL SNe. These had not been identified in the initial WISeREP search because they were not listed as type Ic or Ic-BL SNe. A search of the GRBSN webtool turned up 29 GRB-SNe, 9 of which had already been found in the literature or on WISeREP. Combining events found in multiple data sources, at this point the sample consisted of 113 ordinary type Ic-BL SNe and 29 GRB-SNe.
    
    This sample was further reduced by imposing a minimum of three spectral observations to allow for investigation of velocity evolution, and the need for a pre-determined redshift to take into account any possible cosmological effects. The final sample consists of 61 type Ic-BL SNe without observed GRBs and 13 GRB-SNe. A list of these SNe may be found in Table \ref{tab:sampleIcBL}. 

\subsection{Sample bias}
   The ordinary type Ic-BL supernovae presented in this study were generally discovered  by magnitude limited surveys such as the \textit{ZTF Bright Transient Survey} \citep{Fremling.2020}, which conducts a magnitude limited survey down to $m_{peak} \leq$ 18.5 mag. This results in the sample containing a larger fraction of bright supernovae than exists in reality within the volume sampled by the magnitude limit. This magnitude cut is made more severe by the requirement for these SNe to have spectroscopic confirmation (which is more likely for brighter SNe) and for there to be several spectra with sufficient signal-to-noise ratio to measure a velocity (which requires either a brighter SN or a longer observation/larger telescope).

   There are some benefits to the abundance of bright supernovae in the sample, in that it is often easier to fit the early lightcurve of bright supernovae to determine their $t_0$, since there are likely more observations from surveys earlier in their evolution. This can help to mitigate the issues around uncertainty on $t_0$, including its influence on parameter estimation.

   Another aspect of including an excess of bright SNe is that the brightness of a type Ic-BL supernova is directly related to the amount of nickel produced in the explosion \citep{Arnett.1982}. In GRB-associated SNe, simulations have shown that much of this Nickel production takes place on the jet-axis \citep{Shankar.2021}, possibly making the supernova brighter along this axis. Consequently, an over-abundance of bright, survey detected SNe might bias the sample towards type Ic-BL SNe with a polar viewing angle. This has implications for the interpretation of an absence of a detectable GRB jet in many of these explosions.

\subsection{Redshift distribution}\label{sub:redshift_distribution}

\begin{figure}[t]
    \centering
    \includegraphics[width=\linewidth]{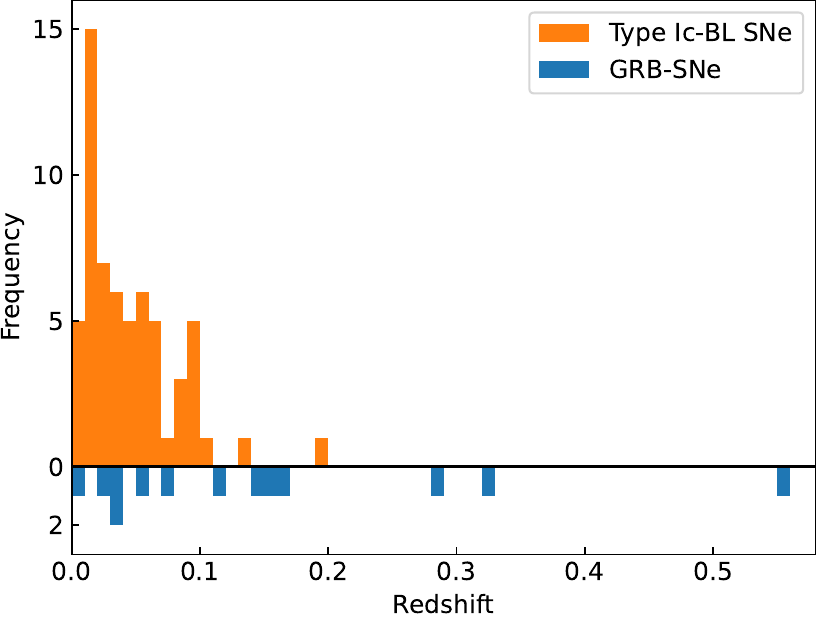}
    \caption[Redshift distribution of GRB-SNe and type Ic-BL SNe.]{Redshift distribution of the supernovae used in this analysis. The most distant supernovae are GRB-SNe, as a result of targeted follow-up campaigns after the discovery of their associated GRB. In contrast, the bulk of the type Ic-BL SN population are discovered within redshift 0.1.}
    \label{fig:redshiftdist}
   \end{figure}

   Figure \ref{fig:redshiftdist} shows the redshift distributions of of ordinary type Ic-BL SNe and GRB-SNe. Ordinary type Ic-BL SNe are detected up to $z\sim$0.2, while GRB-SNe range in redshift up to $z\sim$0.6. This is a result of differences in how these events are discovered. Ordinary type Ic-BL SNe tend to be discovered in magnitude limited surveys, which imposes a redshift limit on the SNe that can be detected. In contrast, GRB-SNe are normally discovered during optical follow-up of a GRB trigger. Knowing where and when to look for the supernova allows us to detect type Ic-BL SNe associated with GRBs at greater distances than those SNe detected by surveys.

   With the differences in redshift distributions of GRB-SNe and ordinary type Ic-BL SNe comes the potential for differences in the environmental conditions of these events, in particular metalicity. However, several studies indicate that the environments of long-GRBs and type Ic-BL supernovae are very similar \citep{Sanders.2012a,Japelj.2018,Modjaz.2008}. As a result, the difference in redshift distributions is not expected to significantly effect any conclusions drawn here. The impact of redshift on the measured spectral velocities has been accounted for by de-redshifting each spectrum before computing the velocity.
    
   \subsection{Data download}
   The supernova spectra and metadata were downloaded from WISeREP and the GRBSN webtool. We also downloaded spectra for GRB180427A-SN2018fip from the ESO Archive\footnote{\url{https://archive.eso.org/cms.html}}. Spectra from WISeREP were downloaded using the WiseRep API\footnote{\url{https://pypi.org/project/wiserep-api/}} \citep{wiserepapi}. This is a third party tool which was modified to provide both the spectra and their metadata from WISeREP and was also used to partially automate the SNID classifications discussed previously. 

   Several instances of duplicate spectra were found in the data downloaded from WISeREP. This was common for telescopes that maintain their own database. WISeREP automatically uploads new spectra from these repositories. In some cases the original observers re-uploaded these spectra following publication of the full dataset, resulting in duplicate spectra. There was also significant overlap between GRBSN Webtool data and WISeREP data, because both tools contain data sourced from the supernova literature.

   To deal with duplicate and similar spectra, a visual examination was performed for all pairs of spectra with the same observation date. This step was also used to remove any near-identical spectral pairs. For example, when galaxy-subtracted spectra were available these were retained instead of the original spectrum. If emission line removal or de-reddening of the spectrum was detected for one of the spectra in a pair, the spectrum with the highest signal to noise ratio and least evidence of artefacts was retained. We found 232 duplicate spectra via this search. These spectra were excluded from velocity measurements. The final sample studied here includes 875 unique spectra, which is the largest to-date for type Ic-BL SNe.
    
    In the case of GRB-SNe, the GRB trigger time was sourced from the GRBSN webtool and used as the explosion time (${t_0}$) (the GRB trigger time is typically accurate to within a few seconds). For non-GRB type Ic-BL SNe, the explosion time needs to be calculated from the rising of their light curve using a low-order polynomial fit. Thus, the precision of ${t_0}$ in these cases depends on how well sampled the light curve is. When available, explosion times were sourced from the literature. Otherwise, ${t_0}$ was estimated from the light curve using publicly available photometric data. If the resulting fit produced uncertainties larger than 10 days, or in the absence of accessible photometric data, the explosion time was assumed to be the halfway point between a non-detection and the first detection, or the first detection itself depending on the available information.

\section{Methods}\label{sec:methods}
\subsection{Visual examination}
    A visual examination was performed for all spectra. The purpose of this review was to remove corrupted data, as well as those which show nebular features or are dominated by continuum emission. Supernovae of type Ic-BL enter the nebular phase on a timescale of a few months. During this transition, the optical depth of the supernova decreases due to its continued expansion, and recombination of elements in the ejecta, rendering it transparent to optical radiation. As a result, the supernova spectrum transitions from an absorption dominated regime to an emission dominated one \citep[see e.g.][]{Sahu.2018}. In a nebular spectrum, the minimum wavelength of absorption features may be shifted due to contamination by flux from adjacent emission lines, or the feature may disappear entirely. To avoid any contamination of the results for velocity with spectra from the nebular phase, each spectrum was examined for nebular features prior to fitting. This avoids any bias that may be introduced by filtering the spectra based on an arbitrary time when the nebular transition is expected to happen. Likewise, spectra which appear to be dominated by continuum are removed, since they have no clearly identifiable features to fit. This is often the case during the early evolution of a GRB-SN, when the contribution of the GRB afterglow is significant.

\subsection{Redshift correction and de-reddening}
   The supernovae analysed in this work span a wide range of redshifts. The expansion of the universe causes spectral features formed in the SN frame to appear at lower velocities in the observer frame. To account for this effect, we divide the observed wavelengths of our spectra by $1+z$, where $z$ is the redshift of the source.

    To perform this correction we use the reported redshift on WISeREP/GRBSN webtool. We initially assumed that the downloaded spectra had not been corrected for redshift, however we then noticed that some spectra from these sources had already undergone redshift correction. If a spectrum that has already been redshift-corrected is corrected a second time, our measured velocities would appear larger than their true values. To try to avoid this issue we performed a visual examination of each spectrum. When we found evidence of prior redshift-correction, we flagged this spectrum to ensure that the correction was not applied a second time.

   One marker we used to determine whether a spectrum had already been redshift-corrected was the 6563\,\AA\,\,H$\alpha$ host-galaxy line. This line was chosen because it can be easily identified, and several of our SNe exhibit this line ($\sim$40\%, though we note that it may have been removed from some spectra prior to their publication). If this line was centred at its rest-wavelength in the downloaded spectrum, then the spectrum has already been corrected for redshift. Since this check was performed visually, it is only capable of catching larger discrepancies. 

   In cases where there was no H$\alpha$ line from the SN host, we compared our redshift-corrected spectrum with the redshift-corrected spectrum in the published literature. When easily identified features were similarly positioned in both the published spectrum and our redshift-corrected spectrum, we considered our spectrum to have been properly redshift-corrected.

   In some cases we were unable to verify whether or not redshift corrections had already been applied to a spectrum. This often happened due to a lack of obvious emission features or because there were no publications available as a cross-check. In these instances we applied the redshift correction ourselves before performing a visual examination of the spectral sequence. With this examination we ensured that similar spectral features remain aligned across the dataset throughout the spectral evolution. This allowed us to unearth some instances in which spectral features were shifted with respect to the same features in the rest of the spectral sequence. This generally indicates that this spectrum had already been redshift-corrected.
    
    As the supernova light propagates through its host galaxy and the Milky Way, it is scattered and absorbed by dust grains in a process known as reddening. A correction for this reddening effect may be applied to the spectra of an SN, for example by following the prescription of \cite{CCM.1989}. However, de-reddening typically scales the continuum of the spectrum, rather than affecting the minimum flux of absorption features, which is a function of the feature's velocity. \cite{Liu.2016} found that there was no difference between the de-reddened velocity and the velocity before reddening in their supernovae sample. In this analysis, no correction for reddening was performed, though it should be noted that some SNe may have been de-reddened prior to being uploaded to WISeREP.
    
\subsection{Emission line removal}
    Emission lines from the host galaxy or the sky, as well as cosmic ray artefacts, are present in some of the sample spectra. These may make it difficult to isolate the continuum, and can affect the location and shape of the absorption lines. Additionally, spectral smoothing is also affected by these phenomena. Since the accurate determination of the wavelength at which the spectral features reach their minimum flux is critical for this analysis, any emission lines in the spectra needed to be removed prior to the fitting of the features. Note that we have not explicitly removed telluric features from our spectra. These absorption features are likely not an issue in the Fe II region of the spectrum but may contribute to the Si II feature. In some cases these lines will have been dealt with during the spectral reduction.
    
    Typically, emission lines are removed during the initial processing of the observational data. However, for most of the SNe in this sample, the original files are not available, and it would be impractical to perform a full data reduction analysis for such a large dataset. For this reason, an interactive interpolation-based emission line removal code, called \texttt{emlineclipper}\footnote{This code is freely available on Github \url{https://github.com/GabrielF98/emlineclipper}}, was developed. This method can be applied directly to the reduced one-dimensional spectra.

\begin{figure}
    \centering
    \includegraphics[width=\linewidth]{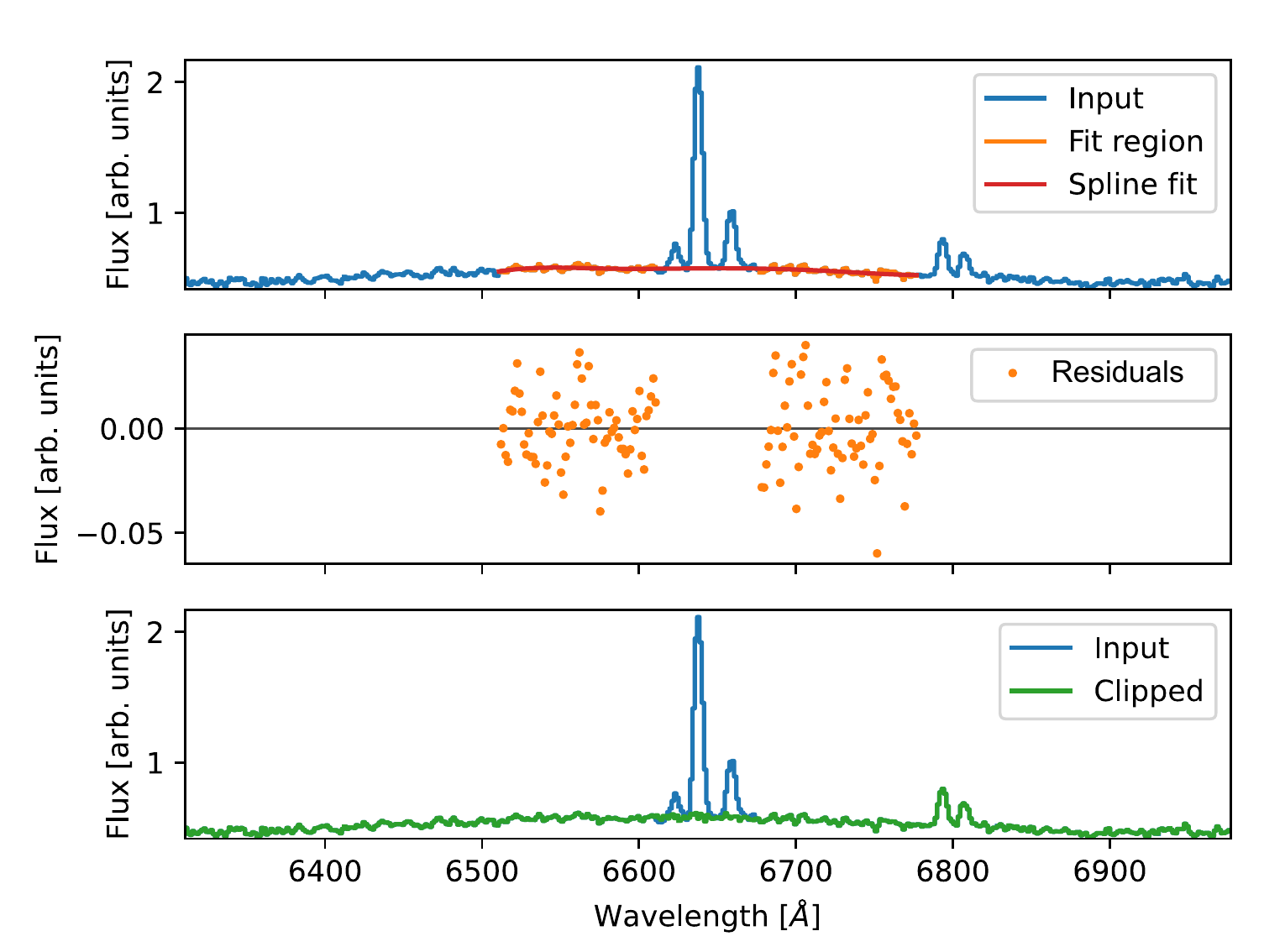}
    \caption[Example of emission line removal using \texttt{emlineclipper}.]{Emission line removal using the \texttt{emlineclipper} Python code. Line selection is performed manually, by selecting the bounding region around the emission lines. The data in a small range either side of the emission line region are fit using a spline. The mean and standard deviation of the spline fit residuals in this region are used to resample the flux in the emission line region.}
    \label{fig:emclip}
\end{figure}

    The \texttt{emlineclipper} program first displays the input spectrum to the user, allowing them to choose boundaries that bracket the emission features that they wish to remove; this range is known as the `bounding region' and is demarcated by the `bounding lines'. The code then fits a spline to the data in two 100\,\AA\,\,windows either side of the bounding region. The residuals between the spline fit and the input spectrum are computed within these windows, followed by computation of the mean and standard deviation of the residual array. These values are then used to resample the spectrum by adding noise to the spline within the bounding region. Results from emission line removal using this code are shown in Fig. \ref{fig:emclip}; the residuals are comparable in magnitude to the noise in the spectrum, and the resampled spectrum follows the evolution of the continuum near the emission lines very closely. The program can also handle cases of multiple nearby emission lines.

\subsection{Smoothing and mitigation of error sources}\label{sec:smoothing}
    Online repositories host spectra from a wide range of instruments and span multiple decades of observations. Thus the dataset contains spectra of varying resolutions and signal to noise ratios. In order to increase the signal to noise ratio, prior to fitting the spectra were smoothed using a Savitzky-Golay (SG) filter \citep{SG.1964}. This increases the likelihood that the minimum wavelength is recovered for an absorption feature, especially if the input data is noisy. An SG filter has two tunable parameters: the degree of the polynomial and the window length. A cubic polynomial was used to reduce the risk of over-fitting the data, especially near the peaks. As there is no formal method to determine the ideal window length, a statistical analysis was conducted to determine the optimum window length \citep{Finneran.2024C}. This analysis suggests that the optimum window length is between 2-5\% of the spectrum length in resolution elements. This limit proved sufficient to avoid loss of information near the peaks when smoothing spectra in the majority of cases. The fine-tuning of smoothing parameters in this method typically introduces an additional uncertainty of $\sim$500-1000\,km/s \citep{Finneran.2024C}.
    
    Despite efforts to reduce the distortion of the spectrum during smoothing, the results are somewhat sensitive to the exact choice of smoothing parameters. Additionally the fitting of the absorption region may be sensitive to the choice of the endpoints of this region. These effects can be minimised by performing the velocity measurement for a large sample of Monte-Carlo noised spectra and averaging the result. 
    
    In general, the spectral uncertainty arrays, which are required for the Monte-Carlo analysis, are not provided by online repositories. These arrays are produced during the sky-subtraction phase of spectral analysis, and cannot be reconstructed without the original data. A method to produce pseudo-uncertainty arrays is described in Appendix \ref{sec:linemethod}. One benefit of the Monte-Carlo approach is the ability to simultaneously quantify the uncertainty introduced by the signal to noise ratio of the data, the choice of the fitting region and the uncertainty introduced by the fitted function.

\subsection{Feature identification}\label{subsec:ident}
   Measuring the velocity of a spectral feature requires a high degree of confidence in the identification of the species responsible for the feature, and hence the rest-frame wavelength used during analysis. Rather than carrying out detailed modelling of the spectra for each SN, line identifications were made based on established patterns within spectra, in combination with comparison to the identifications in the literature where available.

   Feature identification is important for each of the three features that dominate the type Ic-BL spectrum. The Fe II feature is well known to be a blend of at least three spectral lines, with rest-frame wavelengths of 4924\,\AA, 5018\,\AA\,and 5169\,\AA, \citep[e.g.][]{Modjaz.2016}. Depending on the chosen rest-frame wavelength, the Fe II velocity may differ by up to 15000 km/s once the line becomes de-blended later on in the evolution (c.f. \cite{Prentice.2018}). As it is not clear which line is responsible for the minimum in any given spectrum, the 5169\,\AA\,line was chosen in order to maintain consistency with previous studies. Although \cite{Modjaz.2016} developed a method intended to counteract the effects of Fe II blending, they found it to be inapplicable to the Si II feature. Similar methods to those used here have been applied to GRB-SNe and ordinary type Ic-BL SNe before, with the benefit that they can be applied to all the spectral features that we wish to investigate.  In cases where the Fe II feature becomes de-blended, the line which most smoothly continues the decaying velocity trend was followed; this choice was informed by reviewing the overall spectral evolution. This avoids having a discontinuity in the velocity evolution as soon as the lines de-blend \citep[e.g.][]{Prentice.2018}. Applying this identification in a consistent manner to all SNe in the sample should facilitate comparison of the evolutionary trends between SNe without significant issue.

   In the case of the Si II feature, the canonical rest-frame wavelength is 6355\,\AA. However, it has been proposed that blending with the Na I line may be an issue, as observed by \cite{Modjaz.2016} for PTF10qts. There is also an ongoing debate as to whether the feature near 6100\,\AA, commonly identified with Si II, is in fact entirely due to silicon, or whether it is contaminated by other elements including H$\alpha$ in absorption \citep{Parrent.2016}. For Ic-BL SNe the source of this hydrogen would likely be in the material expelled by the WR progenitor star. Once again consistency in identification is the key to studying the evolution.
   
   The Ca II feature is also a triplet, consisting of lines at 8498\,\AA, 8542\,\AA, and 8662\,\AA\,\citep[e.g.][]{Rho.2021}. Following \cite{Rho.2021}, a mean wavelength of 8567\,\AA\,was used as the rest-wavelength for this feature. De-blending of this triplet is not commonly observed for Ic-BL SNe. Indeed, as this triplet is in the near-infrared range of the spectrum it has rarely been studied for a large population of type Ic-BL SNe. Another complication with the Ca II feature is its proximity to the O I 7774\,\AA\,feature, with which it may be blended. Although we attempted to fit this feature, we were not able to obtain robust fits and thus omit these fits from the present analysis.

\subsection{Velocity determination}\label{subsec:velocity} 
\begin{figure}
    \centering
        \centering
        \includegraphics[width=\linewidth, trim={2cm 1cm 1cm 0}]{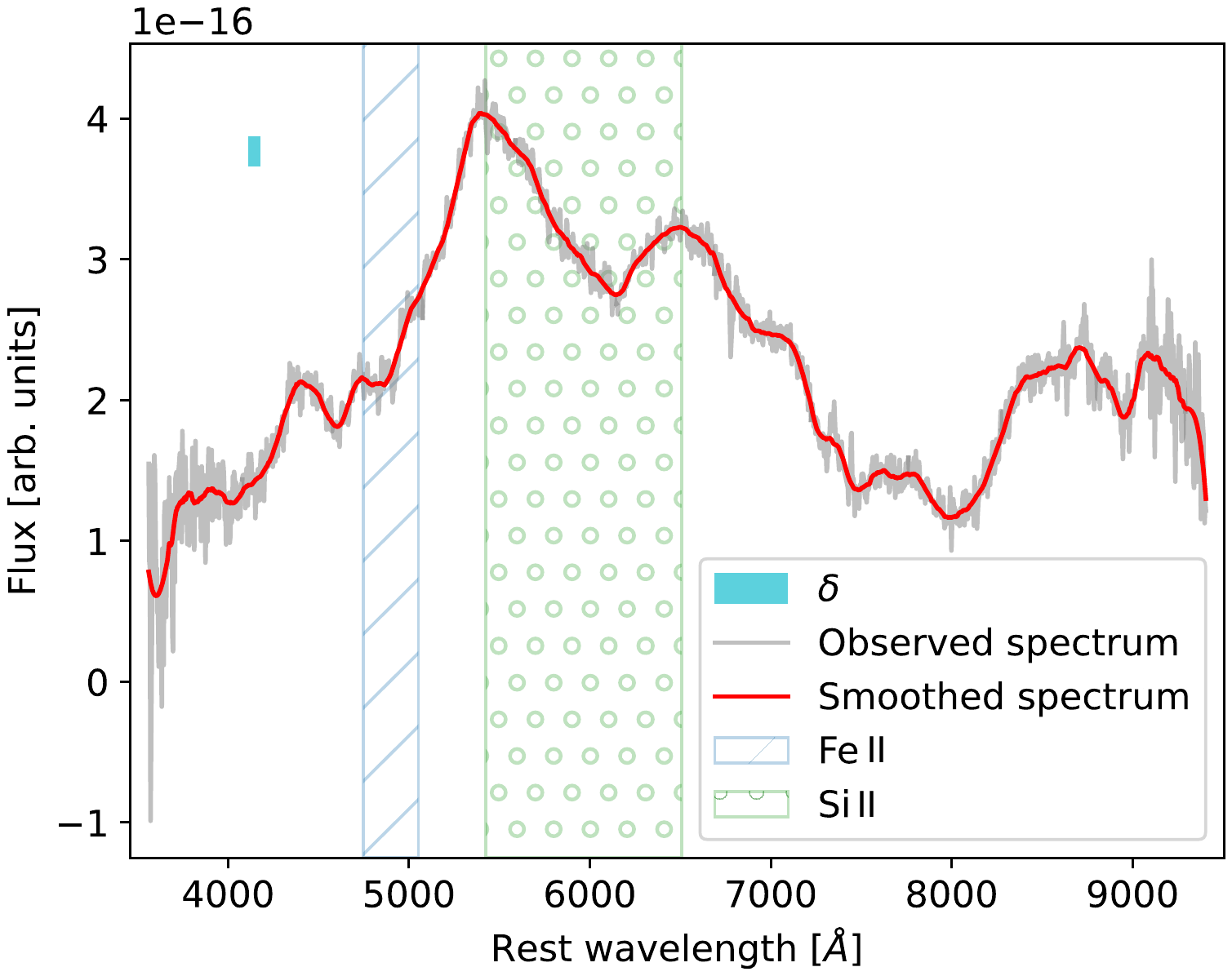}
        \caption[Feature selection for velocity determination.]{Selection of the initial boundaries for the Fe II (5169\,\AA) and Si II (6355\,\AA) features in a spectrum of SN2020bvc. The \textit{``observed spectrum''} (grey) is the raw SN spectrum following redshift and emission line corrections. The \textit{``smoothed spectrum''} (red) is the SG-filtered observed spectrum. The hatched Fe and Si regions denote the user-selected regions in which to perform the spline fit for these features. The final boundaries of the fitting regions are allowed to vary by $\delta$ resolution elements with respect to the initial boundaries.}
        \label{fig:featselect}
        
     \end{figure}

\begin{figure}
    \centering
        \centering
        \includegraphics[width=\linewidth]{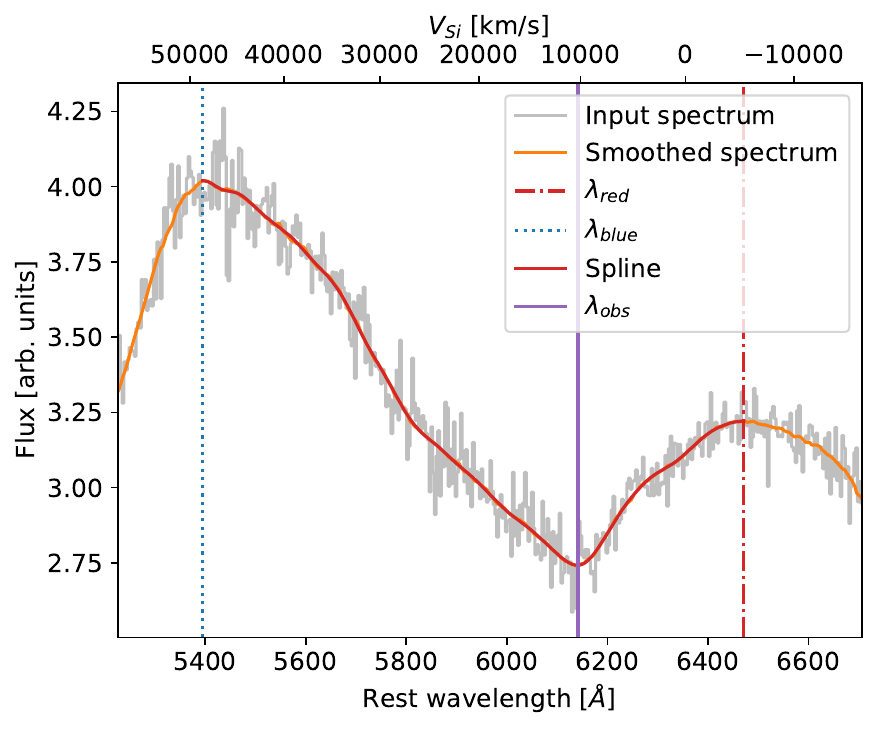}
        \caption[Spline fitting example.]{Spline fit to the Si II feature of 2020bvc. The wavelength at which the spline fit reaches its minimum value is used in the Doppler formula to compute the expansion velocity of the feature. The \textit{``input spectrum''} (shown in grey) was generated by adding random noise to the SG-filtered observed spectrum (i.e. the reduced 1-D spectrum from the source catalogue). The solid orange line represents the \textit{``smoothed spectrum''}, created by applying an SG filter to the input spectrum. A solid red line illustrates the \textit{``spline fit''} applied to the smoothed spectrum between the wavelengths $\lambda_{blue}$ and $\lambda_{red}$. The wavelength at which the spline fit reaches its minimum is taken as the observed wavelength of the feature and is denoted $\lambda_{obs}$ (purple line).}
        \label{fig:splinefit}
     \end{figure}

    The velocities of the three absorption features discussed above were determined using the method discussed in Appendix \ref{sec:linemethod}. This method is very similar to that used by \citet{Silverman.2012} for type Ia supernovae. Similar methods have also been applied to SESNe, including those with associated GRBs \citep[e.g.][]{Liu.2016, Patat.2001q8t}. Selection of the fitting region was performed manually for all features as shown in Fig. \ref{fig:featselect}. Once the observed wavelength has been determined, Eq. \ref{eq:doppler} is used to compute the expansion velocity. An example spline fit is shown in Fig. \ref{fig:splinefit}. The fit is performed on each of the Monte-Carlo spectra and a histogram of velocities is generated. The median and 16th and 84th percentile errors are quoted for all velocities presented in this analysis. 

     A potential source of error is introduced in the case of noisy spectra, where the smoothing by the SG filter may not be sufficient to produce a smooth feature with a clearly defined minimum. While more aggressive smoothing could be performed, this risks removing information from the spectrum. Instead, the smoothness of the cubic spline used in the fit may be adjusted by changing the number of knots used in the spline (hereafter ``knot density''). An analysis of the stability and uncertainty of the measured velocity with different knot densities shows that the optimal knot density is $\sim$10\% of the number of resolution elements within the limits of the feature \citep{Finneran.2024C}. This value was adopted for the majority of spectra in this analysis.
     
\begin{figure}
    \centering\includegraphics[width=\linewidth]{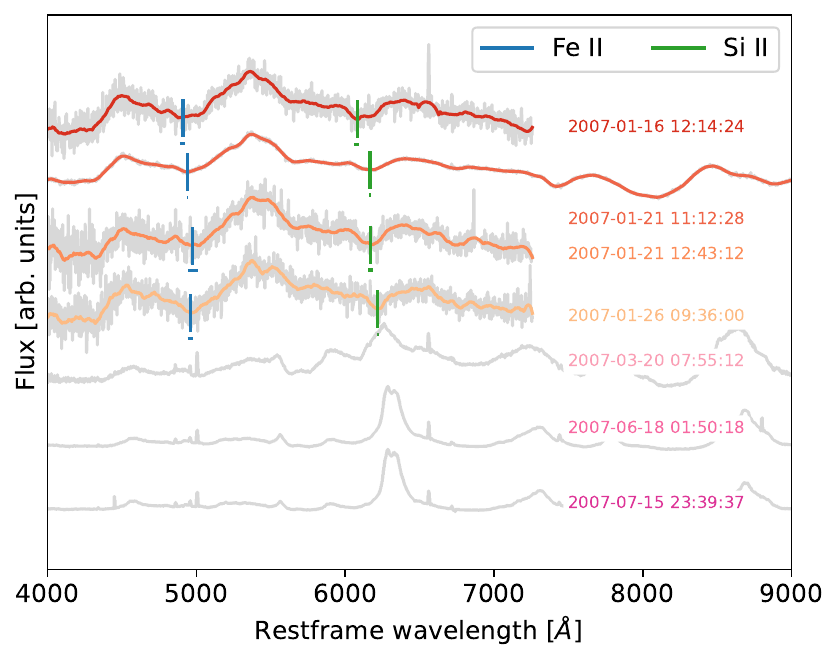}\caption{Spectral sequence of the type Ic-BL supernova SN2007D. The observed spectra (grey) and smoothed spectra (multiple colours) are plotted as solid lines. Also shown are the determined minimum wavelengths of the Fe II (blue) and Si II (green) features, as well as the corresponding uncertainty on the minimum wavelength, which is influenced by the S/N of the spectrum. A trend of increasing minimum wavelength with time is visible for Fe II and Si II; this is a direct result of the deceleration of the supernova ejecta. This plot was generated for each supernova in the sample and was used to verify the feature selection and to classify nebular spectra.  The last three spectra were classified as nebular, as they have a relatively flat continuum and a strong emission feature near 6300\,\AA. These spectra were obtained from WISeREP and come from \cite{Modjaz.2014, Shivvers.2019}.}
        \label{fig:specfit}
    \end{figure}

    Figure \ref{fig:specfit} shows an example of visual examination carried out on the results of the spline fitting. Due to the heterogeneity of the data, this is an important step to confirm that the correct feature has been identified in all spectra for each SN. In cases where some misidentifications or poor fits were noticed, the fits were iteratively improved by: changing the delta parameter; selecting different wavelengths for the bounding maxima so that the fitting region was better constrained; refitting the feature using a spline with an increased knot density; or increasing the spline's resolution.
    
\section{Results}\label{sec:results}

The expansion velocities for each feature measured from the spectra can be found in Table \ref{tab:silverresultsIcBL} and  Table \ref{tab:silverresultsGRB}. In all cases, the velocities are only reported for spectra younger than 60 days post explosion (rest-frame time), in order to avoid the nebular phase, where typical assumptions about the formation of the Fe II and Si II features, and homologous expansion, break down.

\subsection{Trends in the velocity evolution of GRB-SNe and Ic-BL SNe}\label{sec:trends}
   
\subsubsection{Fe II}
\begin{figure*}
   \centering
   \includegraphics[width=0.49\linewidth]{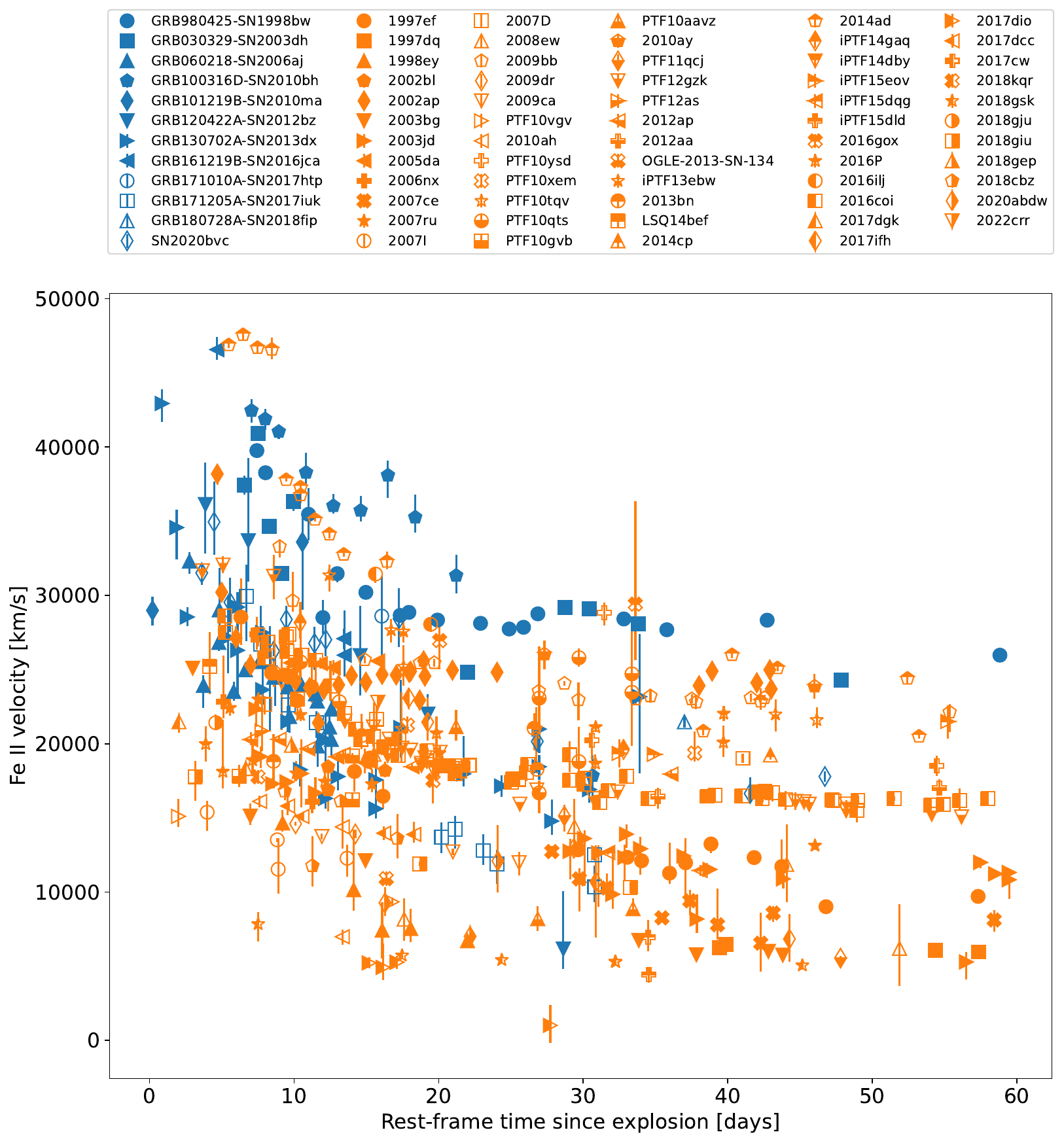}
   \includegraphics[width=0.49\linewidth]{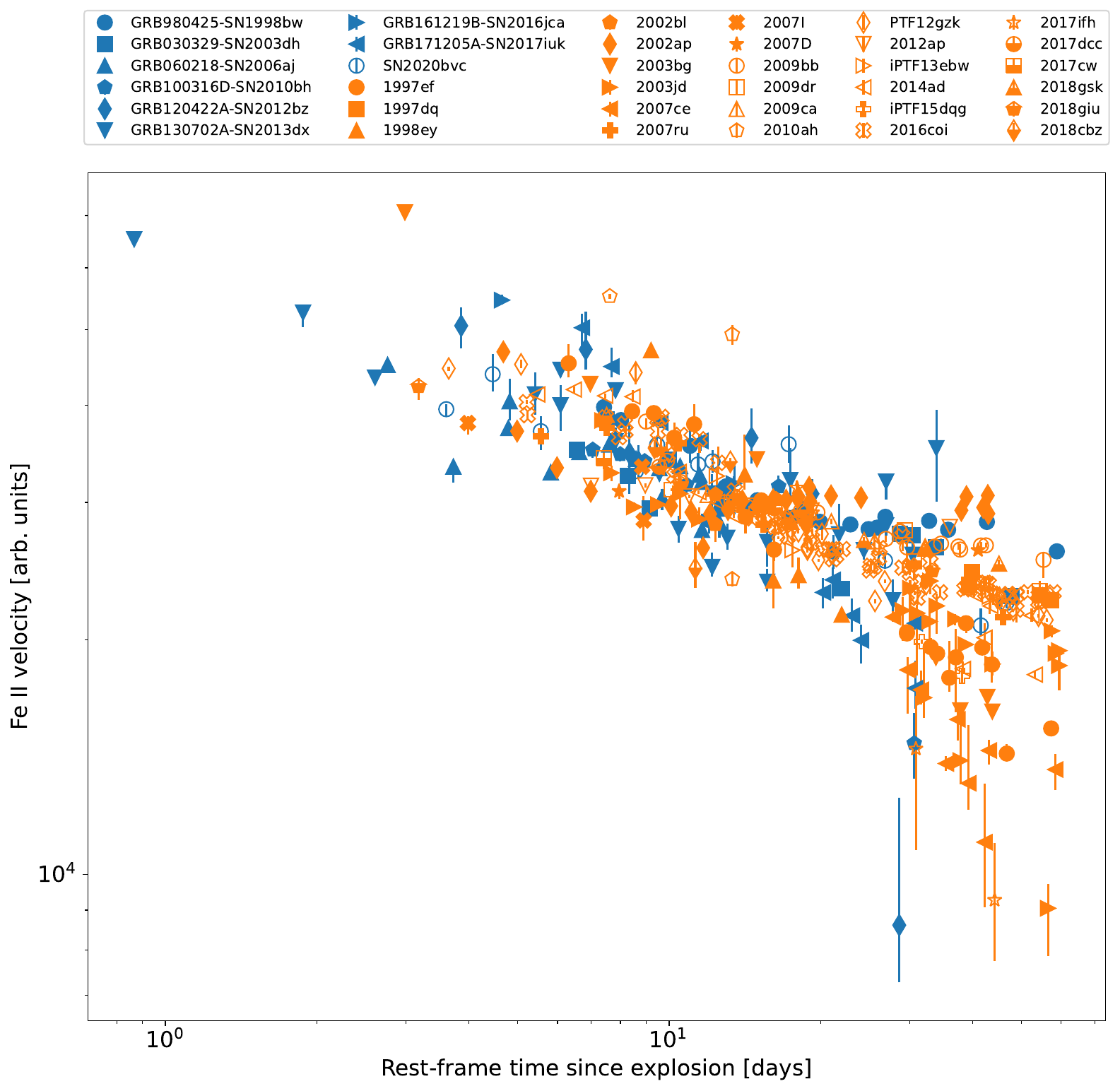}
   \caption[Velocity evolution of the Fe II feature.]{Velocity evolution of the Fe II feature. \textit{Left:} Plot of all SNe in the sample for which velocities were measure for Fe II. All SNe follow a rapid decay at early times (< 20 days) before beginning to plateau. A continuum of velocities exists for ordinary type Ic-BL SNe and GRB-SNe. \textit{Right:} A selection of supernovae for which the velocity evolution has been fitted. The majority show a power-law trend, with others showing broken power-law trends. There are no evolutionary differences between the ordinary type Ic-BL SN and GRB-SN samples. The velocities have been scaled to highlight the overall trend. The scaling factor is computed from the fitted velocity at 15 days divided by the fit velocity of GRB980425-SN1998bw at 15 days.}
              \label{fig:feIIoverall}
    \end{figure*}

Figure \ref{fig:feIIoverall} shows the Fe II velocity plotted against the rest-frame time since the supernova explosion\footnote{To view individual SNe more easily, multiple similar figures with fewer SNe per plot are presented in Appendix \ref{sec:additional_velocity_evolution_figures}}. The velocity of this line was measured in at least one spectrum for 12 GRB-SNe and 59 Ic-BL SNe. The absence of an Fe II velocity measurement for a particular spectrum can be caused by: cases of non-trivial line identification, either due to lack of reference line identification in the literature or multiple features in the region near Fe II; cases where the wavelength range of the SN spectrum did not include the Fe II region;  cases where the Fe II feature did not have a well defined minimum, either as a result of noise or there being no minimum value in this region of the spectrum; and cases in which the removal of an emission line altered the minimum of the feature so dramatically that clipping and smoothing were not feasible.

The left panel of Fig. \ref{fig:feIIoverall} shows that velocities of both GRB-SNe and ordinary type Ic-BL supernovae decline for 15-20 days prior to entering a plateau phase.  There does not appear to be a single velocity which separates GRB-SNe from ordinary type Ic-BL supernovae, as indicated by the significant scatter among the velocities of both populations, which suggests that a continuum of events likely exists, with some GRB-SNe expanding more rapidly than ordinary type Ic-BL SNe. This is quantified in statistical terms in Sect. \ref{sec:compare15dayvels}. Additionally, supernovae that begin at high velocities tend to plateau at high velocities; this might indicate that the plateau is intrinsic to the expansion of a GRB/SN Ic-BL, or indeed to SNe in general, rather than being a distinguishing feature of either class. 

Figure \ref{fig:feIIoverall} also shows that some SNe (e.g. GRB980425-SN1998bw) exhibit a rise in velocity during their evolution around 30 days post explosion. It is not clear what has caused this fluctuation in velocity. For all SNe presented in our plots of velocity we verified from the spectral sequences that our fits give reasonable results for the locations of feature minima during this time period for all SNe. This type of behaviour has been observed in the analysis presented by \cite{Modjaz.2016} (their Fig. 5) and \cite{Taddia.2019} (their Fig. 15), however no physical rationale for this was given by either paper. This trend is highlighted in this analysis because the uncertainties of the method are small with respect to these fluctuations.

The right panel of Fig. \ref{fig:feIIoverall} reveals that a large fraction of the supernova sample, both type Ic-BL SNe and GRB-SNe, follow power-law decays. It also shows that the velocities, though scaled, still have some scatter, around 1000 km/s at ${t_0}$+10 days rest-frame. The scatter becomes larger after 15-20 days partially due to some SNe showing evolutions consistent with a broken power-law decay. These decays tend to follow a steep-shallow evolution. Examples of likely broken power-law decays are GRB980425-SN1998bw, SN2002ap and SN2009bb. The scatter at late times could also be due to the effects of line blending \citep[see][]{Prentice.2018}. The lack of early observations for many SNe makes it difficult to determine whether power-laws emerge early in the evolution, or appear later.
    
\subsubsection{Si II}
 
\begin{figure*}
   \centering
   \includegraphics[width=0.49\linewidth]{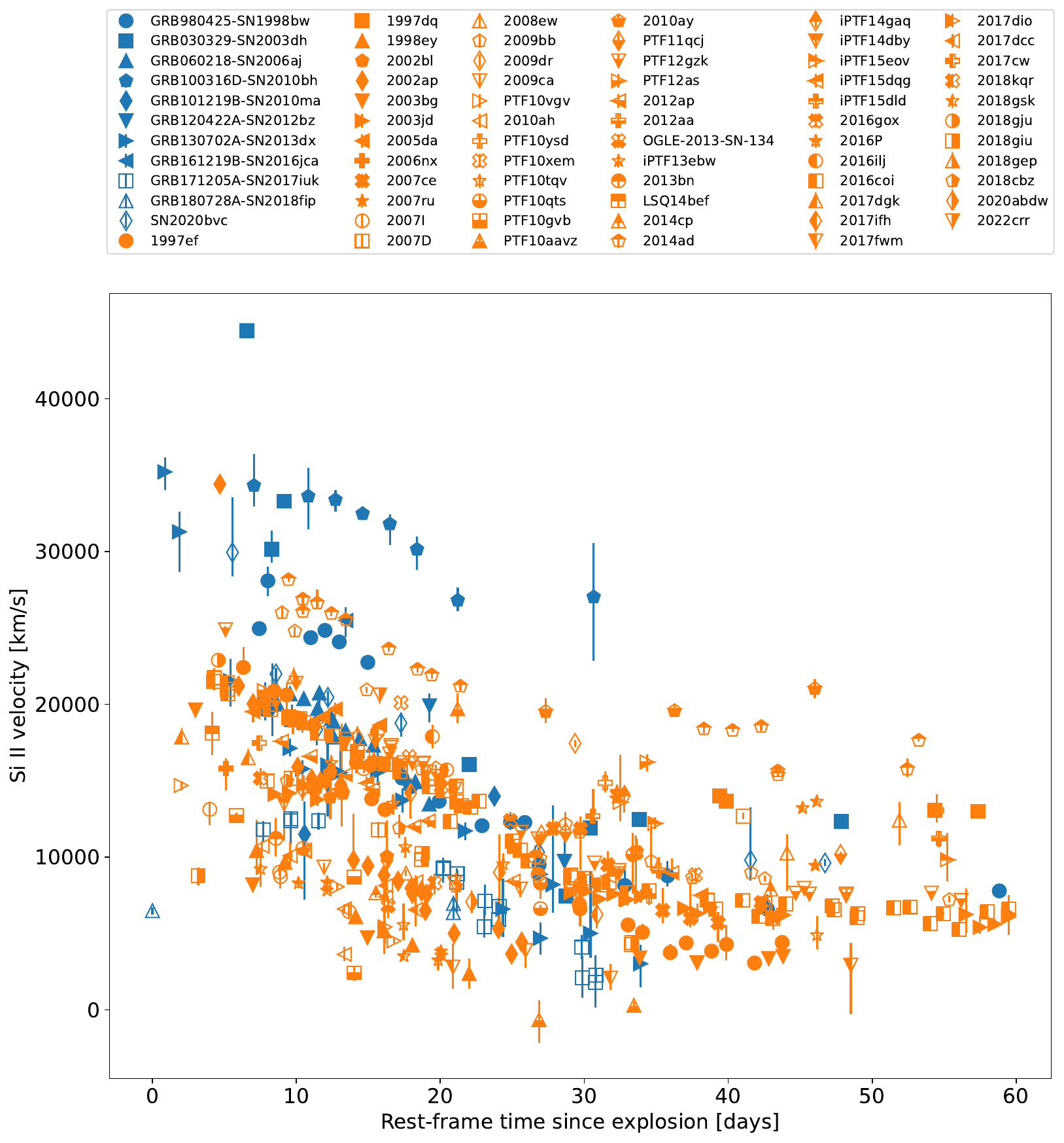}
   \includegraphics[width=0.49\linewidth]{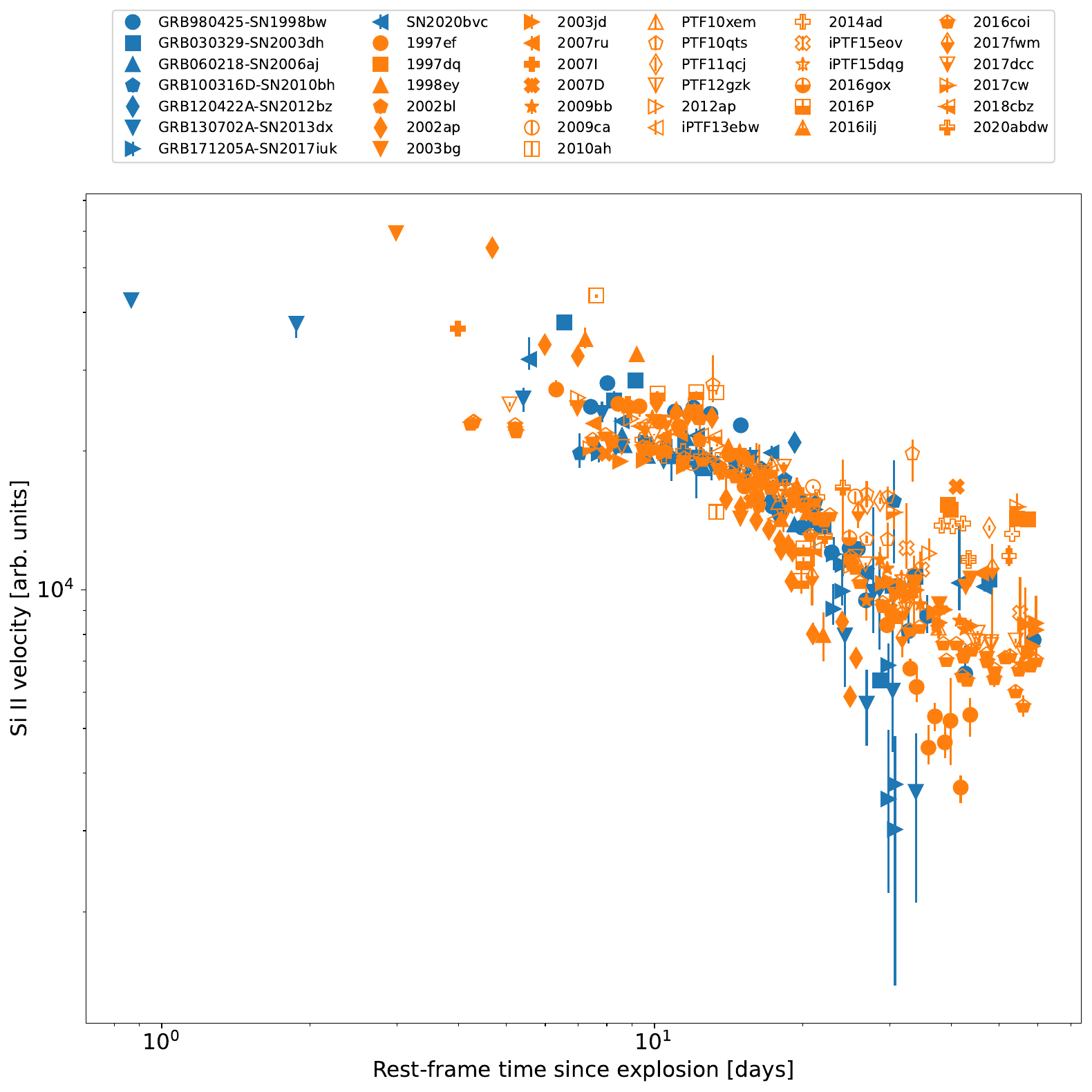}
   \caption[Velocity evolution of the Si II feature.]{Velocity evolution of the Si II feature; same sample criteria and graphs as Fig. \ref{fig:feIIoverall}. \textit{Left:} Many ordinary type Ic-BL SNe and GRB-SNe evolve rapidly at early times (< 20 days); following this they then begin to plateau. A continuum of velocities exists for ordinary type Ic-BL SNe and GRB-SNe. GRB100316D-SN2010bh has a much higher velocity at all epochs studied than the rest of the sample. \textit{Right:} Several ordinary type Ic-BL SNe and GRB-SNe can be fitted by power-law decays; supernovae which evolve more rapidly tend to have a higher initial velocity and vice-versa. Some type Ic-BL SNe and GRB-SNe show broken power-law decays, with the breaks all being around 15 days.}
    \label{fig:siIIoverall}
    \end{figure*}

Figure \ref{fig:siIIoverall} shows the Si II velocity plotted against the rest-frame time since the supernova explosion\footnote{To view individual SNe more easily, multiple similar figures with fewer SNe per plot are presented in Appendix \ref{sec:additional_velocity_evolution_figures}}. The velocity of this line was measured in at least one spectrum for 11 GRB-SNe and 60 type Ic-BL supernovae.

The left panel of Fig. \ref{fig:siIIoverall} shows a steady decline of the Si II velocity from 20000-25000 km/s to below 10000 km/s in the first 30 days, though there are some SNe with velocities larger than this. Some supernovae appear to show a plateau phase after $\sim$25 days, while others simply keep decaying to very low velocities.  Similar to the evolution of the Fe II feature, there does not seem to be a clear distinction between GRB-SNe and ordinary type Ic-BL SNe. This is investigated statistically in Sect. \ref{sec:compare15dayvels}.

The right panel of Fig. \ref{fig:siIIoverall} shows that, based on their Si II velocity, the supernovae in the sample can be divided between those with broken power-law evolutions and those with power-law decays. Many of those with broken power-law evolution follow a shallow-steep decay, unlike the broken power-laws for Fe II. In the case of SNe with power-law decays, their slopes appear to match the decay rates of the first segment of the broken power-laws quite well. The breaks in the power-laws appear to occur between 10-20 days rest-frame.

There is some scatter in the velocity when viewed in log-log scale, which can be attributed to intrinsic differences between the supernovae. The scatter is similar in magnitude to that found for Fe II over the same time periods, and appears to be similar throughout the evolution, though the effect of the log-log scale serves to visually enhance the scatter after 20 days when it the velocity falls below 10000 km/s.

Figure \ref{fig:siIIoverall} includes four well-sampled high-velocity events: GRB100316D-SN2010bh, GRB980425-SN1998bw, GRB030329-SN2003dh and SN2014ad. The evolution of GRB980425-SN1998bw and SN2014ad seems quite similar, with both declining from a similar initial velocity and plateauing at around 20-22 days. However, the final plateau velocity of SN2014ad is larger by nearly a factor of two ($\sim$9000 km/s vs $\sim$18000 km/s). GRB100316D-SN2010bh appears to have a shallow velocity decrease before 10 days, followed by a steepening of its decay. GRB030329-SN2003dh achieves the highest Si II velocity ($\sim$45000 km/s), which rapidly declines to velocities that are more in line with the general population by day 10.

\subsection{Parameterising the velocity evolution}\label{sec:evoparams}
Motivated by the evolutionary trends visible in the log-log velocities of Fe II and Si II, their velocity evolution was fit with either a power-law or broken power-law function. While this is the first quantitative analysis of velocity evolution for type Ic-BL SNe, similar analysis has been carried out for Ia supernovae by \cite{PLBPL.2017} and for Ib supernovae in \cite{Branch.2002}.

For the power-law model, the function adopted was
\begin{equation}\label{eq:pl}
    v = at^b,
\end{equation}

where $t$ is the rest-frame time since explosion, $a$ is the velocity of the SN one day after the explosion and $b$ is the slope of the velocity evolution; a negative $b$ value implies a velocity that decreases over time.

The broken power-law adopted is similar to those used in GRB afterglow fitting \citep[e.g.][]{AMC.2014} and SN studies \citep[e.g][]{PLBPL.2017}, and can be expressed as:

\begin{equation}\label{eq:bpl}
    v = A \left[\left(\frac{t}{t_b}\right)^{-\alpha_1 s}+\left(\frac{t}{t_b}\right)^{-\alpha_2 s}\right]^{-1/s},
\end{equation}

where $A$ is the SN velocity at the break time, ${\alpha_1}$ and ${\alpha_2}$ are the slopes of the power-law segments, ${t_b}$ is the rest-frame time of the break, and $s$ is a parameter which controls the smoothness of the break.

While the power-law fit was performed for all SNe with at least three velocity measurements, the broken power-law fit was only performed if there were five or more velocity measurements, and only if the data showed clear deviation from the single power-law decay. For each SN fitted with a broken power-law model, an F-test was used to determine whether that model provided a better fit than the simple power-law model, requiring a three sigma confidence level to accept the broken power-law fit.

Marginalisation of the fit parameters was performed using the \texttt{emcee} Python package\footnote{\url{https://emcee.readthedocs.io/en/stable/tutorials/line/}} \citep{Foreman-Mackey.2013}. A basic check for convergence, based on the autocorrelation time of the walkers was performed for each fit, as recommended by the \texttt{emcee} documentation. Although the uncertainties on the computed feature velocities are asymmetric, the maximum error value was used for the fit as a conservative estimate of the uncertainty.

Similarly to \cite{PLBPL.2017}, the value of the smoothness parameter, $s$, was fixed during the fitting process, since the fits seem to be only weakly dependent on it. In general, a value of ${s\,=\,\pm\,20}$ seemed to produce reasonable results for the majority of fits. However, in a few cases, the $s$ parameter had to be changed to $s\,=\,\pm\,5$, indicating a smoother break. The sign of $s$ dictates whether the power-law evolution is steep-shallow or shallow-steep.

\begin{figure*}[t]
      \centering
      \includegraphics[width=0.33\linewidth]{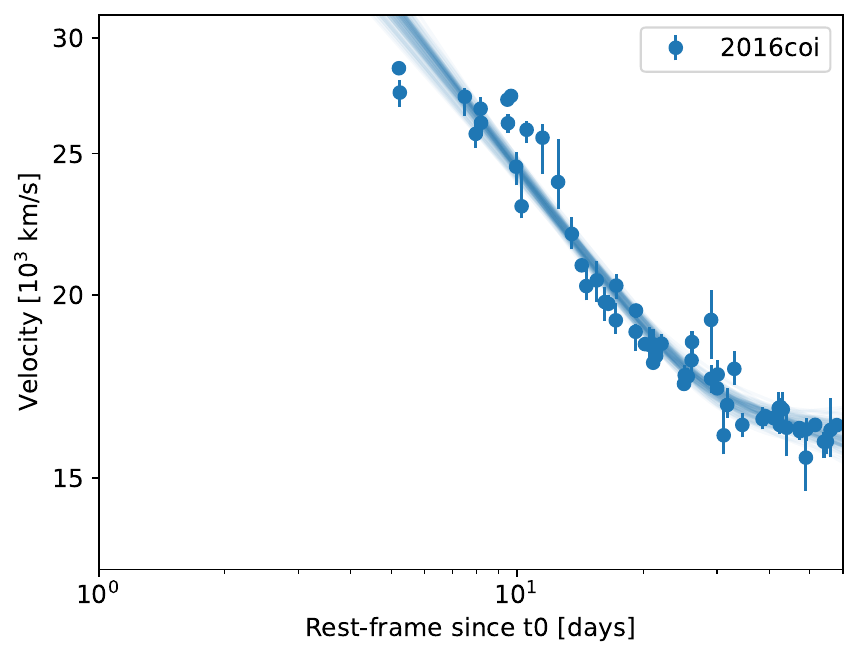}
      \includegraphics[width=0.33\linewidth]{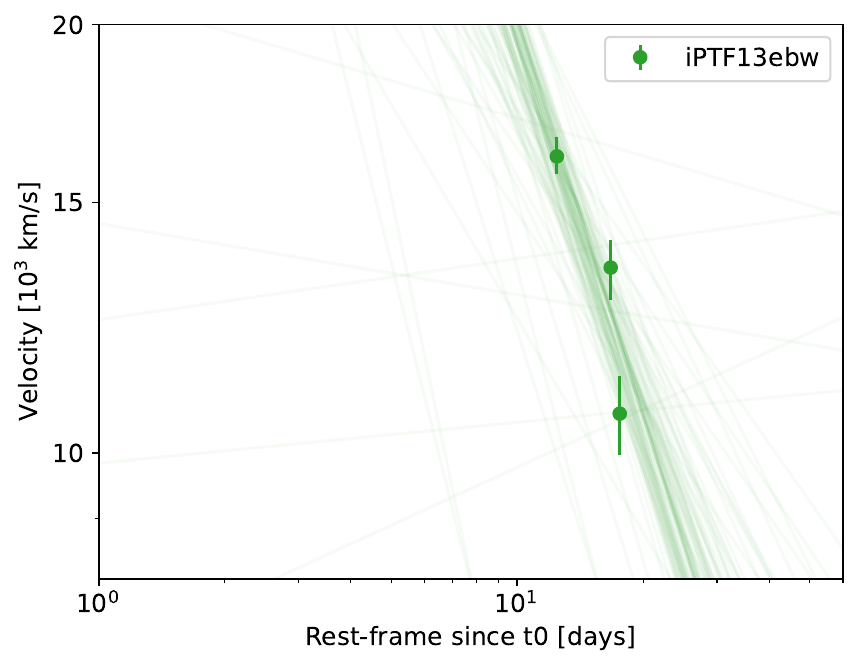}
      \includegraphics[width=0.33\linewidth]{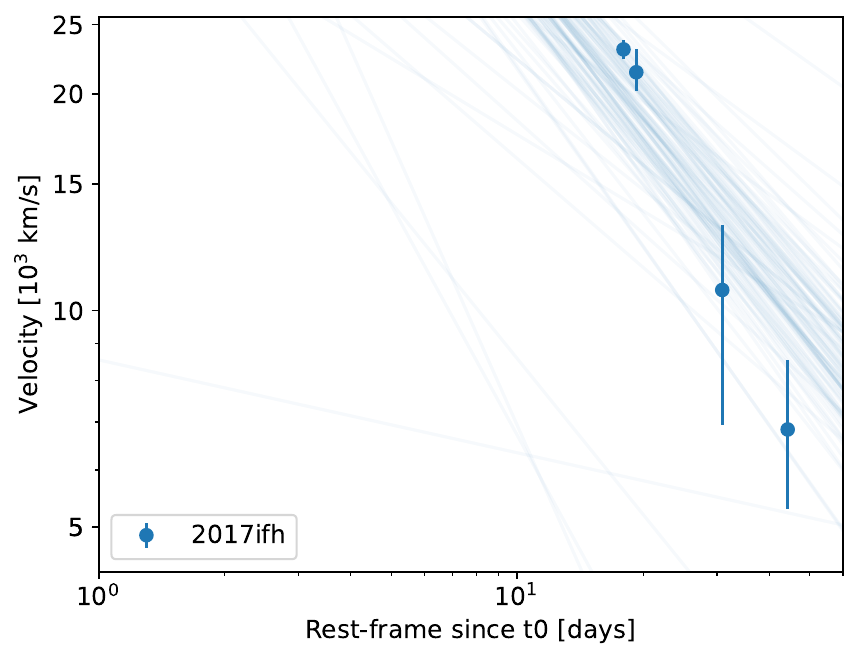}
      \caption{A sample of SNe representative of the \textit{Gold}, \textit{Silver} and \textit{Bronze} samples. This figure shows their velocities as circles and a random sample of MCMC power-law/broken power-law fits as solid lines. \textit{Left:} The ordinary type Ic-BL SN 2016coi, a \textit{Gold} sample SN. \textit{Centre:} The ordinary type Ic-BL SN iPTF13ebw, a \textit{Silver} sample SN due to a fit which did not converge for one of the parameters. \textit{Right:} The ordinary type Ic-BL SN 2017ifh, a \textit{Bronze} sample SN due to the scatter on the data points and the quality of the fit.}
      \label{fig:sampleexamples}
   \end{figure*}

The quality of the fit was estimated based on the posterior distributions generated by \texttt{emcee}. Using this information, the data were cast into one of three samples: \textit{Gold}, \textit{Silver}, or \textit{Bronze}. Figure \ref{fig:sampleexamples} shows examples of SNe belonging to each of these samples, along with their best fits. The best fit curves for each SN may be viewed in the figures presented in Appendix \ref{sec:additional_velocity_evolution_figures}.

\textit{Gold sample}: Fits for which the velocity evolution shows little scatter and which have converged for all parameters\footnote{For GRB-SNe the majority of the fits converged, probably because their evolution tended to be better sampled than that of the ordinary type Ic-BL SNe; consequently the convergence criterion was ignored when determining the sample group for a GRB-SN.}, showing near-Gaussian posterior distributions \textbf{- AND -} Supernovae for which the $t_0$ time is constrained via modelling or non-detections to within 3 days.

\textit{Silver sample:} Fits which have less clear velocity evolution than the \textit{Gold} sample, but which have sufficient number of data points to produce a good fit. The requirements on scatter and error size are relaxed compared with the \textit{Gold} sample \textbf{- OR -} Fits which have not converged for all parameters, but whose corner plots show good marginalisation of the posterior parameter distributions \textbf{- OR -} Supernovae for which the ${t_0}$ time is constrained via modelling or non-detections to within 5 days.

\textit{Bronze sample:} Fits which have a less clear velocity evolution than the \textit{Silver} sample, or very few data points. The requirements on scatter and error size are relaxed compared with the \textit{Silver} sample \textbf{- OR -} Supernovae for which the ${t_0}$ time is constrained via modelling or non-detections with an error of more than 5 days, or where the ${t_0}$ time is taken as 48 hours before the last non-detection.

A final \textit{Exclude} sample was created for SNe that seemed to have no clear evolutionary trend in their data; most often due to a very low number of scattered velocities or unrealistic values for the scaling constants (${a}$ in Eq. (\ref{eq:pl})). This sample was not included in the analysis of the evolution parameters.

 \subsubsection{Fe II evolution parameters}

   \begin{table}[t]
   \caption{Cardinality of the sample groups for the Fe II and Si II features.}\label{tab:cardinality}
   \centering
   \begin{tabular}{lccccc}
   \toprule
    & \textbf{Gold} & \textbf{Silver} &  \textbf{Bronze} & \textbf{Excluded} & \textbf{Total} \\
   \midrule
   \textbf{Fe II}  & 19 & 7 & 10 & 8 & 44 \\
   \textbf{Si II} & 17 & 10 & 12 & 9 & 48\\
   \bottomrule
   \end{tabular}
   \end{table}

   A total of 44 SNe satisfied our criterion for power-law fits in the case of the velocities measured from the Fe II feature\footnote{Although PTF10qts and SN2017dio met the criteria for number of observations, the could not be fit because the large uncertainty on their late-time velocities produced negative values which the fitting pipeline could not handle; they are not included in Table \ref{tab:cardinality}.}. Table \ref{tab:cardinality} shows the number of SNe assigned to each sample for the Fe II evolution fits. More than half of the events are in the \textit{Gold} or \textit{Silver} samples.

   As we will demonstrate in Appendix \ref{sec:testrobustness}, shifts in $t_0$ alter the marginalised fit parameters, particularly the power-law index. The potential magnitude of this effect grows as the uncertainty in $t_0$ increases. Among our samples, the \textit{Gold} sample has the lowest uncertainty in $t_0$ ($<$3 days). Statistical tests presented here are only performed for the \textit{Gold} sample. This will reduce the likelihood that the conclusions drawn from these tests are biased due to uncertainty in $t_0$. All figures shown in this subsection show the data form all three samples (\textit{Gold, Silver} and \textit{Bronze}) to provide an overview of the whole population.

\begin{figure*}[t]
    \centering
    \includegraphics[width=0.49\linewidth]{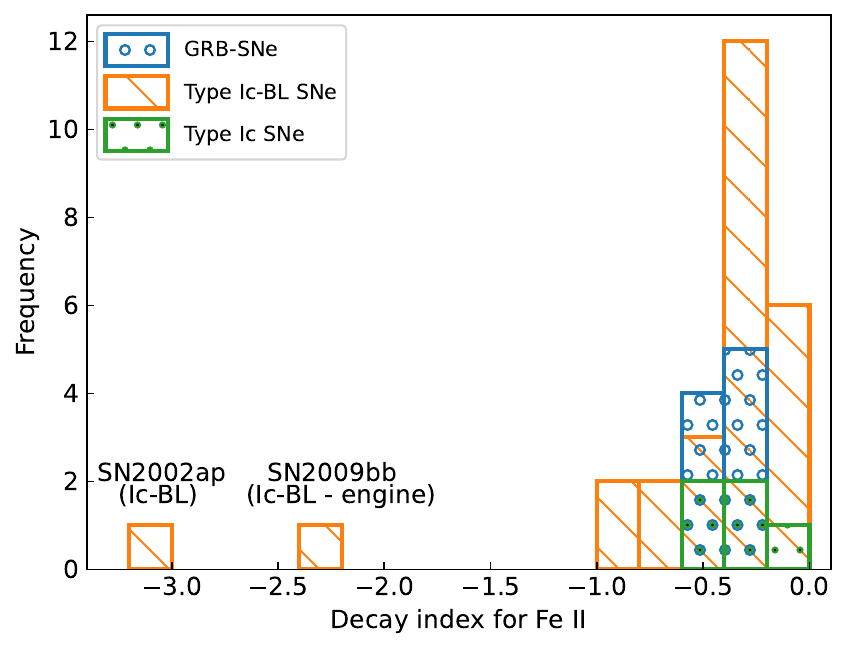}
   \includegraphics[width=0.49\linewidth]{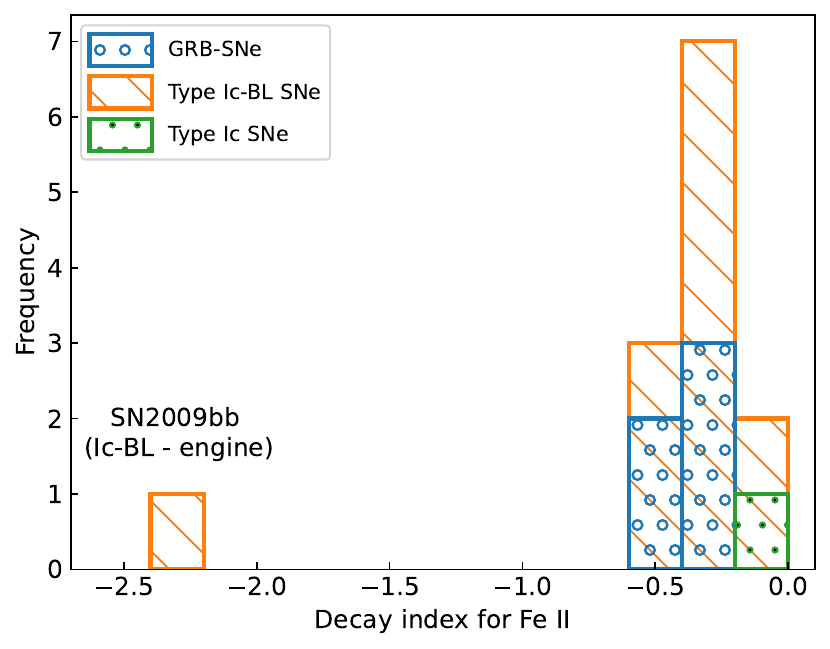}
    \caption{Distributions of the power-law index for the Fe II velocity of GRB-SNe and ordinary type Ic-BL SNe. For broken power-laws the decay index of the first segment is plotted. Fit results for sample of type Ic supernovae from \cite{Modjaz.2016} are shown for comparison purposes. \textit{Left:} Histograms of Fe II decay indices including data from the \textit{Gold}, \textit{Silver} and \textit{Bronze} samples. \textit{Right:} Histograms of Fe II decay indices including data from the \textit{Gold} sample only. Allowing for the low population statistics, the decay index appears to be very similar for type Ic SNe, ordinary type Ic-BL SNe and GRB-SNe. This indicates that the presence or absence of a central engine or a relativistic jet has no impact on the rate of velocity evolution of a supernova.}
    \label{fig:feslopehistogram}
   \end{figure*}

   Figure \ref{fig:feslopehistogram} shows the distributions of the decay indices for the Fe II features of GRB-SNe and ordinary type Ic-BL SNe\footnote{For those SNe that are best fit by broken power-laws, the figure shows the index of the first segment of the broken power-law; this paradigm is used for all other plots of the decay index unless stated otherwise.}. A sample of velocities for Ic supernovae taken from \cite{Modjaz.2016} is shown for comparison purposes. Regardless of subtype, the Fe II velocities of all of these SNe exhibit similar decay indices, suggesting that their decay rates are intrinsic to (stripped-envelope) supernovae, rather than being a distinguishing feature of type Ic-BL SNe or GRB-SNe. Since jet activity is not generally invoked in descriptions of type Ic supernovae, the similarity of these distributions suggests that jets may not play a significant role in determining the pre-break evolution of the Fe II velocity. The median decay indices for GRB-SNe and ordinary type Ic-BL SNe from the \textit{Gold} sample are presented in Table \ref{tab:decaymedians}. Their similarity supports the theory that GRB-SNe and ordinary type Ic-BL supernovae undergo similar early decline.
 
   \begin{table}[h]
   \caption{Median Fe II and Si II decay indices for GRB-SNe and ordinary type Ic-BL SNe in the \textit{Gold} sample. The quoted uncertainties are derived from the 16th and 84th percentiles. No value is reported for Ic supernovae, since the \textit{Gold} sample only has one instance of a Ic supernova.}             
   \label{tab:decaymedians}
   \centering
   \begin{tabular}{lcc}
   \toprule
    & \textbf{GRB-SNe} & \textbf{Type Ic-BL SNe} \\   
   \midrule
   \textbf{Fe II} & ${-0.27_{-0.27}^{+0.04}}$ & ${-0.30_{-0.25}^{+0.09}}$  \vspace{0.1cm}\\ 
   \textbf{Si II} & ${-0.4_{-0.3}^{+0.2}}$ & ${-0.6_{-0.2}^{+0.3}}$ \\
   \bottomrule
   \end{tabular}
   \end{table}

   The observation of visually similar decay-index distributions motivated the use of a KS-test to validate the null-hypothesis that these decay indices are drawn from the same underlying distribution. This test was performed using the \texttt{scipy} \textit{Python} library \citep{SciPy}, using the \texttt{ks\_2samp} function \citep[method is based on][]{Hodges.KSTest}. A two-sample KS-test compares the cumulative distribution functions (CDFs) of two discrete empirically observed distributions, computing the maximum difference between the two CDFs. This statistic is then compared to the KS distribution to determine the $p$-value. In this test, the null-hypothesis is typically that the two distributions are drawn from the same population, with the alternative hypothesis being that the distributions are drawn from distinct populations. In this work, we reject the null-hypothesis if $p<0.01$.
   
   \begin{table*}[t]
   \caption{KS-test results comparing the decay indices of the Fe II and Si II features in GRB-SNe and ordinary type Ic-BL SNe from the \textit{Gold} sample. The results of these KS-tests suggest that there is insufficient evidence to reject the null-hypothesis that the decay indices of GRB-SNe and ordinary type Ic-BL SNe are drawn from the same underlying distribution. No comparisons can be made with type Ic supernovae, since the \textit{Gold} sample only has one instance of a type Ic supernova.}             
   \label{tab:kstestslopes}
   \centering
   \begin{tabular}{lcc}
   \toprule
    & \textbf{Fe II} & \textbf{Si II} \\ 
   \midrule
   \textbf{KS-statistic} & 0.29 & 0.31 \\
   \textbf{$p$-value} & 0.85 & 0.39  \\
   \textbf{Sample sizes}  & 13 type Ic-BL SNe - 5 GRB-SNe & 11 type Ic-BL SNe - 6 GRB-SNe \\
   \bottomrule
   \end{tabular}
   \end{table*}

   A KS-test was used to compare the decay-index distribution of GRB-SNe to that of ordinary type Ic-BL SNe in the \textit{Gold} sample. The results of this test are presented in Table \ref{tab:kstestslopes}. The test yielded a $p$-value of 0.85, and so we cannot reject the null-hypothesis that the decay indices of GRB-SNe and ordinary type Ic-BL SNe are drawn from the same underlying distribution. This conclusion is in line with the significant overlap of the median decay indices presented in Table \ref{tab:decaymedians}. It should be noted that the \textit{Gold} sample has a limited number of samples (for Fe II there are 5 GRB-SNe and 13 ordinary type Ic-BL SNe), and that these results may be altered if additional observations are made in future.

\begin{figure}
    \centering
    \includegraphics[width=\linewidth]{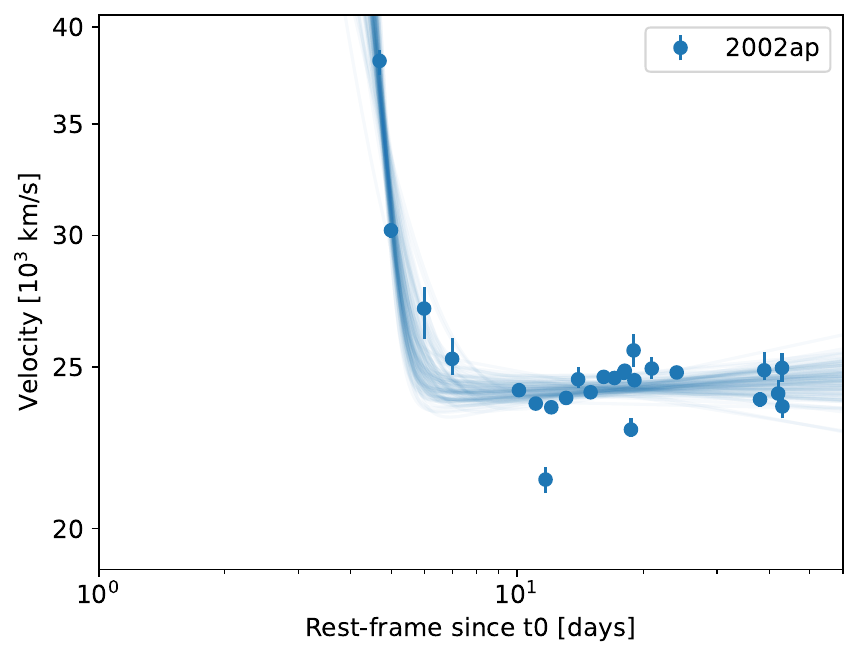}
    \caption[Fe II velocity evolution of SN2002ap.]{Fe II velocities of SN2002ap. The best fit is a broken power-law, showing the steep decay at early times. The decay index is poorly constrained, and the break region has been poorly fit; as a result of these issues, SN2002ap was assigned to the \textit{Bronze}  sample.}
    \label{fig:2002apbplfit}
   \end{figure}

   There are two outliers among the SNe in Fig. \ref{fig:feslopehistogram}, both of which are type Ic-BL supernovae. SN2002ap belongs to the \textit{Bronze}  sample and shows evidence of a very extreme power-law decay at early times, as shown in Fig. \ref{fig:2002apbplfit}, with a decay index of ${-3^{+1}_{-1}}$, based on a broken power-law fit. Although there is clear evidence that SN2002ap is not a pure power-law decay, the fit does not put stringent constrains on the break time (assuming a smooth break, $s$=-5). The velocity of the first datapoint is almost 40000 km/s, which as shown in Fig. \ref{fig:feIIoverall} is not particularly unusual. However, there is a decay of almost 10000 km/s to the next datapoint less than 1 day later, which results in the high decay index. It should be noted that, although the high decay index of SN2002ap is abnormal compared with the rest of the sample, it is still consistent within its 2$\sigma$ error. In the literature SN 2002ap was found to have a spectral shape similar to GRB980425-SN1998bw and SN1997ef \citep{Gal-Yam.2002, Mazzali.2002}. \cite{Mazzali.2002} also found that the spectral evolution is more rapid than SN1997ef, and that its kinetic energy was similar to those of GRB-SNe, though no GRB was detected. They also found no evidence for asymmetry, suggesting a weak or absent jet. Evidence for helium in the spectrum was also found by \cite{Mazzali.2002}, making it interesting that the velocity evolved so rapidly, given the larger mass available within the ejecta to slow down the recession of the photosphere.
 
\begin{figure}
    \centering
    \includegraphics[width=\linewidth]{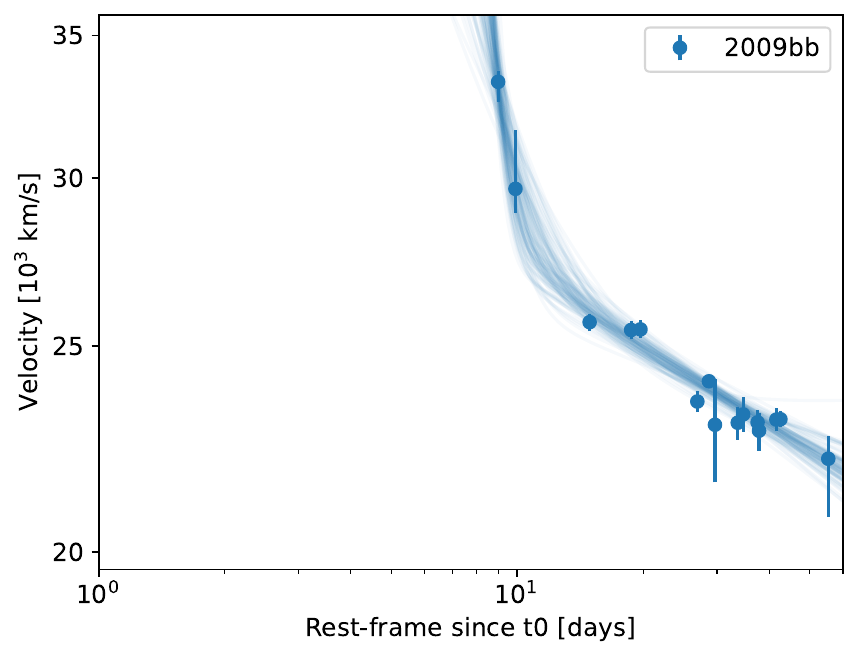}
    \caption[Fe II velocity evolution of SN2009bb.]{Fe II velocities of SN2009bb. The best fit is a broken power-law, showing the steep decay at early times. The decay index is poorly constrained, and the break region has been poorly fit; as a result of these issues, SN2002ap was assigned to the \textit{Gold} sample.}
    \label{fig:2009bbbplfit}
   \end{figure}

   The second outlier is SN2009bb, with a decay index of ${-2^{+1}_{-1}}$ (see Fig. \ref{fig:2009bbbplfit}). The large uncertainty on the decay index means that it is consistent with the rest of the sample at 2$\sigma$. This SN is a member of the \textit{Gold} sample, and has well marginalised parameters, with the fit appearing to capture the break behaviour quite well. Similar to SN2002ap, this SN was fit with $s$=-5, as $s$=-20 produced an unrealistically sharp break. Notably SN2009bb is an example of an engine driven type Ic-BL SN \citep{Pignata.2011}, but with no evidence of an aspherical explosion.

\subsubsection{Si II evolution parameters}
\begin{figure*}
    \centering
      \includegraphics[width=0.49\linewidth]{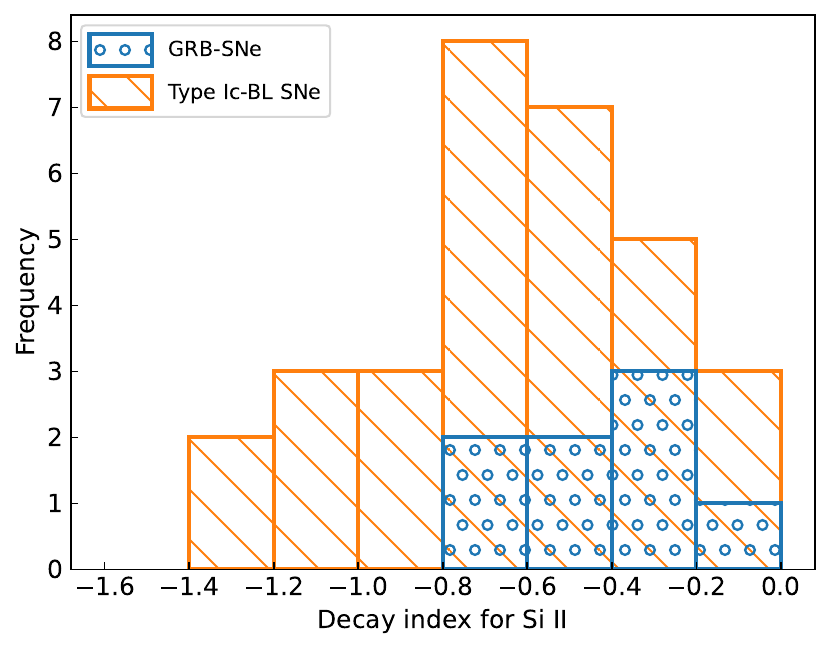}
      \includegraphics[width=0.49\linewidth]{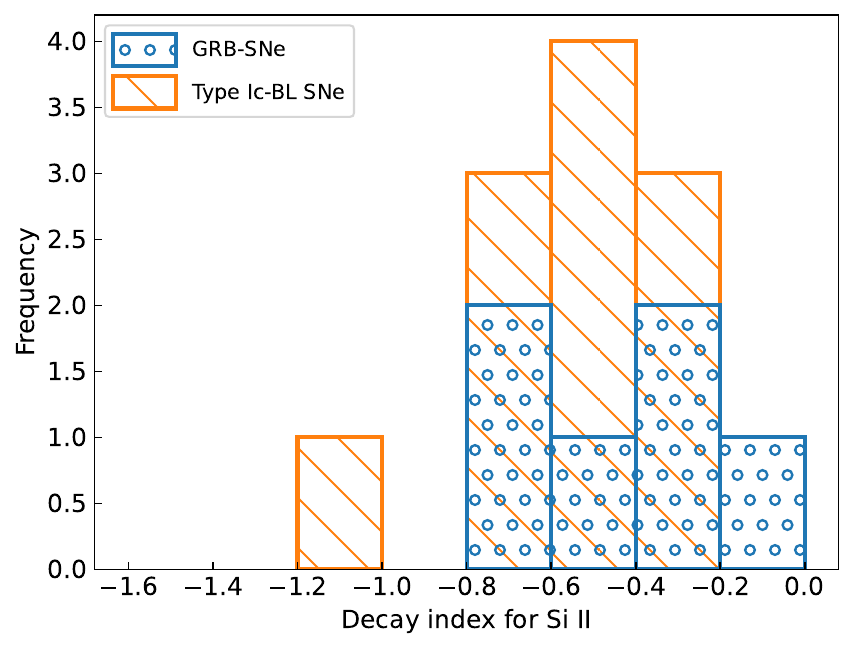}
    \caption[Si II power-law index histogram.]{Same as Fig. \ref{fig:feslopehistogram}, but for Si II. The decay indices of \textit{Gold} sample GRB-SNe are similar to those of \textit{Gold} sample ordinary type Ic-BL SNe. This could be an indication that the decay index of Si II is not affected by central engine or jet activity.}
    \label{fig:sislopehistogram}
   \end{figure*}

   The distributions of the Si II decay indices are shown in Fig. \ref{fig:sislopehistogram}, with the median values for GRB-SNe and ordinary type Ic-BL SNe from the \textit{Gold} sample shown in Table \ref{tab:decaymedians}. The Si II decay indices of \textit{Gold} sample GRB-SNe and ordinary type Ic-BL SNe show considerable overlap, and their median values are consistent within uncertainty. There is a larger difference between the decay index distributions when comparing data from the \textit{Gold}, \textit{Silver} and \textit{Bronze} samples. To test the similarity of the distributions, a KS-test was performed for the \textit{Gold} sample. The results of this test are presented in Table \ref{tab:kstestslopes}. With a $p$-value of 0.21, there is insufficient evidence to reject the null-hypothesis, that the observed decay indices come from the same underlying distribution. This strengthens the case that GRB-SNe and ordinary type Ic-BL SNe undergo similar velocity evolution at early times (pre-break). However, the population statistics are still very low at this time (for Si II there are 6 GRB-SNe and 11 type Ic-BL SNe) and so this result will need to be re-tested with a larger sample size in future.

   The lack of positive decay indices, associated with increasing velocities over time, aligns with assumptions that Si II is linked to the photospheric evolution. In homologous expansion, the greatest velocities are found in material at the edge of the expansion. This means that as the ejecta becomes optically thin and the photosphere recedes, the velocity of the features decrease. If an increasing velocity was detected, it would mean that the photosphere is moving to higher velocities, which would be difficult to explain given the expansion and cooling of the ejecta (it could be true during the very early evolution however, see \cite{Liu.2018}).

\subsection{Broken power-law fits}
   As mentioned previously, a sub-sample of well sampled SNe, show velocity evolutions that can be better constrained using a broken power-law model. While theoretical studies suggest that the evolution of the photospheric velocity should follow a power-law \citep{Branch.2002}, observation of broken power-law evolution may indicate that the spectra are sensitive to the evolution of more than just the photospheric velocity. It may also indicate that for these SNe, the commonly used lines are not appropriate tracers of the photospheric velocity. The results of the broken power-law fits can be found in Table \ref{tab:Feevoresults} and Table \ref{tab:Sievoresults}. The best fit curves for each SN may be viewed in the figures presented in Appendix \ref{sec:additional_velocity_evolution_figures}.

   We also considered the possibility that a large uncertainty in $t_0$ measurements for our type Ic-BL SN sample may have been responsible for our observation of broken power-law evolution. This is because shifting the velocity data in time may introduce conditions under which power-law evolution may be best fit (erroneously) by a broken power-law. We quantified this using simulated data from a broken power-law, and then artificially shifting this data in time and re-fitting it using power-law and broken power-law fits. We then determined the best-fit using an F-test. We did this for multiple shifts of $t_0$ (from -6 to +10 days) and concluded that errors in the fit parameters grow as uncertainty in $t_0$ increases. However, we also found that no amount of uncertainty on $t_0$ can cause a power-law fit to become the preferred fit compared to a broken power-law fit and vice-versa. It is based on these conclusions that we decided to present statistical tests of our measured parameters using only the \textit{Gold} and \textit{Silver} samples, which have well-constrained $t_0$. We present our full analysis in Appendix \ref{sec:testrobustness}.

\subsubsection{Fe II}
\begin{figure}
    \centering
    \includegraphics[width=\linewidth]{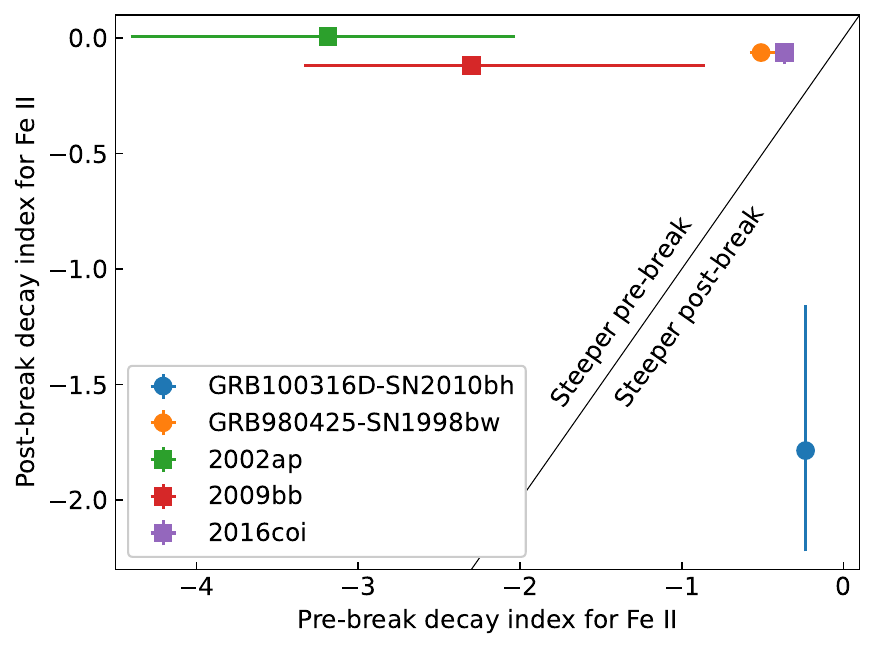}
    \caption[Pre and post-break decay indices for Fe II.]{Fe II decay indices pre and post-break. A solid black line indicates the set of points for which the pre-break and post-break Fe II decay indices are equal. Points above this line indicate that the event has a steeper pre-break velocity evolution than post-break. The majority of Fe II features studied undergo a steep-shallow break.}
    \label{fig:FeBPLslopes}
   \end{figure}

   There are five SNe consistent with broken power-law evolution for the Fe II feature. Figure \ref{fig:FeBPLslopes} shows the decay indices pre and post-break for these cases. A solid black line indicates the set of points for which the pre-break and post-break Fe II decay indices are equal. Points above this line indicate that the event has a steeper pre-break velocity evolution than post-break. Four out of five supernovae follow a steep-shallow evolution (only GRB100316D-SN2010bh follows a shallow-steep decay), and there is no distinction between GRB-SNe and ordinary type Ic-BL SNe in this regard. This late shallow evolution indicates that these SNe experience a near-plateau phase, which could potentially suggest that the photosphere is stationary relative to the velocity stratification of the expanding ejecta. In the case of GRB100316D-SN2010bh, the shallow-steep decay is mostly based on the final observation, which shows a greatly reduced Fe II velocity in comparison to the rest of the evolution.

\begin{figure}
    \centering
    \includegraphics[width=\linewidth]{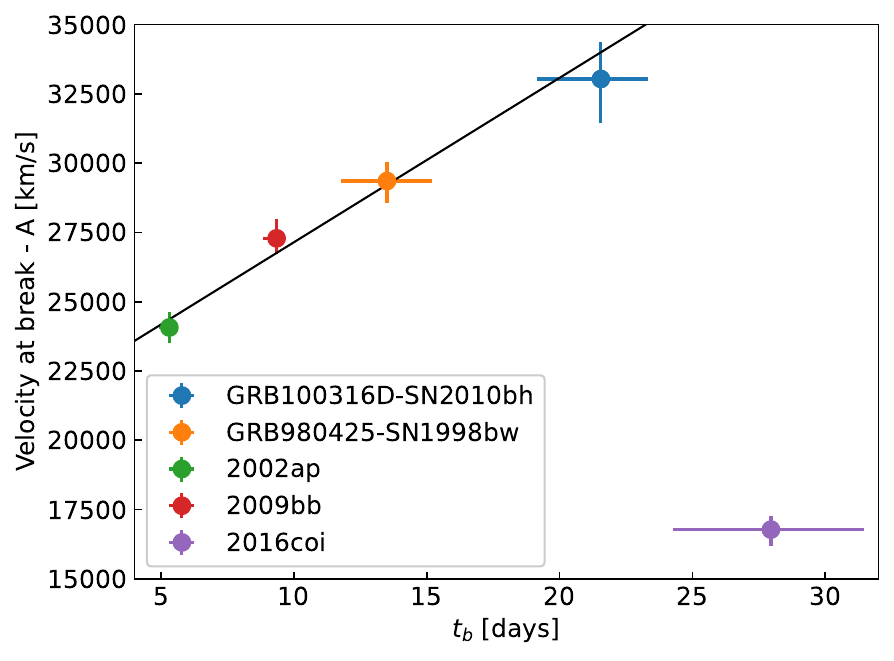}
    \caption[Correlation between $A$ and $t_b$ for Fe II.]{Break velocity ($A$) vs rest-frame break time ($t_b$) for Fe II. The black line shows the line of best fit. A later break time implies a higher break velocity. Note that we have excluded SN2016coi from this fit.}
    \label{fig:FeAvstb}
   \end{figure}

   A comparison between the velocity at break, $A$, and the break time, ${t_b}$, (see Fig. \ref{fig:FeAvstb}) shows that the break velocity is strongly correlated with the break time for 4 of the 5 broken power-laws; with larger break times implying higher velocities at break. The outlier is SN2016coi, which shows a much lower break velocity than expected. A linear fit (excluding SN2016coi) shows a positive correlation (see Table \ref{tab:Avstbparams} for details on the resulting parameters). This differs from the negative correlation between these parameters seen in the corner plots for the individual fits. It seems that the correlation between the break velocity and the time of the break shown in Fig. \ref{fig:FeAvstb} is intrinsic to the population, rather than a result of the fitted models. However, a larger dataset is needed to confirm this result. Both ordinary type Ic-BL SNe and GRB-SNe lie along this correlation, so it cannot be used as a discriminating test without more data.

   \begin{table}[h]
   \caption{Best fit parameters for the A vs $t_b$ plots.}             
   \label{tab:Avstbparams}
   \centering
   \begin{tabular}{l c c}
   \toprule
    & \textbf{$A$ [km/s]} & \textbf{$t_b$ [km/s/day]} \\ 
   \midrule
      \textbf{Fe II} & 21200$\pm$700 & 590$\pm$70  \\
      \textbf{Si II} & 32000$\pm$5000 & -900$\pm$300 \\
   \bottomrule
   \end{tabular}
   \end{table}

\subsubsection{Si II}
\begin{figure}[t]
    \centering
    \includegraphics[width=\linewidth]{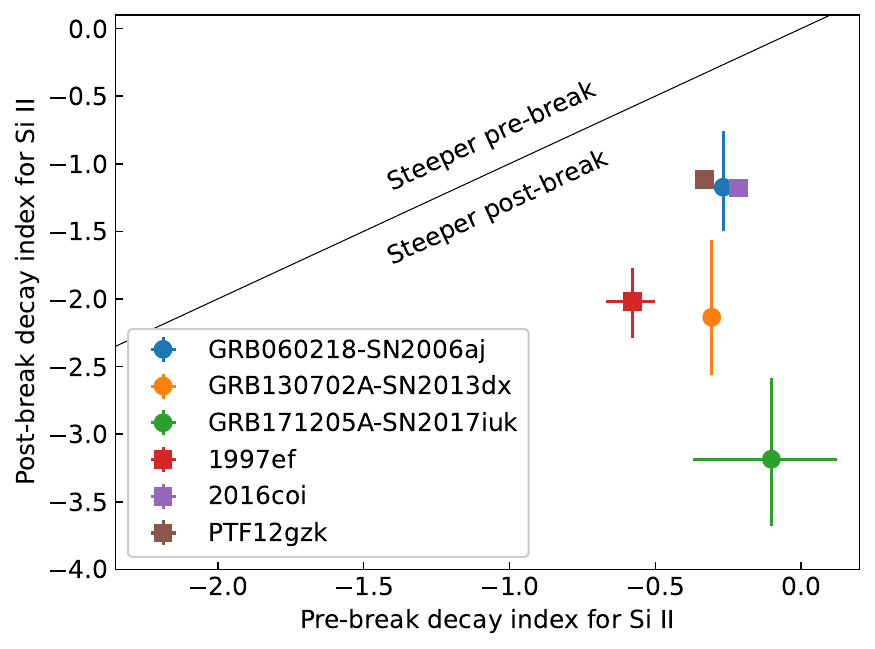}
    \caption[Plot of the pre and post-break decay indices for Si II]{Same as Fig. \ref{fig:FeBPLslopes}, but for Si II feature. A solid black line indicates the set of points for which the pre-break and post-break Fe II decay indices are equal. Points below this line indicate that the event has a shallower pre-break velocity evolution than post-break. All SNe have a shallow-steep decay. There is no clustering of GRB-SNe or ordinary type Ic-BL SNe that can be used to distinguish the populations.}
    \label{fig:SiBPLslopes}
\end{figure}

   Figure \ref{fig:SiBPLslopes} shows that all Si II features for which a broken power-law was the best fit undergo a shallow-steep decay. There appears to be a larger spread in the post-break decay index compared with the pre-break index. Notably, GRB171205A-SN2017iuk has the steepest post-break power-law, and the most extreme change of index at the break.

   \begin{figure*}
   \centering
   \includegraphics[width=0.33\linewidth]{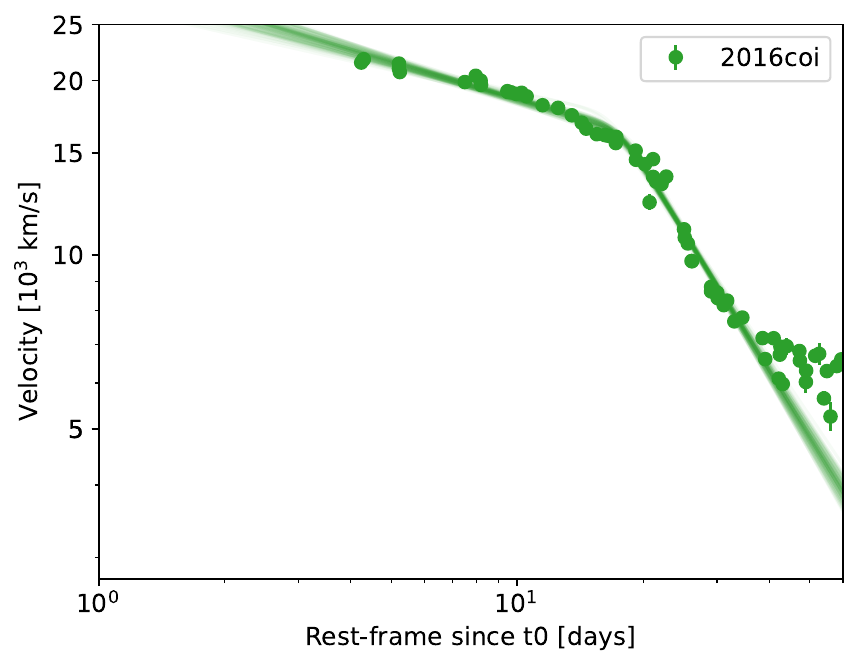}
   \includegraphics[width=0.33\linewidth]{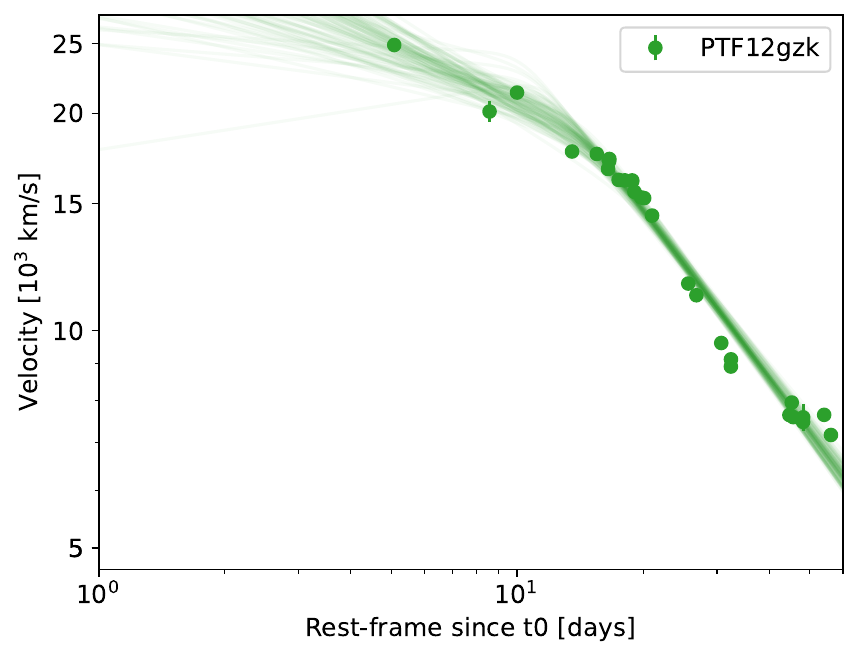}
   \includegraphics[width=0.33\linewidth]{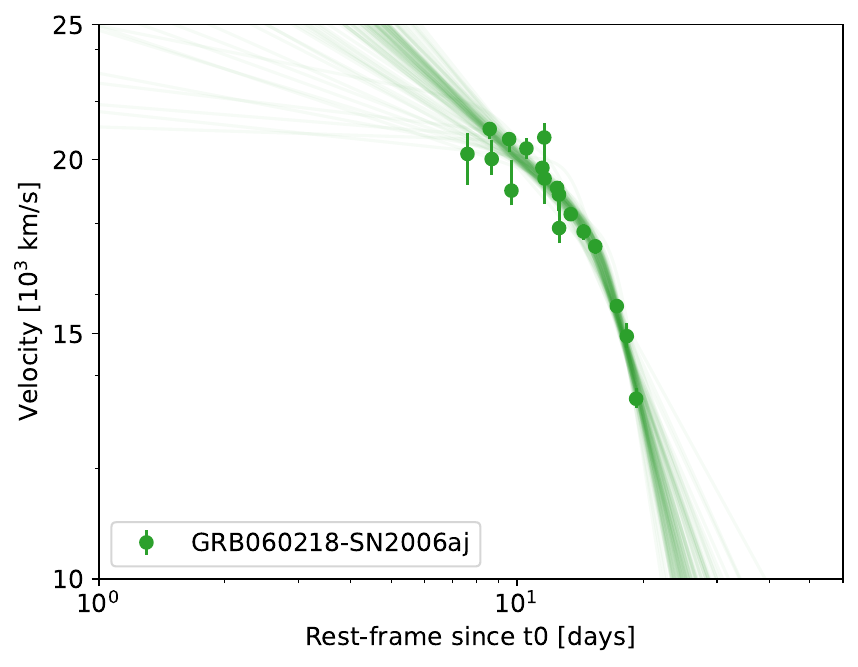}
   \caption[Broken power-law fits for SN2016coi, PTF12gzk and GRB060218-SN2006aj.]{Broken power-law fits to the Si II velocities of SN2016coi, PTF12gzk and GRB060218-SN2006aj. All supernovae evolve from a similar maximum velocity of 20000 km/s, initially following a shallow decay before a break at around 18 days. A second break is visible at around 40 days for SN2016coi and PTF12gzk; there is no data beyond 20 days for GRB060218-SN2006aj, so it is not possible to prove or disprove the existence of a second break for this event.}
    \label{fig:2016coibpl-PTF12gzk}
    \end{figure*}

   As shown in Fig. \ref{fig:2016coibpl-PTF12gzk}, in contrast with the rest of the sample, SN2016coi and PTF12gzk show evidence for a third power-law segment in their Si II velocity evolution. PTF12gzk was found to have a relativistic velocity component by \cite{Horesh.2013}, who suggested that it may be similar to GRB-SN events, with similar explosion parameters \citep{Ben-Ami.2012}\footnote{While PTF12gzk is not truly a type Ic-BL SN, \cite{Ben-Ami.2012} found that its photospheric evolution is similar to that of most GRB-SNe. However, this SN was included in the sample due to the suggestion that it may be engine driven, an intermediate case between a type Ic SN and a GRB-SN/Ic-BL.}. In the case of SN2016coi, it has a similar radio and X-ray emission to SN2009bb and SN2012ap, two relativistic supernovae; however the velocity of the shock-wave is sub-relativistic \citep{Terreran.2019}. SN2016coi's classification as a type Ic-BL SN \citep{Prentice.2018, Terreran.2019} is robust, though evidence of helium in the early spectrum was also found \citep{Prentice.2018}. Like PTF12gzk, this event  appears to be a transitional object between type Ic-BL SNe and GRB-SNe.

   Evidence for the third power-law segment appears at ${t_0}$+40 days for both events. As such, the fits for these broken power-laws were restricted to data prior to 40 days, and no attempt was made to fit the velocities beyond 40 days. Figure \ref{fig:SiBPLslopes} shows that these SNe have a similar set of power-law characteristics to GRB060218-SN2006aj. The break times of these SNe are all consistent within their uncertainties with a break at 18 days. We were unable to find spectra for GRB060218A-SN2006aj between $\sim$20 and 60 days on WISeREP or GRBSN, and so it was not possible to confirm the presence or absence of this third decay component in this GRB-SN. The third decay component could indicate the existence of a separate ejecta component that has not yet been studied in detail, and whose influence becomes apparent only after maximum light. The similarity with respect to GRB060218A-SN2006aj supports the conclusion that these events are similar to (some) GRB-SNe.

\begin{figure}[h]
    \centering
    \includegraphics[width=\linewidth]{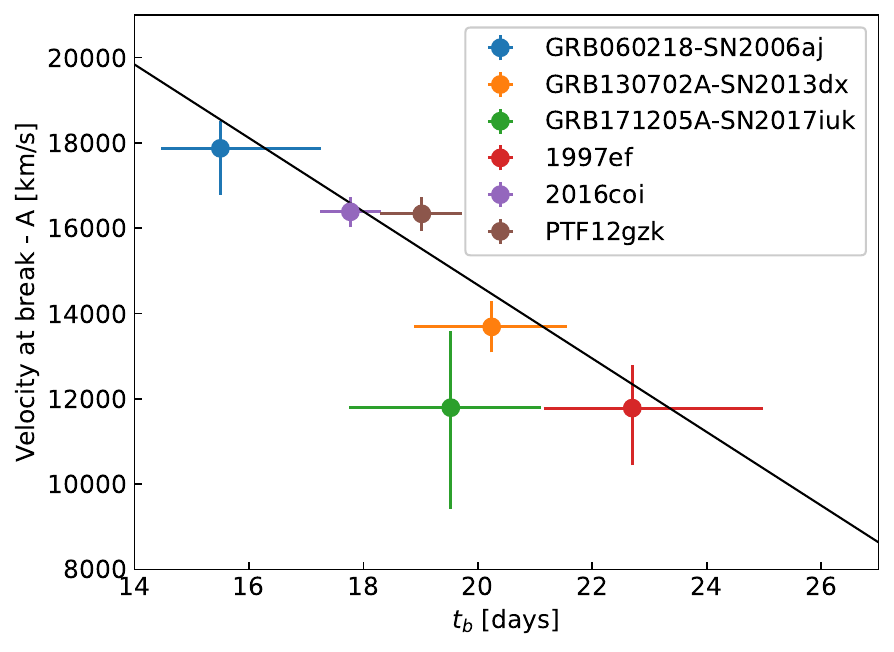}
    \caption[Correlation between $A$ and $t_b$ for Si II.]{Same as \ref{fig:FeAvstb} but for Si II. The black line shows the line of best fit. A later break time implies a lower break velocity.}
    \label{fig:SiAvstb}
   \end{figure}

   Figure \ref{fig:SiAvstb} shows that there is a negative linear correlation between the break velocity and break time for the Si II feature. The slope and intercept values can be found in Table \ref{tab:Avstbparams}. The slope appears to be similar to the slope of the correlation visible in the corner plots of the posterior parameter distributions for the individual fits. Given this, it is difficult to know whether the correlation seen here is something intrinsic to the supernovae considered in this analysis, or whether the result is strongly influenced by the behaviour of the fitted function. Nonetheless, all SNe with Si II velocity measurements exhibit this correlation, regardless of the type of event. Additionally, it can be seen from Fig. \ref{fig:SiBPLslopes} that there are vastly different slopes on either side of the break for different events, so it is not clear why the break time should correlate so well with the velocity at the break. Finally, this correlation may also be influenced by the decision to fix the smoothness parameter during fitting. If future SNe exhibit broken power-law evolution then this analysis may produce a more definitive answer as to whether the correlation is intrinsic, or a result of the modelling.
    
\subsection{Comparing the 15-day velocities of ordinary Ic-BL SNe and GRB-SNe}\label{sec:compare15dayvels}
If there is energy injection in GRB-SNe that is not present in (all) Ic-BL SNe, then there should be observable differences in the velocities of the two populations. Based on the results described  in Sect. \ref{sec:trends}, this section provides a more quantitative answer to this question.

For this analysis, the velocities computed from the (broken) power-law fits at ${t_0}$+15 days were used, since many of the SNe in this sample have observations around this time, and thus their fits are well constrained\footnote{This well covered epoch time is likely a direct consequence of the light-curve of type Ic-BL SNe, that typically peaks at $\sim$15 days \citep[e.g.][]{Taddia.2019, Cano.201723, Taddia.2015}}.

In the case of those SNe with broken power-law fits, the power-law index was determined based on the break time of the broken power-law. Thus, if an SN has a break after ${t_0}$+15 days, then the index of the first power-law segment was used; conversely, if the break occurs before the chosen epoch, the index of the second power-law segment was used.

Only SNe from the \textit{Gold} and \textit{Silver} samples were included in this analysis. This decision is justified by the following arguments: 
\begin{itemize}
    \item The velocity evolution begins at ${t_0}$; if ${t_0}$ is not well known, comparisons based on time periods after ${t_0}$ are not valid.
    \item SNe where the evolution is not clear may have poorly constrained fits at the chosen epoch, which would make it hard to compare their velocities
    \item SNe in the \textit{Gold} and \textit{Silver} samples tend to have better sampled evolutionary curves, increasing the level of confidence in the prediction of velocity for any given epoch.
\end{itemize}

   \begin{table}
   \caption{Median velocities for the \textit{Gold} sample of GRB-SNe and type Ic-BL SNe at ${t_0}$+15 days (rest-frame). No value is reported for Ic supernovae, since the \textit{Gold} sample only has one instance of a Ic supernova.}             
   \label{tab:vel15medians}
   \centering 
   \begin{tabular}{l c c} \toprule
    & \textbf{GRB-SNe [km/s]} & \textbf{Type Ic-BL SNe [km/s]} \\
    \midrule
      \textbf{Fe II} & ${{21000}^{+5000}_{-2000}}$ &  ${21000}^{\mathrm{+5000}}_{\mathrm{-3000}}$   \vspace{0.1cm} \\
      \textbf{Si II} & ${17000}^{\mathrm{+1000}}_{\mathrm{-3000}}$ & ${15000}^{\mathrm{+3000}}_{\mathrm{-4000}}$ \\
   \bottomrule
   \end{tabular}
   \end{table}

\subsubsection{Fe II}

\begin{figure*}[h]
   \centering
   \includegraphics[width=0.49\linewidth]{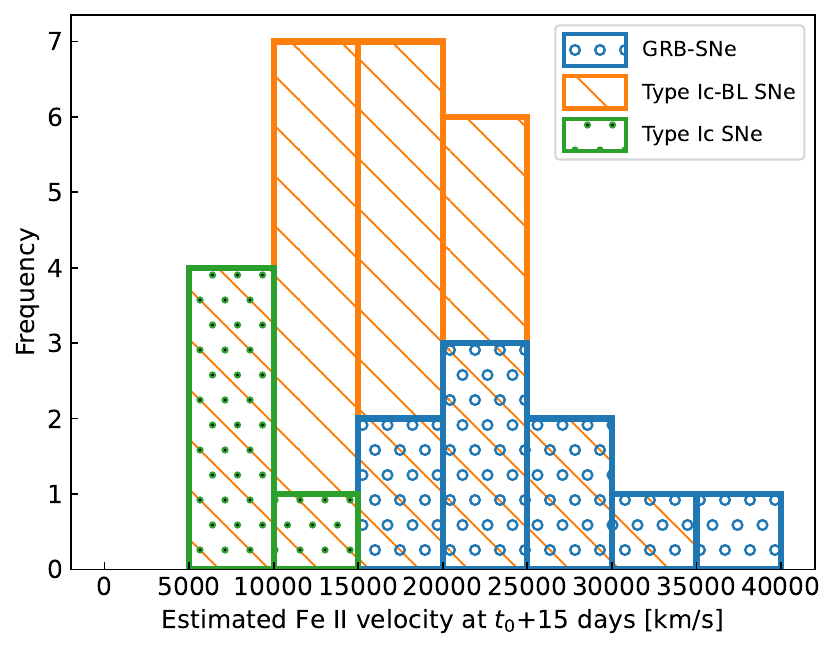}
   \includegraphics[width=0.49\linewidth]{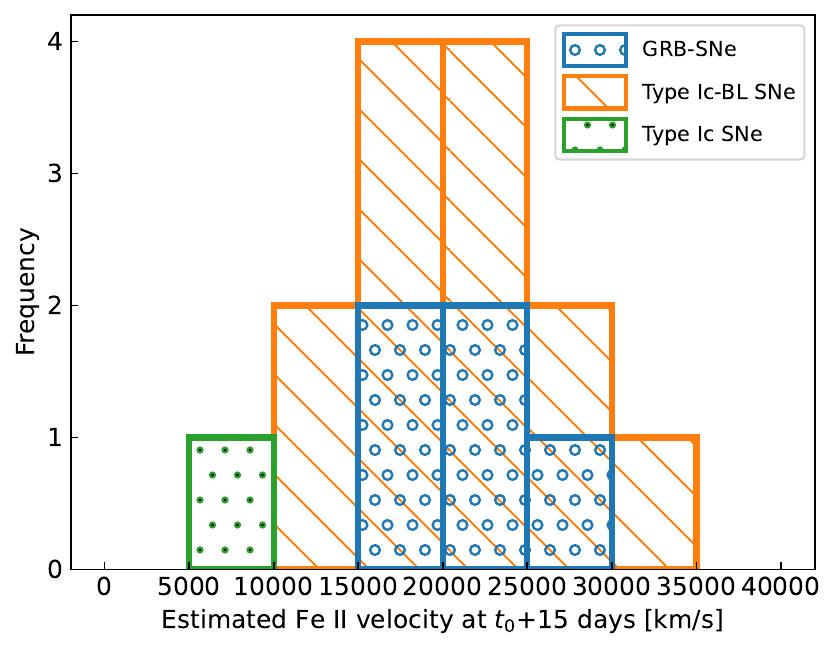}
   \caption[Histograms of Fe II velocities at ${t_0}$+15 days]{Histograms of Fe II velocities at ${t_0}$+15 days. The histograms are shown with bins of 5000 km/s. \textit{Left:} Histograms of Fe II velocities including data from the \textit{Gold}, \textit{Silver} and \textit{Bronze} samples. \textit{Right:} Histograms of Fe II velocities including data from the \textit{Gold} sample only.}
   \label{fig:fe15velshist}
   \end{figure*}

   Figure \ref{fig:fe15velshist} shows the distribution of velocities predicted from the power-law fits to the GRB-SN, type Ic-BL SN and type Ic SN samples at ${t_0}$+15 days. The left panel of this figure shows the data from all three samples. In this plot, the distribution of type Ic-BL SN velocities appears to peak at a lower velocity than that of GRB-SNe. Ordinary type Ic-BL SN velocities are intermediate between those of type Ic SNe and GRB-SNe. The GRB-SN with the maximum velocity is only around 5000 km/s faster than the fastest ordinary type Ic-BL SN. In the right panel of Figure \ref{fig:fe15velshist}, the \textit{Gold} sample is presented, and the situation is somewhat different. In this case, an ordinary type Ic-BL SN is the highest velocity SN, and the GRB-SN and type Ic-BL SN peak velocities are quite similar. It remains the case that Ic supernovae expand less rapidly than both type Ic-BL SNe and GRB-SNe. 

\begin{figure}[h]
   \centering
   \includegraphics[width=\linewidth]{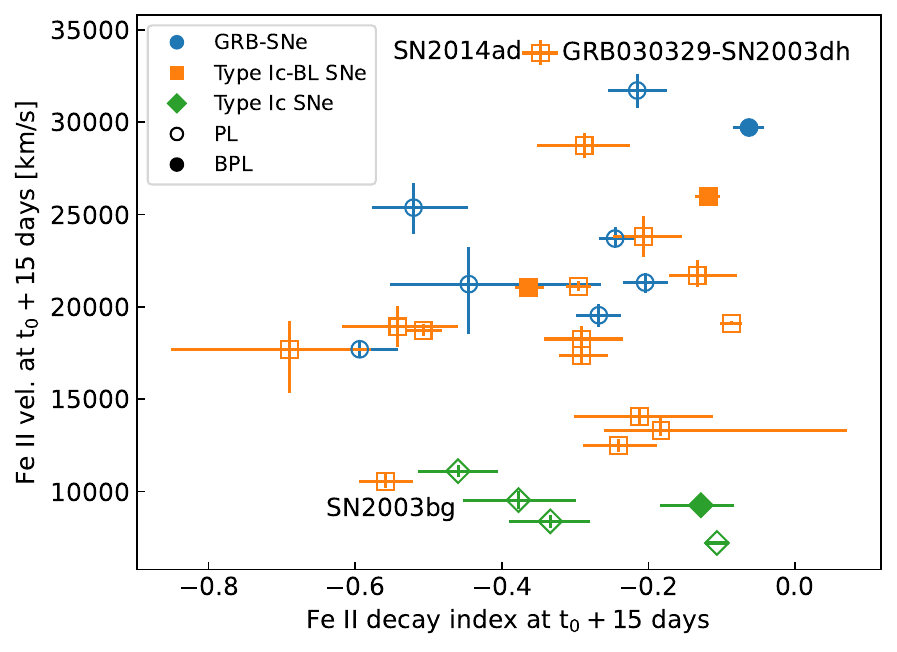}
      \caption{Fe II velocity vs decay index at ${t_0}$+15 days for GRB-SNe and ordinary type Ic-BL SNe from the \textit{Gold} and \textit{Silver} samples. Velocities have been computed from the power-law (PL) and broken power-law (BPL) fits. A sample of type Ic supernovae from \cite{Modjaz.2016} are shown for comparison purposes. Both ordinary type Ic-BL SNe and GRB-SNe have higher velocities than the type Ic SNe population, but there is no clear distinction between the GRB-SN and ordinary type Ic-BL SN velocities.}
      \label{fig:fe15dayvels}
   \end{figure}

   Figure \ref{fig:fe15dayvels} shows the relationship between the Fe II velocity and the decay index at ${t_0}$+15 days\footnote{We show Ic supernovae from the \textit{Bronze}  sample for comparison; several Ic supernovae are in the \textit{Bronze}  sample due to uncertainties in ${t_0}$, but have good fits.}. The velocities of GRB-SNe and ordinary type Ic-BL SNe show significant overlap, although some ordinary type Ic-BL SNe are located at lower velocities. This is also reflected by the overlapping median values of the populations shown in Table \ref{tab:vel15medians}. There appears to be no correlation between decay index and velocity for ordinary type Ic-BL SNe or GRB-SNe. The three SNe presented here that are consistent with broken power-law fits seem to have velocities consistent with the other, single power-law SNe. These broken power-law SNe have relatively shallow decay indices, indicating that they have all transitioned to the slow decay segment of their velocity evolution.

   A KS-test was performed on the \textit{Gold} sample to test the null-hypothesis that the GRB-SN and ordinary type Ic-BL SN velocities at $t_0$+15 days are drawn from the same underlying velocity distribution. The results of the KS-test are shown in Table \ref{tab:kstestvels}. The null-hypothesis (that the distributions of GRB-SN and ordinary type Ic-BL SN velocities come from the same underlying distributions) cannot be rejected based on the high $p$-value of 0.89. This is also reflected by the overlapping median values of the \textit{Gold} sample velocities, shown in Table \ref{tab:vel15medians}. It is therefore not possible to identify a type Ic-BL SN with an associated GRB based on its Fe II velocity alone.

   \begin{table*}[h]
   \caption{KS-test results comparing the velocities of the Fe II and Si II features in GRB-SNe and ordinary type Ic-BL SNe from the \textit{Gold} sample. These results suggest that there is insufficient evidence to reject the null-hypothesis that velocities of GRB-SNe and ordinary type Ic-BL SNe are drawn from the same underlying distribution. No comparisons could be made with type Ic supernovae, since the \textit{Gold} sample only has one instance of a type Ic supernova.}
   \label{tab:kstestvels}
   \centering
   \begin{tabular}{lcc}
   \toprule
    & \textbf{Fe II} & \textbf{Si II} \\ 
   \midrule
   \textbf{KS-statistic} & 0.26 & 0.39 \\
   \textbf{$p$-value} & 0.89 & 0.46  \\
   \textbf{Sample sizes}  & 13 type Ic-BL SNe - 5 GRB-SNe & 11 type Ic-BL SNe - 6 GRB-SNe \\
   \bottomrule
   \end{tabular}
   \end{table*}

   Figure \ref{fig:fe15dayvels} shows the relationship between the Fe II velocity and the decay index at ${t_0}$+15 days for GRB-SNe and ordinary type Ic-BL SNe from the \textit{Gold} and \textit{Silver} samples. These samples have $t_0$ constrained to within 5 days, and so predictions of their decay indices and velocities at 15 days are likely reasonable. There is no correlation between decay index and velocity for ordinary type Ic-BL SNe or GRB-SNe. The three SNe presented here that are consistent with broken power-law fits seem to have velocities consistent with the other, single power-law SNe. These broken power-law SNe have relatively shallow decay indices, indicating that they have all transitioned to the slow decay segment of their velocity evolution. 

   There is a population of low-velocity type Ic-BL SNe, with similar velocity to the Ic population. Among these is SN2003bg, whose type is ambiguous. Some authors have proposed that it evolves from a BL-like SN to a type IIb, before becoming similar to a Ibc supernova in the nebular phase \citep{Hamuy.2009, Soderberg.2006,Mazzali.2009}. Its spectral evolution is certainly very different to classical type Ic-BL supernovae, with more narrow features than expected after the early epochs. This SN may be more similar to Ic supernovae, both in terms of its decay index and velocity at ${t_0}$+15 days. The other members of this group are: SN2018giu, which was classified using SNID as a type Ic-BL SN based on a single spectrum, so the classification may not definitive; SN2017dcc, which both \cite{Taddia.2019} and \cite{Prentice.2019} found to be a type Ic-BL SN; and SN2009ca, which may be an SLSN according to \citet{Stritzinger.2023}. The fastest SNe are SN2014ad and GRB030329-SN2003dh; both of these supernovae also have similar decay indices. 

\subsubsection{Si II}
\begin{figure*}[h]
   \centering
   \includegraphics[width=0.49\linewidth]{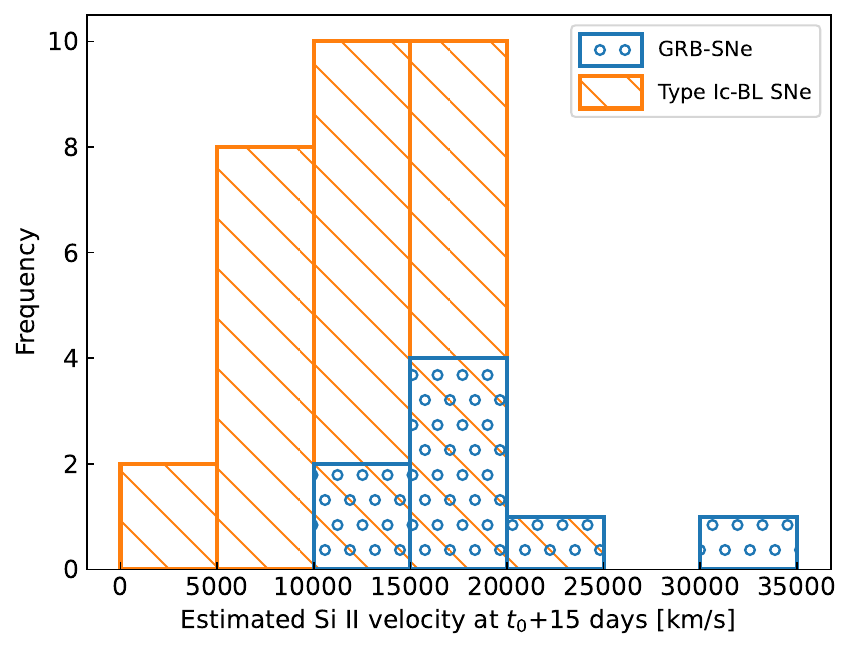}
   \includegraphics[width=0.49\linewidth]{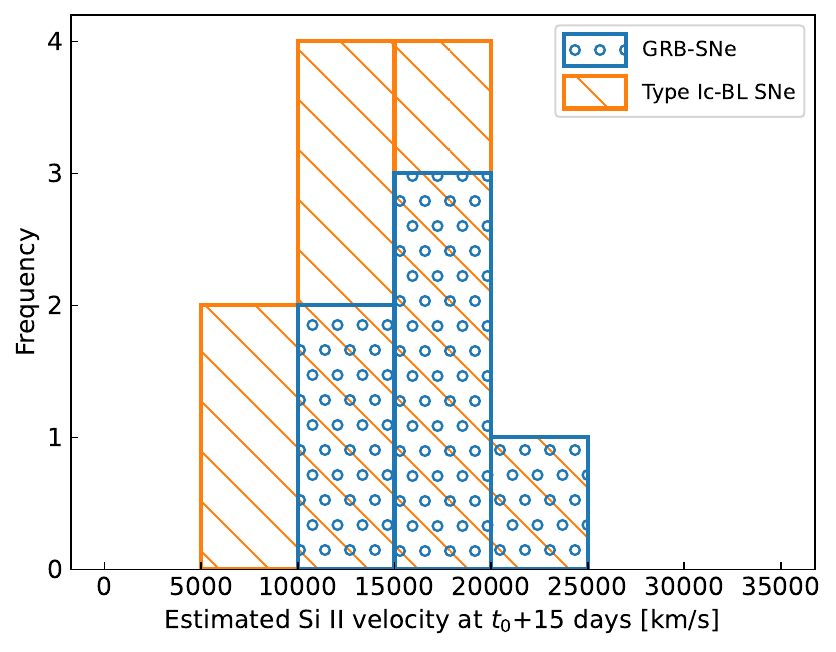}
   \caption[Histograms of Si II velocities at ${t_0}$+15 days]{Histograms of Si II velocities at ${t_0}$+15 days. The histograms are shown with bins of 5000 km/s. \textit{Left:} Histograms of Si II velocities including data from the \textit{Gold}, \textit{Silver} and \textit{Bronze} samples. \textit{Right:} Histograms of Si II velocities including data from the \textit{Gold} sample only.}
   \label{fig:sivelshist}
   \end{figure*}

   Figure \ref{fig:sivelshist} presents the distributions of velocity computed from the power-law/broken power-law fits to Si II velocities at $t_0$+15 days. When comparing data from all samples (left panel of Fig. \ref{fig:sivelshist}), it appears that GRB-SNe exhibit higher Si II velocities than ordinary type Ic-BL SNe. However, given the uncertainties in $t_0$ among the \textit{Bronze} and \textit{Silver} samples, and the impact of such an uncertainty on the velocities computed from the fitted models, it is certain that these velocities are accurate. Thus, the right panel of Fig. \ref{fig:sivelshist} presents the \textit{Gold} sample only, showing that the velocity distributions of GRB-SNe and type Ic-BL SNe are visually quite similar. Their similarity was confirmed using a KS-test, the results of which are shown in Table \ref{tab:kstestvels}. The large $p$-value produced by this test ($p=0.46$) means that we cannot reject the null-hypothesis that these two velocity samples are drawn from a common distribution. Additionally, there is significant overlap in the median velocities of both populations from the \textit{Gold} sample, which are presented in Table \ref{tab:vel15medians}. This result suggests that the Si II velocities of supernovae with well determined $t_0$ cannot be used to identify a type Ic-BL SN associated with a GRB.

\begin{figure}[h]
    \centering
      \includegraphics[width=\linewidth]{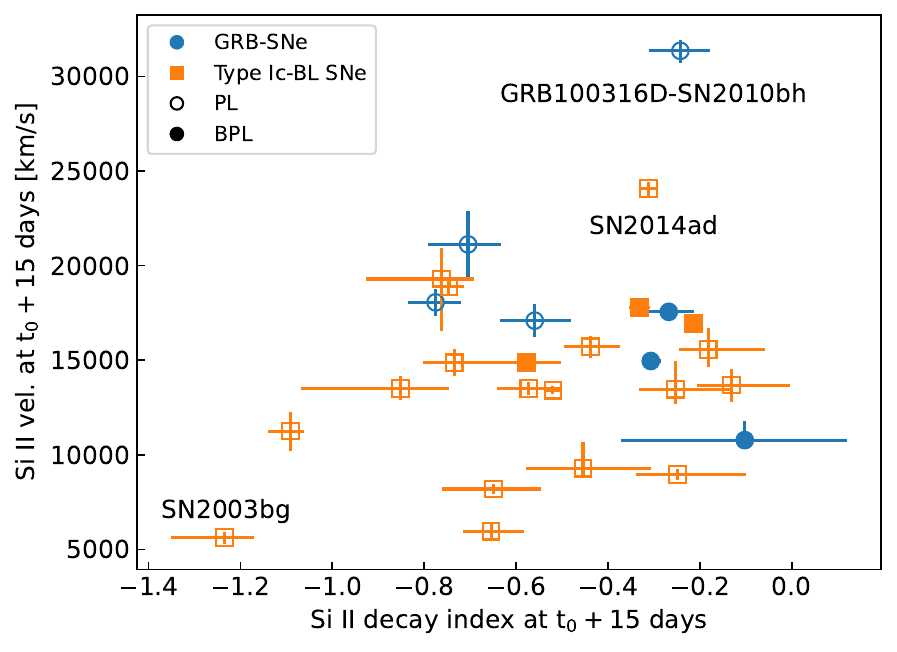}
    \caption{Si II velocity vs decay index at ${t_0}$+15 days for GRB-SNe and ordinary type Ic-BL SNe from the \textit{Gold} and \textit{Silver} samples. Velocities have been computed from the power-law (PL) and broken power-law (BPL) fits. GRB-SNe show similar velocities to some ordinary type Ic-BL SNe, but evolve more slowly.}
    \label{fig:si15dayvels}
   \end{figure}

   Figure \ref{fig:si15dayvels} shows the relationship between the Si II velocity and the decay index at ${t_0}$+15 days for the \textit{Gold} and \textit{Silver} samples. It does not appear to be possible to identify GRB-associated Ic-BLs purely on velocity or slope within this plot. Once again the fastest SNe are SN2014ad and GRB030329-SN2003dh; both of these supernovae also have similar decay indices. SN2003bg is once again the slowest SN in this sample, which could be due to the same reasons as listed in the case of the Fe II feature. 
 
   Many of the SNe with broken power-law fits show a shallow index Figure \ref{fig:si15dayvels}. Since the majority of the Si II feature fits show a shallow-steep decay, this indicates that many supernovae have a break after 15 days. There is no evidence that SNe that follow broken power-law evolution have higher velocities than SNe that are consistent with a single power-law model. 
 
\subsection{Comparison of the Fe II and Si II decay indices}
   The range of the decay index distribution for ordinary type Ic-BL SNe is larger for both Fe II and Si II compared with the distributions for GRB-SNe. This could perhaps be intrinsic, but the more likely explanation is that there are a larger number of ordinary type Ic-BL SNe than GRB-SNe in this sample, and so it is more likely that events with a wider range of indices may be observed.

\begin{figure}
    \centering
   \includegraphics[width=\linewidth]{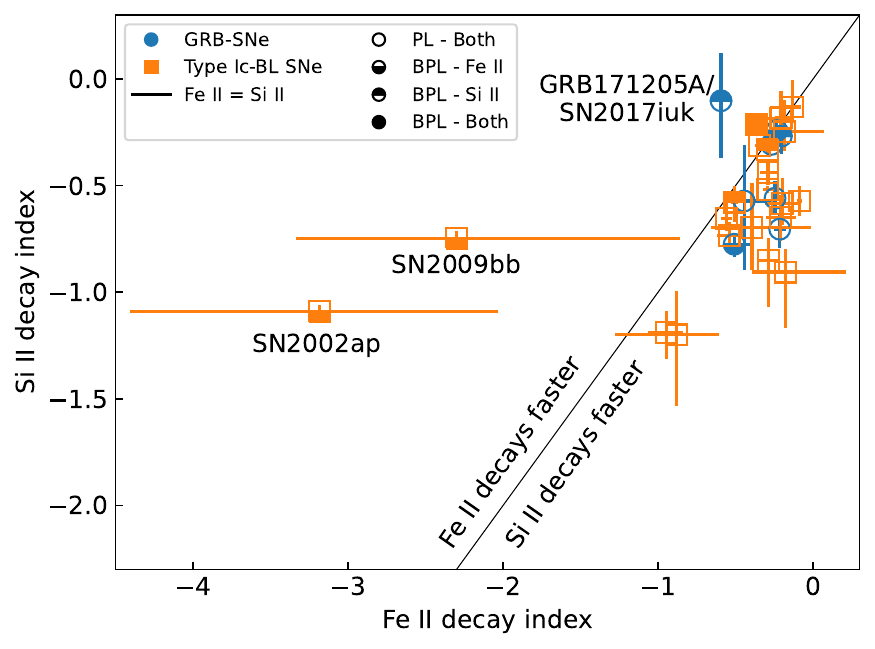}
    \caption{Si II decay index vs Fe II decay index for SNe from the \textit{Gold}, \textit{Silver} and \textit{Bronze} samples. The decay indices have been taken from either the power-law (PL) slope or the pre-break slope for a broken power-law (BPL) fit. For the majority of supernovae, the silicon velocity decreases more rapidly than the iron velocity. There appears to be no significant differences between GRB-SNe and ordinary type Ic-BL supernovae in this regard. For broken power-laws the decay index of the first segment is plotted.}
    \label{fig:FevsSiSlopes}
   \end{figure}

   In Figure \ref{fig:FevsSiSlopes} supernovae from the \textit{Gold}, \textit{Silver} and \textit{Bronze}  samples are presented, showing the decay indices of both features. The caveats around determining the decay index when $t_0$ is uncertain also influence this plot, however it is still useful to include \textit{Silver} and \textit{Bronze} events here to understand the behaviour of the whole sample. Additionally, the impact of $t_0$ on decay index is likely to shift both the Fe II index and Si II index simultaneously, so this plot is still useful in determining which feature slows down more rapidly. The majority of the supernovae show an Si II decay index above -1 and an Fe II index larger than -0.8. Within this cluster there is substantial overlap between GRB-SNe and ordinary type Ic-BL SNe. The velocity of the iron feature decays faster than the silicon feature velocity in just three cases: GRB171205A-SN2017iuk, SN2009bb and SN2002ap.

   \cite{Izzo.2019} propose that GRB171205A-SN2017iuk has a spectrum characterised by an excess of high velocity material (>30000 km/s in the first few days); with the high velocity material attributed to a cocoon around the GRB jet. They find that the early spectra are consistent with Si II velocities of up to 115000 km/s one day after the burst, and declining to around 20000 km/s at ${t_0}$+10 days in the rest frame of the GRB-SN. The spectra in this sample for this event begin at ${t_0}$+8 days, at which time the velocity is in a plateau at 10000 km/s. The velocity then reduces in the following weeks. \cite{Wang.2018} also fit the Si II velocity, and find a result similar to that presented here around the same epoch. As the spectrum is relatively featureless, the discrepancy between the velocities reported here and those of \cite{Izzo.2019} are difficult to explain.

\begin{figure}
    \centering
    \includegraphics[width=\linewidth]{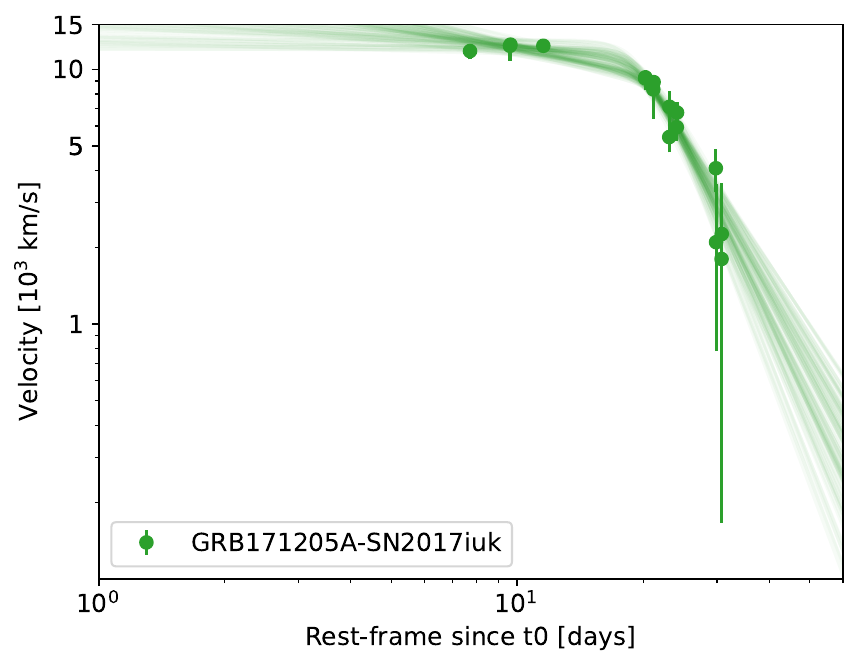}
    \caption{Broken power-law fit to the Si II velocities of GRB171205A-SN2017iuk.}
    \label{fig:2017iuk_bpl}
   \end{figure}

   The Si II velocities of GRB171205A-SN2017iuk are best fit with a broken power-law (Fig. \ref{fig:2017iuk_bpl}). This is a GRB-SN and thus has a well constrained $t_0$, so conclusions drawn from its decay indices are robust. In contrast with many SNe in the sample, the decay index is nearly flat for the first segment of the fit, with a value of ${-0.1^{0.2}_{-0.3}}$. The presence of a cocoon was also suggested by \cite{DElia.2018}. If the cocoon is active during the first few days of the explosion, it is unlikely that the (relatively) late-time spectra presented here are influenced by its activity. Additionally the velocity space of these observations is well below the 30000 km/s which may be associated with cocoons.

\subsection{Co-evolution of Fe II and Si II}\label{sec:fesicoevo}
   Figure \ref{fig:fesicoveo1} shows the difference between the Fe II and Si II velocity over time\footnote{The propagation of asymmetric uncertainties for this plot where computed using the \texttt{asymmetric\_uncertainties} Python package, available at: \url{https://github.com/muryelgp/asymmetric_uncertainties}}. There is no clear difference between the GRB-SN and ordinary type Ic-BL SN populations at early times, however from around 20 days it appears that GRB-SNe show a larger difference in this metric. The Fe II lines are thought to be a good tracer of the photospheric velocity \citep[e.g.][]{Branch.2002}. Figure \ref{fig:fesicoveo1} shows that the Fe II (photospheric) velocity is larger than the velocity in the Si II forming region for the majority of SNe, regardless of their type; with the difference normally in the range of 0-10000 km/s. As discussed earlier, some supernovae have uncertain $t_0$ measurements. This will not influence the y-axis distribution of this plot since we are comparing the directly measured velocities of Fe II and Si II, whose difference is independent of $t_0$. The only noticeable effect would be a movement of the velocity difference in time by at most a few days. Of course, this scatter may occur in either direction, and so overall this plot is still useful for gaining an overview of the evolution of the two populations.

   \begin{figure*}[h]
   \centering
   \includegraphics[width=0.45\linewidth]{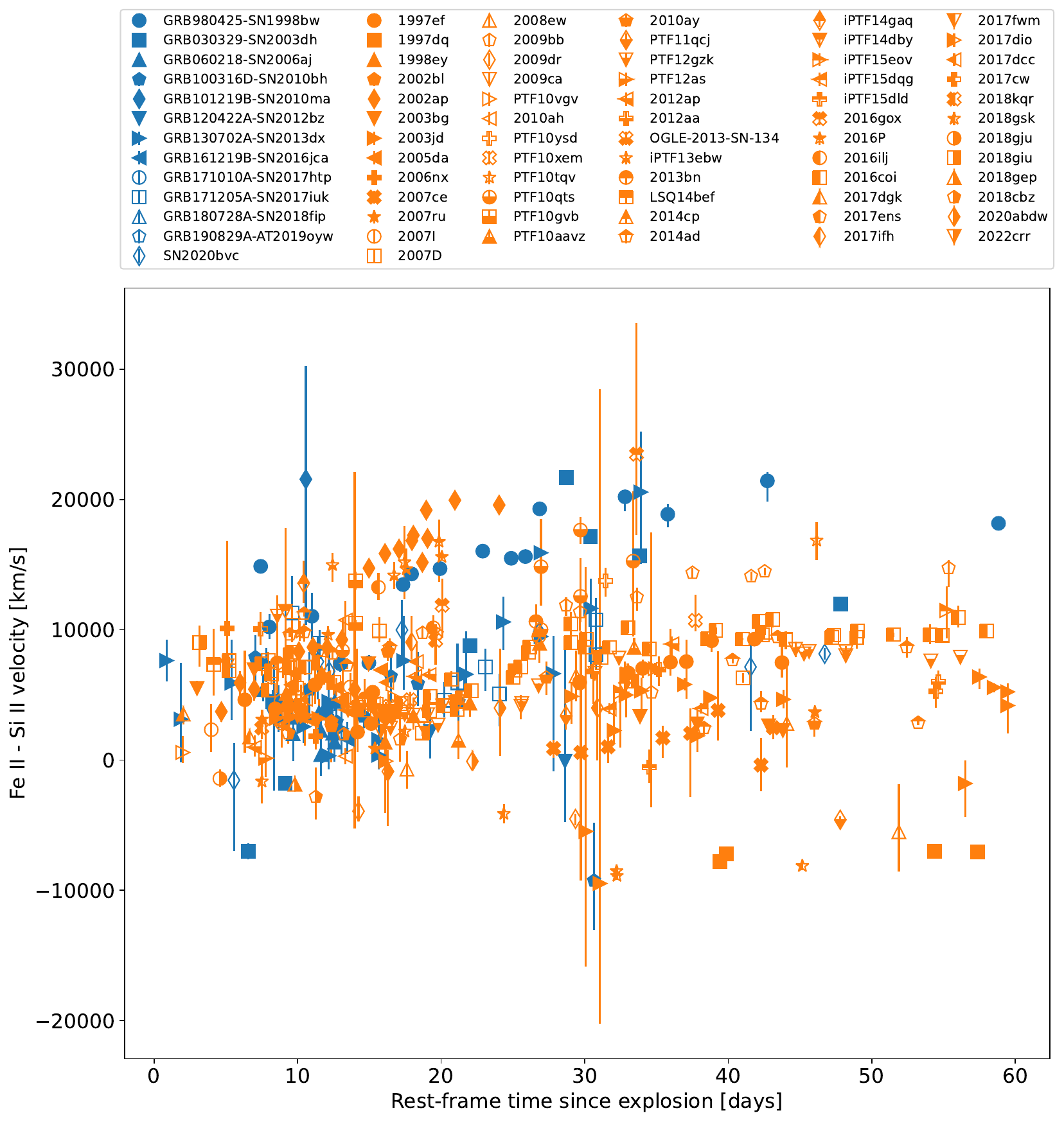}\
   \includegraphics[width=0.49\linewidth]{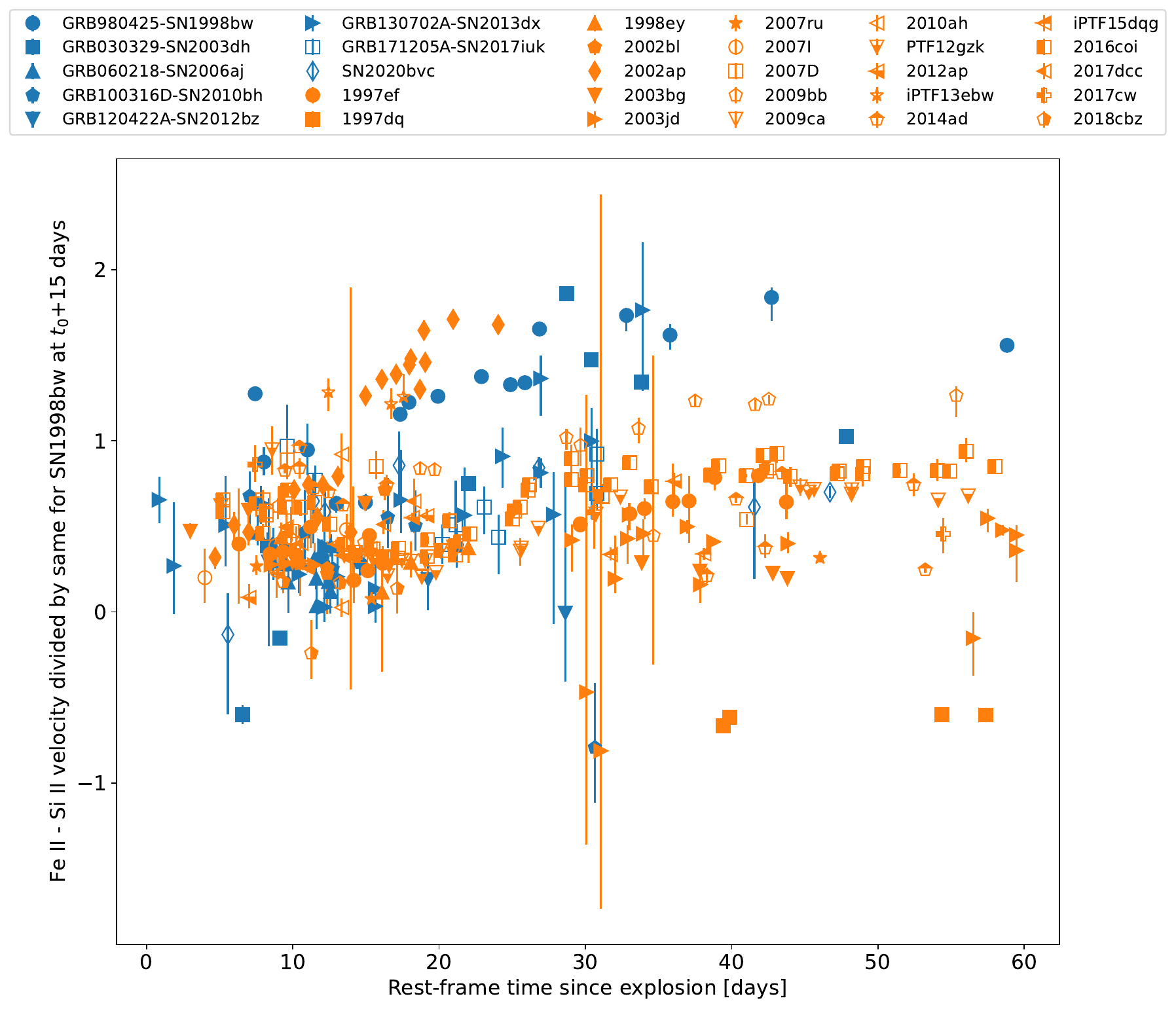}
   \caption{Difference between the Fe II and Si II velocities over time for ordinary type Ic-BL SNe and GRB-SNe from the \textit{Gold}, \textit{Silver} and \textit{Bronze} samples. \textit{Left:} Absolute velocity differences for each SN. \textit{Right:} Velocity differences scaled by the velocity difference for GRB980425-SN1998bw at $t_0$+15 days.}
   \label{fig:fesicoveo1}
   \end{figure*}

   Figure \ref{fig:fesicoveo1} suggests that the Fe II feature often maintains a high velocity relative to the Si II feature at late times, and also that this velocity difference is relatively constant throughout the evolution. This is in tension with the idea that the Fe II line traces the photospheric velocity. Since the Si II feature must be formed in a region above the photosphere, its velocity should be higher than that of Fe II, assuming that Fe II is a good tracer of photospheric velocity. The relative difference between the features could be due to misidentification of the underlying feature responsible for the emission in the Fe II region, likely due to the effects of blending. It has been shown by \cite{Prentice.2018} that large velocity differences can emerge between the velocity of a blended feature and the velocities of its component lines at late times when de-blending occurs. The choice to use 5169\,\AA\, as the rest-wavelength of the Fe II feature could explain the large velocity discrepancies seen at late time. Alternatively it may be the case that Fe II is not a suitable proxy for photospheric velocity as suggested by \cite{Prentice.2018}.

   As shown in Fig. \ref{fig:fesicoveo1}, for many SNe there is an early period prior to 10 days where the Fe II velocity is declining more rapidly than the Si II velocity; this manifests as a decrease from a delta of around 20000 km/s at early times to below 10000\,km/s between 10-20 days. Following this, the difference between the two velocities rises again. For some SNe this velocity difference exceeds 20000 km/s after 30 days, while for others a range of around 5000-10000 km/s is maintained out to day 60. The behaviour of both the GRB-SNe and Ic-BL SNe appears to be very similar in this regard.

   Some of the SNe in Fig. \ref{fig:fesicoveo1} show evolution that is unusual. GRB130702A shows an increasing delta between Fe II and Si II out to around 35 days. This appears to be driven by a sharp decline in the Si II velocity around this time. On the opposite side of the distribution around this time is GRB100316D-SN2010bh, whose sharp decline in Fe II velocity results in a negative delta at this time.

\section{Discussion}\label{sec:discussion}
\subsection{Comparisons to previous analyses}
   In order to verify that our method gives reliable results, we compared the Fe II and Si II velocities of a subset of our SNe to those in the literature. The SNe in this subset are: SN1997ef \citep[data from][]{Patat.2001q8t}, GRB980425-SN1998bw \citep[data from][]{Modjaz.2016,Patat.2001q8t} and GRB130702A \citep[data from][]{DElia.2015}. These SNe were chosen as they had well-studied velocity evolution. The spectral sequences of these SNe may be found in Appendix \ref{sec:specsequences}. Figure \ref{fig:velocitycomparisons} shows the results of this investigation. The Si II velocity measurements of SN1997ef and GRB980425-SN1998bw appear to be reliable, as they show strong similarities to the literature values in magnitude and evolutionary trend. This is to be expected, since both the method applied in this analysis and the methods used in the literature rely on the determination of the wavelength of minimum flux within a spectral feature. We conclude that our estimates of Si II velocities for type Ic-BL SNe presented here are reliable.

   In contrast to the results for Si II velocities, the picture for Fe II velocities is mixed. In recent years, a method of velocity determination proposed by \cite{Modjaz.2016} has become commonplace for measuring the Fe II velocities of type Ic-BL SNe. This method blue-shifts and convolves a type Ic SN template spectrum in order to try and match a type Ic-BL SN spectrum. To determine the Fe II velocity of the type Ic-BL SN, the fitted blue-shift velocity is added to the velocity of the Fe II 5169\,\AA\,line in the template spectrum. This is intended to deal with the issue of line-blending within the Fe II feature, which can corrupt the results of velocity measurements which assume that the minimum flux of the Fe II feature is due to the Fe II 5169\,\AA\,line (as our method does).

   We observe in Fig. \ref{fig:velocitycomparisons} that the Fe II velocities of GRB980425-SN1998bw are approximately 12000\,km/s higher when measured from the minimum of this feature using our method than they are when measured using the method of \cite{Modjaz.2016}. The magnitude of this discrepancy suggests that Fe II 4924\,\AA\,or 5018\,\AA\,may be the dominant lines in this feature, and it is their velocities that are being (erroneously) inferred from the minimum flux wavelength using our method. We investigate the validity of both methods in much greater detail in \cite{Finneran.2024C}. Figure \ref{fig:velocitycomparisons} also shows the Fe II velocities of GRB130702A-SN2013dx determined by \cite{DElia.2015} using the \cite{Modjaz.2016} method. In this case, there is excellent agreement between our result and those of \cite{DElia.2015}. This implies that the minimum of this Fe II feature is strongly influenced by Fe II 5169\,\AA. These results suggest that Fe II velocities presented here may not always reflect the Fe II 5169\,\AA\,velocity, but rather the average velocity of the Fe II feature. Notably however, Fig. \ref{fig:velocitycomparisons} shows that the velocity evolution obtained by both methods is similar, and so we conclude that the observed evolutionary trends are robust.

   \begin{figure*}
      \centering
      \includegraphics[width=0.49\linewidth]{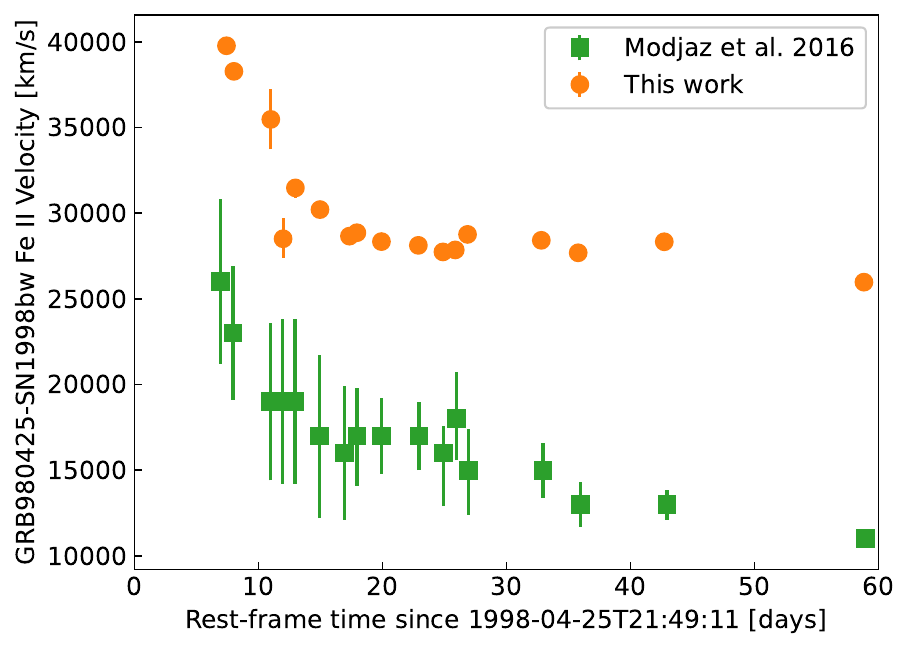}
      \includegraphics[width=0.49\linewidth]{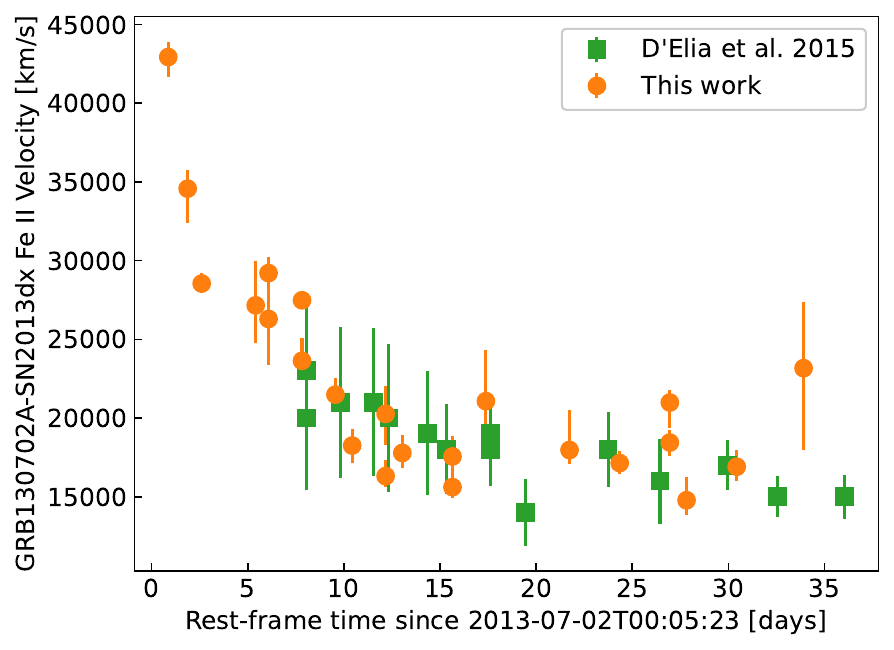}
      \ContinuedFloat
      \includegraphics[width=0.49\linewidth]{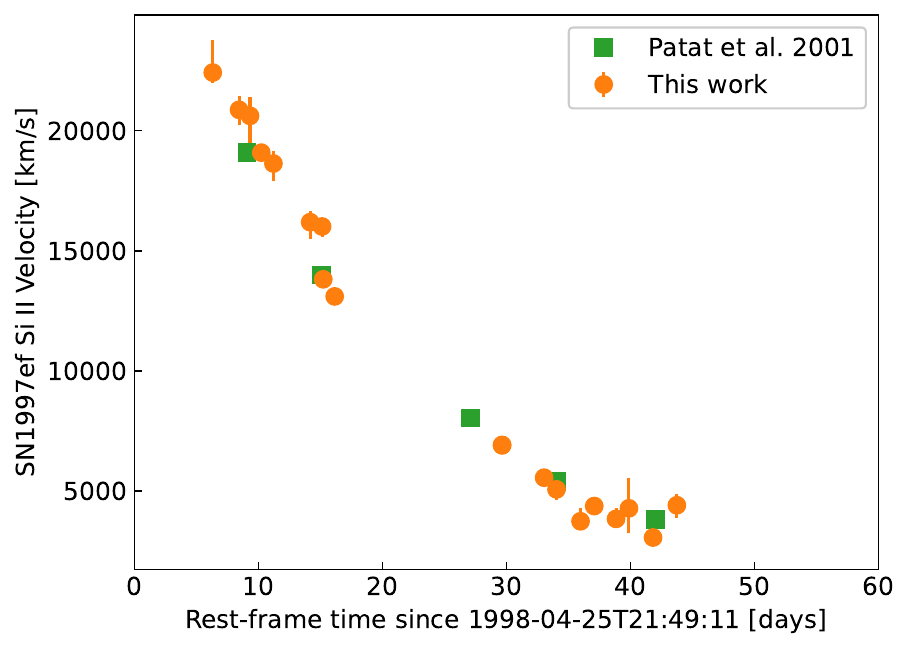}
      \includegraphics[width=0.49\linewidth]{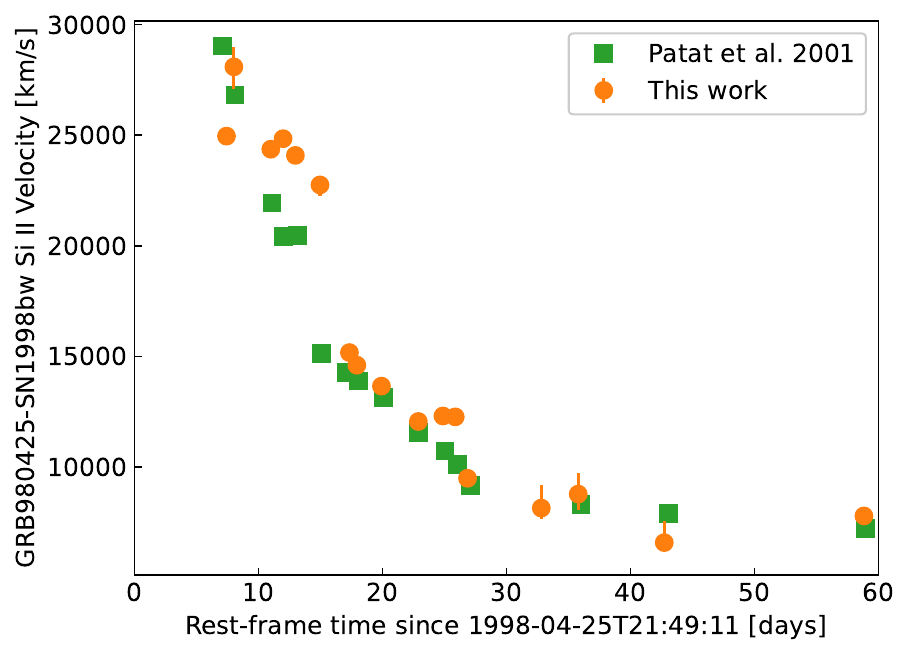}
      \caption{Comparison of SN velocities derived from the minima of absorption features in this work (orange circles) with SN velocities from the literature (green squares). \textit{Top left:} Fe II velocities of GRB980425-SN1998bw. The literature data is from \cite{Modjaz.2016}, who use a custom velocity determination method in which a blue-shifted and convolved type Ic template spectrum is fit to the type Ic-BL spectrum in order to determine the Fe II velocity. This is intended to handle the issue of rest-wavelength assignment for a heavily blended feature. In this instance, there is a clear disparity between the velocities measured by both methods. This type of disparity was observed by \cite{Modjaz.2016}. A possible explanation for this is that Fe II 5169\,\AA\,may not be the dominant Fe line in this feature for this SN. Consequently, using the wavelength of the minimum flux of the feature with Eq. \ref{eq:doppler} leads to an over-estimate of the velocity. The discrepancy in this case is approximately 12000\,km/s, which suggests that Fe II 4924\,\AA\,or 5018\,\AA\,may be the dominant lines in this feature, and it is their velocities that are being (erroneously) inferred from the minimum flux wavelength. \textit{Top right:} Fe II velocities of GRB130702A-SN2013dx. The literature data is from \cite{DElia.2015}, who use the method proposed by \cite{Modjaz.2016}. A much better agreement between methods is observed in this case, implying that the minimum of this Fe feature is strongly influenced by Fe II 5169\,\AA. \textit{Bottom left:} Si II velocities of SN1997ef. Literature data are from \cite{Patat.2001q8t}, who obtained the Si II velocity from the minimum flux wavelength of the Si II 6355\,\AA\, feature. As expected, the results of both methods show excellent agreement. \textit{Bottom right:} Si II velocities of GRB980425-SN1998bw. Literature data are from \cite{Patat.2001q8t}, who obtained the Si II velocity from the minimum flux wavelength of the Si II 6355\,\AA\, feature. As expected, the results of both methods show excellent agreement.}\label{fig:velocitycomparisons}
   \end{figure*}

   \cite{Modjaz.2016} studied the Fe II velocity evolution of 11 GRB-SNe and 10 type Ic-BL supernovae. For ordinary type Ic-BL SNe, this sample represents a six-fold increase in the number of type Ic-BL supernovae with measured Fe II velocity, compared to \cite{Modjaz.2016}. This is in part due to the explosion in type Ic-BL SN classification rates since 2018 with the advent of ZTF \citep{Dekany.2020}. Differences between the GRB-SNe in the sample and those of the \cite{Modjaz.2016} sample can be explained by the selection criteria: only those SNe which had at least 3 epochs of spectral data in online repositories are presented here.

   Qualitatively, the evolutionary trend for Fe II feature velocity seen by \cite{Modjaz.2016} is confirmed. However, the quantitative analysis reveals that there is no velocity which divides GRB-SNe from ordinary type Ic-BL SNe with no associated gamma-ray emission, nor are the median velocities of the two groups at $t_0$+15 days statistically different (the opposite conclusion to that of \cite{Modjaz.2016}). The statistical results presented here were based on supernovae in the \textit{Gold} sample, which mitigates the effects of uncertainty in $t_0$ on the modelled parameters and the velocities. It should be noted that \cite{Modjaz.2016} do not account for or quantify the uncertainty on $t_0$ in their sample, even though it will influence their comparison of velocities at $t_0$+15 days, by shifting the observed velocities in time. The differences between the conclusions presented here and in \cite{Modjaz.2016} may be due to the different methods used to determine velocity (template fitting vs. spline fitting) or they could be due to issues with line blending. Overall the range of velocities for both studies seems to be quite similar, with all SNe lying between 5000-50000 km/s. In contrast with the results of \cite{Modjaz.2016}, the velocities of GRB100316D-SN2010bh are similar to GRB980425 -SN1998bw, though it is still the highest velocity GRB-SN.  The remaining GRB-SNe are surrounded by Ic-BLs, in a band that evolve from 25-35000 km/s to 5-20000 km/s. This could be evidence for jet activity in at least some Ic-BLs, or it could mean that jet activity has no impact on the magnitude or evolution of the velocities.

   SN2014ad seems to be an outlier among ordinary type Ic-BL SNe; its velocities and their evolution are similar to those of GRB980425-SN1998bw and GRB100316D-SN2010bh. \cite{Sahu.2018} corroborate this finding, observing that SN2014ad is characterised by higher expansion velocities than typical GRB-SNe and ordinary type Ic-BL supernovae (measured from the Si II line); they also concluded that parameters derived from its lightcurve make it similar to GRB-SNe.
  
   The sample of Si II features presented here is the largest of its kind for GRB-SNe and Ic-BL SNe. As such, no direct population-level comparisons can be drawn with previous research. Statistical results for the Si II feature are in line with those found for Fe II and suggest that the velocities and decay rates of velocity for GRB-SNe are similar to those of ordinary type Ic-BL SNe. Overall the behaviour of the Si II features is consistent with the idea that GRB-SNe and ordinary type Ic-BL supernovae arise from a population of similar progenitors.
  
\subsection{Are jets active in all type Ic-BL supernovae?}
   It appears that neither the velocity, nor its evolution are distinguishing characteristics between GRB-SNe and ordinary type Ic-BL supernovae, at least for supernovae with well-constrained $t_0$ values. Previous studies have ruled out GRBs (or put very stringent constraints on the phase space of the explosion) for several SNe in this sample \citep[e.g.][]{Soderberg.2006, Corsi.2016, Corsi.2023}. Despite this, there is significant overlap between the ordinary type Ic-BL SNe and GRB-SNe studied here, including for some which are known not to harbour a GRB. This overlap exists for all lines studied here, and therefore cannot purely be attributed to issues with line-identification or blending, since these issues do not affect every event or every feature in the spectrum.

   Jetted emission is inferred to power the afterglow lightcurves of GRB-SNe, thus the supernovae observed in association with GRBs must contain a jet. To form a GRB the jet must have sufficient energy to break out of the stellar envelope without choking \citep[e.g.][]{Corsi.2021}. It is thought that energy injection from a central engine may be the source of the high velocities observed in GRB-SNe. However, models of the spectra of GRB-SNe suggest that the velocity increases due to a jet are small, and are present at early times \citep{Barnes.2018}. The results of this study suggest that jet energy injection does not have a significant influence on the velocity of the ejecta. Thus, it is not possible to conclude that jets are present in all Ic-BL SNe, or infer the frequency with which these jets may become choked and be seen as non-GRB type Ic-BL SNe.

\subsection{Factors influencing the velocity evolution}
   The velocities of both absorption features studied show evidence for a plateau phase beginning around 20 days. \cite{Long.2023} showed that a similar plateau observed for SN2020bvc may be modelled assuming an r-process component within the ejecta, rather than the classical model of a supernova powered by the decay of nickel produced by the supernova shock or a GRB jet. However, a study by \cite{Anand.2024} ruled out r-process enrichment in a sample of 25 type Ic-BL supernovae discovered by ZTF. Given the weight of this large sample it is difficult to conclude that the plateaus are all due to r-process elements. If r-process SNe are indeed rare then plateaus are not anticipated for many SNe. A possible explanation for some of these plateaus may then be that the decay indices of these SNe are impacted by the uncertainty on $t_0$. However, some GRB-SNe exhibit this plateau also, so this cannot be the only explanation.

   It is also possible that plateaus may be a natural result of the underlying ejecta's density profile. \cite{Liu.2018} presents the theoretical framework for the evolution of the photospheric velocity. Out of all the density profiles studied in this work, only the broken power-law model shows evidence for a rapid decline in velocity followed by a slower decline phase. In this model there is an inner core where the density falls off more slowly with increasing radius, and an outer shell where the density declines more rapidly. \cite{Liu.2018} points out that the actual photospheric radius evolves more slowly than their theoretical prediction; it is then possible that plateaus could be a consequence of this.

   Mixing of the radioactive nickel in the ejecta can also impact the photospheric velocity evolution and the magnitude of those velocities. \cite{Moriya.2020} performed simulations of SESN progenitors, finding that the greater the degree of mixing in the ejecta, the smoother and more rapid the decline in the photospheric velocity is.  They found evidence for plateaus in the photospheric velocity around 15 days prior to the bolometric luminosity peak. There is no evidence for early plateaus such as these in this sample, and so qualitatively it appears that many of the supernovae studied here are highly mixed, in line with the results of studies performed on supernova lightcurves.

\subsection{Physical interpretation of power-law and broken power-law fits}
   A power-law decline is anticipated for the photospheric velocity, and likely also for absorption features formed near the photosphere \citep[e.g.][]{Branch.2002}. The majority of supernovae in this sample can be fit with a power-law decline during their photospheric evolution. Differences between the decay indices for the Fe II and Si II lines may be put down to differences in the optical depths for these lines, though in most cases they appear to follow similar declining trends,  with the Si II feature appearing to decline more rapidly than the Fe II one in most cases.

   Interpretation of the broken power-law fits is more difficult. The analysis presented in Appendix \ref{sec:testrobustness} strengthens the case for these broken power-law fits, showing that uncertainty in $t_0$ is not responsible for the observation of broken power-law fits. Broken power-law evolution has not been studied for a large sample of type Ic-BL SNe before, and there has been limited evidence for this type of evolution in other supernova populations. The Si II velocities of type Ia supernovae were investigated by \cite{PLBPL.2017}, who found some limited evidence for broken power-laws in type Ia supernovae. However, it is unlikely that this was due to the presence of jets, since type Ia supernovae are not typically associated with jet-generated emission. It is therefore possible that the broken power-laws seen here are not due to the jet. As mentioned in the preceding section, they may be also be linked to the density profile of the ejecta, however many GRB-SNe have been found to reside in relatively constant density, ISM-like media \citep[e.g.][and references therin]{Cano.201723}. In addition, these fits cannot be purely due to identification or blending issues, as they are observed for both the Fe II feature (often heavily blended) and the Si II feature (usually not heavily blended) in a handful of SNe. \\

   It is possible that these fits reveal the presence of additional ejecta components, whose velocity evolution may differ due to differences in temperature, composition, or orientation relative to the observer. Broken power-laws may also be linked to aspherical explosion geometry, however in this case it is difficult to explain why they are not observed in all GRB-SNe. One possibility is that the cocoon formed by the GRB jet as it tunnels through the star may be responsible for this break. However, the times of the breaks, and their velocities, are not consistent with the proposed cocoon properties \citep[e.g][]{Izzo.2019}. Additionally, there is only one observation of cocoon emission for any of the events in this sample: GRB171205A-SN2017iuk \citep{Izzo.2019}, which does not stand out significantly from the overall population. For all Si II features, the broken power-law fits follow a shallow-steep decay, whereas for Fe II the decays tended to be steep-shallow. It is difficult to explain why different velocity components would be present for different lines in the same SN. Only SN2016coi showed a similar broken power-law evolution for both Fe II and Si II, though the break times and slopes are not compatible.

\section{Conclusions}\label{sec:conclusions}
   This paper has presented measurements of the expansion velocities for a large sample of type Ic-BL supernovae. These events were drawn from online archives, and are split between a population of type Ic-BL SNe with an associated GRB detection, and ordinary type Ic-BL SNe where no GRB component was found. This sample of 61 ordinary type Ic-BL SNe and 13 GRB-SNe represents the largest sample of velocity measurements for these types of events to date. The velocities have been measured using a spline fitting method.
    
   The presence of a GRB component has no impact on the expansion velocity of a type Ic-BL supernova. For all features, there is no separation between the two populations in velocity-time space. This has been confirmed statistically and by visual examination of the data for Fe II and Si II. This is in line with expectations from theory which show that the interaction of the jet with the stellar mantel should have little impact on the supernova observables. For this reason the analysis presented here is not capable of ruling out a GRB component for type Ic-BL SNe where no such component has been observed.
    
   The velocity evolution of a type Ic-BL SN may be fit with either a power-law or broken power-law model. A single power-law fit is in agreement with the expectations for a single component photosphere, while those with broken power-law fits may indicate the presence of a second (or even third) velocity component in the ejecta of some type Ic-BL supernovae. This could be evidence for asphericity in the explosion. However, there is no particular preference for broken power-laws among type Ic-BL SNe with an associated GRB, again making it difficult to rule out jets in non-GRB associated type Ic-BL SNe. The decay slopes of the Fe II and Si II features are also identical for both populations. This is likely further evidence that the jet has no direct impact on the expansion velocity of the supernova.
     
    There is some evidence that broken power-law fits may be related to engine driven explosions. Both PTF12gzk and SN2016coi show strong similarities to GRB060218-SN2006aj in terms of their velocity and velocity evolution. Emission from a GRB cocoon is expected to peak within $\sim$1 day of the explosion \citep{Izzo.2019}. Since breaks are observed at around 15 days post explosion, this raises the question of what the origin of this additional velocity component could be.
    
    These results demonstrate there is structure within the velocity evolution which has not yet been explored before now. It is likely that a greater number of broken power-law fits may be found in data with a higher sampling rate. If all type Ic-BL supernovae harbour a GRB jet, the impacts of this may only be found at very early times (<10 days), which are not probed in detail by this study. 

   Both of these issues can be addressed by changes to observing strategies, with greater emphasis placed on rapid classification and multi-wavelength follow-up of new type Ic-BL SNe. Reliable classifications are ideally needed within hours to days rather than weeks as is currently the case for type Ic-BL SNe without a GRB detection. With upcoming facilities such as the \textit{Vera Rubin Observatory} \citep{LSST2019} expected to observe many more type Ic-BL SNe, rapid classification will become even more important. To this end, the authors are investigating the possibility of classifying new type Ic-BL SNe using a machine learning algorithm applied to the pre-peak lightcurve (Cotter et al. 2025, in prep.). 

   In tandem with improving classification speed and accuracy, efforts should be made to perform multi-wavelength follow-up of all type Ic-BL SNe, particularly at X-ray and radio wavelengths. This will allow us to constrain the frequency of off-axis GRB jets in these explosions, and more clearly differentiate between ordinary type Ic-BL SNe and GRB-SNe.

\begin{acknowledgements}
      The authors would like to thank the anonymous reviewer for their thorough feedback which has greatly improved this manuscript. GF and AMC acknowledge support from the UCD Ad Astra programme. LC and AMC acknowledge support from the Irish Research Council Postgraduate Scholarship No GOIPG/2022/1008. This work has made extensive use of WISeREP (available at: \url{https://www.wiserep.org}) as one source for our type Ic-BL spectra. In order to access data from WISeREP we used the WISeREP API, created by Tomás E. Müller Bravo; we thank them for providing a modified code from which we built our data-collection pipeline. GRB-SN data was gathered from the GRBSN webtool, which can be found at \url{https://grbsn.watchertelescope.ie/}. We also cross checked the list of type Ic and type Ic-BL supernovae using the Transient Name Server (available at: \url{https://www.wis-tns.org/}). We have made use of the Rochester Astronomy List of Bright Supernovae to gather data about the detection dates of OGLE-2013-SN-134 and LSQ14bef. We also make use of the Central Bureau for Astronomical telegrams (available at: \url{http://tamkin2.eps.harvard.edu/cbet/RecentCBETs.html}) to locate some of the legacy data for SNe.
\end{acknowledgements}

\bibliographystyle{aa} 
\bibliography{references.bib} 

\begin{appendix}
\onecolumn
\begin{landscape}
\section{Supernova sample}
\begin{tiny}
\begin{longtable}{llccccp{9cm}}
\caption{Table of the 61 ordinary type Ic-BL SNe and 13 GRB-SNe used for velocity measurements, showing the event name, SN type, redshift ($z$), explosion time ($t_{exp}$) and its error ($\Delta t_{exp}$), the number of spectra in the sample and the data source. The instruments used to obtain these spectra have a range of resolving powers of 200-20000.}\\
\label{tab:sampleIcBL}\\
\toprule
\textbf{Event name} & \textbf{Type\tablefootmark{*}} & \textbf{$z$} & \textbf{$t_{exp}$ [UTC]\tablefootmark{$\dagger$}} & \textbf{$\Delta t_{exp}$ [days]} & \textbf{Num. Spectra\tablefootmark{$\ddagger$}} & \textbf{Sources\tablefootmark{$\dagger\dagger$}} \\
\midrule
\endfirsthead

\caption[]{List of Ic-BLs and GRB-SNe used for velocity measurements, continued.}\\
\toprule
\textbf{Event name} & \textbf{Type\tablefootmark{*}} & \textbf{$z$} & \textbf{$t_{exp}$ [UTC]\tablefootmark{$\dagger$}} & \textbf{$\Delta t_{exp}$ [days]} & \textbf{Num. Spectra\tablefootmark{$\ddagger$}} & \textbf{Sources\tablefootmark{$\dagger\dagger$}} \\
\midrule
\endhead

\midrule
\multicolumn{7}{r}{Continued on next page} \\
\midrule
\endfoot
\bottomrule
\endlastfoot
{GRB980425-SN1998bw} & GRB-SN & {0.008499} & {1998-04-25T21:49:11} & {} & 29 (5) & \cite{Patat.2001q8t,Pian.1999} \\
{GRB030329-SN2003dh} & GRB-SN & {0.1685} & {2003-03-29T11:37:14} & {} & 16 (0) & \cite{Matheson.2003,Kosugi.2004} \\
{GRB060218-SN2006aj} & GRB-SN & {0.033023} & {2006-02-18T03:34:31} & {} & 34 (3) & \cite{Modjaz.2014,Modjaz.2006,Pian.2006,Sonbas.2008} GRBNASASwift \\
{GRB100316D-SN2010bh} & GRB-SN & {0.0593} & {2010-03-16T12:44:50} & {} & 16 (0) & \cite{Bufano.2012} GRBNASASwift \\
{GRB101219B-SN2010ma} & GRB-SN & {0.55185} & {2010-12-19T16:27:53} & {} & 6 (0) & \cite{Sparre.2011} GRBNASASwift \\
{GRB120422A-SN2012bz} & GRB-SN & {0.283} & {2012-04-22T07:12:03} & {} & 13 (0) & \cite{Melandri.2012} GRBNASASwift \\
{GRB130702A-SN2013dx} & GRB-SN & {0.15} & {2013-07-02T00:05:23} & {} & 29 (0) & \cite{DElia.2015,Taddia.2019,Schulze.2021,Volnova.2017} \\
{GRB161219B-SN2016jca} & GRB-SN & {0.1475} & {2016-12-19T18:48:39} & {} & 3 (0) &  GRBNASASwift \\
{GRB171010A-SN2017htp} & GRB-SN & {0.328} & {2017-10-10T19:00:50} & {} & 5 (0) & \cite{Bright.2019} \\
{GRB171205A-SN2017iuk} & GRB-SN & {0.0368} & {2017-12-05T07:20:43} & {} & 26 (0) &  GRBNASASwift \\
{GRB180728A-SN2018fip} & GRB-SN & {0.117} & {2018-07-28T17:29:00} & {} & 14 (0) & GRBNASASwift \\
{GRB190829A-AT2019oyw} & GRB-SN & {0.0785} & {2019-08-29T19:56:44} & {} & 10 (0) &  GRBNASASwift \\
{SN2020bvc} & GRB-SN & {0.025235} & {2020-02-03T16:04:48} & {} & 14 (0) & \cite{Ho.2020} \\
{1997ef} & Ic-BL & {0.011805} & {1997-11-20T00:00:00} & {} & 38 (7) & \cite{Modjaz.2014,Matheson.2001,Mazzali.2000} \\
{1997dq} & Ic-BL & {0.003196} & {1997-09-29T00:00:00} & {} & 15 (5) & \cite{Modjaz.2014,Matheson.2001,Mazzali.2004} \\
{1998ey} & Ic-BL & {0.016} & {1998-12-02T15:37:26} & {2.0} & 6 (0) & \cite{Arbour.1998} \\
{2002bl} & Ic-BL & {0.016} & {2002-02-22T10:48:00} & {8.4} & 4 (0) & \cite{Shivvers.2019,Armstrong.2002} \\
{2002ap} & Ic-BL & {0.002108} & {2002-01-27T00:00:00} & {2.0} & 43 (12) & \cite{Modjaz.2014,Foley.2003,Chornock.2013,Gal-Yam.2002} \\
{2003bg} & Ic-BL/IIb\tablefootmark{a} & {0.004403} & {2003-02-25T00:00:00} & {} & 14 (6) & \cite{Hamuy.2009,Mazzali.2009} \\
{2003jd} & Ic-BL & {0.01886} & {2003-10-20T12:00:00} & {4.5} & 29 (4) & \cite{Modjaz.2014,Shivvers.2019,Valenti.2008} \\
{2005da} & Ic-BL & {0.01501} & {2005-07-16T08:38:24} & {2.0} & 4 (0) & \cite{Modjaz.2014,Shivvers.2019,Lee.2005} \\
{2006nx} & Ic-BL & {0.137} & {2006-11-11T05:02:24} & {0.71} & 3 (0) & \cite{Taddia.2015} \\
{2007ce} & Ic-BL & {0.046} & {2007-04-10T02:24:00} & {5.2} & 11 (0) & \cite{Modjaz.2014,Shivvers.2019,Gomez.2022} \\
{2007ru} & Ic-BL & {0.015464} & {2007-11-25T12:00:00} & {3.0} & 5 (0) & \cite{Modjaz.2014,Shivvers.2019,Sahu.2009} \\
{2007I} & Ic-BL & {0.021638} & {2007-01-12T10:33:36} & {2.0} & 7 (3) & \cite{Modjaz.2014,Shivvers.2019,Lee.2007} \\
{2007D} & Ic/Ic-BL\tablefootmark{b} & {0.023146} & {2007-01-05T07:12:00} & {0.3} & 4 (0) & \cite{Modjaz.2014,Shivvers.2019,Drout.2011,Shan-Qin.2019} \\
{2008ew} & Ic/Ic-BL\tablefootmark{c} & {0.02026} & {2008-08-08T05:16:48} & {2.0} & 4 (0) & \cite{Shivvers.2019,Griffith.2008}, SNID\\
{2009bb} & Ic-BL (pec.)\tablefootmark{d} & {0.0104} & {2009-03-19T02:24:00} & {0.6} & 28 (2) & \cite{Stritzinger.2023,Pignata.2011} \\
{2009dr} & Ic-BL & {0.199} & {2009-04-15T06:00:00} & {2.0} & 6 (1) & \cite{Quimby.2009} \\
{2009ca} & Ic-BL/SLSN\tablefootmark{e} & {0.0957} & {2009-03-25T08:38:24} & {3.35} & 4 (0) & \cite{Stritzinger.2023,Taddia.2018} \\
{PTF10vgv} & Ic-BL & {0.015} & {2010-09-13T23:16:48} & {2.0} & 10 (3) & \cite{Corsi.2012,Taddia.2019} \\
{2010ah} & Ic-BL & {0.0498} & {2010-02-21T00:00:00} & {} & 4 (0) & \cite{Corsi.2011,Taddia.2019,Mazzali.2013} \\
{PTF10ysd} & Ic-BL & {0.0963} & {2010-09-25T12:00:00} & {2.0} & 3 (0) & \cite{Taddia.2019} \\
{PTF10xem} & Ic-BL & {0.0567} & {2010-09-22T03:36:00} & {2.0} & 5 (2) & \cite{Taddia.2019} \\
{PTF10tqv} & Ic-BL & {0.0795} & {2010-08-14T04:04:48} & {2.0} & 3 (0) & \cite{Taddia.2019} \\
{PTF10qts} & Ic-BL & {0.0907} & {2010-08-03T14:24:00} & {2.0} & 13 (2) & \cite{Walker.2014,Taddia.2019} \\
{PTF10gvb} & Ic/Ic-BL\tablefootmark{f} & {0.098} & {2010-04-29T14:24:00} & {2.0} & 6 (0) & \cite{Taddia.2019,Quimby.2018} \\
{PTF10aavz} & Ic-BL & {0.062} & {2010-11-07T11:31:12} & {2.0} & 4 (1) & \cite{Taddia.2019} \\
{2010ay} & Ic-BL & {0.0671} & {2010-02-21T07:12:00} & {1.3} & 3 (0) & \cite{Sanders.2012,Shivvers.2019,Sanders.2012a} \\
{PTF11qcj} & Ic-BL & {0.028} & {2011-10-08T12:00:00} & {} & 6 (1) & \cite{Corsi.2014} \\
{PTF12gzk} & Ic/Ic-BL\tablefootmark{g} & {0.01377} & {2012-07-23T07:12:00} & {0.6} & 55 (12) & \cite{Childress.2016,Shivvers.2019,Ben-Ami.2012} \\
{PTF12as} & Ic-BL & {0.033} & {2011-12-25T12:14:24} & {2.0} & 3 (1) & \cite{Taddia.2019} \\
{2012ap} & Ic-BL (pec.)\tablefootmark{h} & {0.01224} & {2012-02-07T17:02:23} & {2.5} & 15 (4) & \cite{Milisa.2015,Liu.2015} \\
{2012aa} & Ic-BL/SLSN\tablefootmark{i} & {0.0799} & {2011-12-26T12:00:00} & {10.0} & 4 (1) & \cite{Shivvers.2019,Roy.2016} \\
{OGLE-2013-SN-134} & Ic/Ic-BL\tablefootmark{j} & {0.039} & {2013-12-08T07:29:48} & {2.0} & 2 (0) & \cite{Wyrzykowski.2013}, SNID \\
{iPTF13ebw} & Ic-BL & {0.069} & {2013-11-15T17:02:24} & {2.0} & 6 (0) & \cite{Schulze.2021,Taddia.2019} \\
{2013bn} & Ic-BL & {0.054} & {2013-04-03T23:02:24} & {2.0} & 5 (0) & \cite{Taddia.2019,Shivvers.2019} \\
{LSQ14bef} & Ic-BL & {0.05} & {2014-04-20T00:00:00} & {2.0} & 1 (0) & \cite{Polshaw.2014} \\
{2014cp} & Ic-BL & {0.016164} & {2014-06-23T00:00:00} & {} & 10 (6) & \cite{Childress.2016,Zheng.2022} \\
{2014ad} & Ic-BL & {0.005} & {2014-03-08T12:00:00} & {3.0} & 31 (5) & \cite{Sahu.2018} \\
{iPTF14gaq} & Ic-BL & {0.0826} & {2014-09-19T17:02:24} & {2.0} & 3 (0) & \cite{Taddia.2019} \\
{iPTF14dby} & Ic-BL & {0.074} & {2014-06-19T13:55:12} & {2.0} & 4 (1) & \cite{Taddia.2019} \\
{iPTF15eov} & Ic-BL & {0.0535} & {2015-12-02T05:16:48} & {2.0} & 17 (0) & \cite{Taddia.2019} \\
{iPTF15dqg} & Ic-BL & {0.065} & {2015-11-02T05:31:12} & {2.0} & 4 (0) & \cite{Taddia.2019} \\
{iPTF15dld} & Ic-BL & {0.047} & {2015-10-03T00:00:00} & {1.0} & 6 (0) & \cite{Pian.2017} \\
{2016gox} & Ic-BL & {0.042} & {2016-09-10T02:09:36} & {2.0} & 4 (0) & \cite{Taddia.2019} \\
{2016P} & Ic/Ic-BL\tablefootmark{k} & {0.0146} & {2016-01-16T00:00:00} & {2.0} & 19 (1) & \cite{Prentice.2019,Gangopadhyay.2020} \\
{2016ilj} & Ic-BL & {0.039711} & {2016-11-08T18:00:00} & {2.0} & 3 (0) & \cite{Taddia.2019} \\
{2016coi} & Ic-BL (pec.)\tablefootmark{l} & {0.0036} & {2016-05-23T21:36:00} & {1.5} & 100 (26) & \cite{Prentice.2018,Terreran.2019} \\
{2017dgk} & Ic/Ic-BL\tablefootmark{m} & {0.065} & {2017-04-21T22:31:42} & {2.0} & 2 (0) & \cite{Gutierrez.2017}, SNID \\
{2017ens} & Ic-BL/IIn\tablefootmark{n} & {0.1086} & {2017-06-03T19:12:00} & {1.5} & 17 (14) & \cite{Chen.2018} \\
{2017ifh} & Ic-BL & {0.039} & {2017-10-30T00:15:50} & {7.5} & 5 (0) & \cite{Prentice.2019} \\
{2017fwm} & Ic-BL & {0.016} & {2017-07-26T18:37:26} & {0.2} & 6 (0) &  WISeREP, SNID\\
{2017dio} & Ic-BL/IIn\tablefootmark{o} & {0.037} & {2017-04-14T06:00:00} & {0.25} & 15 (0) & \cite{Kuncarayakti.2018,Shi.2024} \\
{2017dcc} & Ic-BL & {0.0245} & {2017-04-12T11:45:38} & {3.96} & 9 (2) & \cite{Tonry.2017dcc} \\
{2017cw} & Ic-BL & {0.093} & {2016-12-29T21:36:00} & {2.0} & 3 (0) & \cite{Taddia.2019} \\
{2018kqr} & Ic-BL & {0.045} & {2018-12-06T12:07:19} & {3.0} & 2 (0) & \cite{De.2020}, SNID \\
{2018gsk} & Ic-BL & {0.0116} & {2018-09-15T15:15:50} & {2.0} & 13 (1) & \cite{Tonry.2018gsk}, SNID \\
{2018gju} & Ic-BL & {0.05} & {2018-09-13T12:30:14} & {2.0} & 1 (0) & \cite{Tonry.2018gju}, SNID \\
{2018giu} & Ic-BL & {0.026} & {2018-09-13T21:23:45} & {1.5} & 7 (0) & \cite{Tonry.2018giu}, SNID \\
{2018gep} & Ic-BL (pec.)\tablefootmark{p} & {0.033} & {2018-09-09T02:24:00} & {0.95} & 7 (0) & \cite{Pritchard.2021} \\
{2018cbz} & Ic-BL & {0.0223} & {2018-05-29T08:54:14} & {2.0} & 6 (0) & \cite{Prentice.2019,Tonry.2018cbz} \\
{2020abdw} & Ic/Ic-BL\tablefootmark{q} & {0.0327790007} & {2020-11-21T06:02:28} & {4.0} & 3 (0) & \cite{ZTF.2020abdw}, SNID \\
{2022crr} & Ic-BL & {0.0188} & {2022-02-08T06:11:31} & {0.94} & 3 (0) & WISeREP, SNID \\
\midrule
\multicolumn{7}{r}{\tablefoot{
    \tablefoottext{*}{For cases where the SN type is ambiguous, the second-most likely type is shown. Refer to the lettered footnotes for further information.}
    \tablefoottext{$\dagger$}{In some instances, $\Delta t_{exp}$ was not known for an SN, while for the GRB-SNe $t_{exp}$ is taken from the GRB trigger time, meaning that the error is just a few seconds.}
    \tablefoottext{$\ddagger$}{The total number of spectra for each SN is shown in this column, with the number in brackets indicates the number of these spectra that show evidence of emission/nebular features.}
    \tablefoottext{$\dagger\dagger$}{In the \textit{Sources} column, SNID indicates that an SN was classified by the authors using the SNID tool \citep{Blondin.2007} and GRBNASASwift indicates that the redshift and explosion times of a GRB-SN was sourced from \url{https://swift.gsfc.nasa.gov/archive/grb_table/}.}
    \tablefoottext{a}{SN2003bg is suggested by \citep{Hamuy.2009} to be a type IIb hypernova, because the late time nebular spectra look somewhat like a type Ib/c SN (lacking hydrogen). However, \citep{Soderberg.2006} suggest it started as a type Ic-BL SN before evolving to a type II SN before becoming a type Ib/c SN. This SN is also radio loud and X-ray luminous \citep{Soderberg.2006}.} 
    \tablefoottext{b}{SN2007D Possible transition object between a type Ic SN and a type Ic-BL SN. It has clear type Ic-BL SN broad features, however the velocities are much lower than typical type Ic-BL SNe \citep{Shan-Qin.2019}. Additionally it has produced significantly more Ni than type Ic/Ic-BL SNe.} 
    \tablefoottext{c}{SN2008ew was classified for this publication using SNID. Both type Ic SN templates and type Ic-BL SN templates provide reasonable fits.}
    \tablefoottext{d}{SN2009bb is an engine-driven type Ic-BL SN with strong radio emission \citep{Pignata.2011}. A choked jet scenario has been proposed for this SN \citep{Margutti.2014}.}
    \tablefoottext{e}{SN2009ca was initially claimed as a type Ic-BL SN by \cite{Taddia.2019}, however more recent analysis by \cite{Stritzinger.2023} suggested it may be a SLSN.}}}\\

\multicolumn{7}{r}{\tablefoot{
    \tablefoottext{f}{PTF10gvb was initially claimed as a type Ic-BL SN by \cite{Taddia.2019}, however analysis by \cite{Quimby.2018} suggested it may be a SLSN.} 
    \tablefoottext{g}{PTF12gzk has high velocities, similar to type GRB-SNe \citep{Modjaz.2016}. However, its spectra show narrow more consistent with type Ic SNe \citep{Ben-Ami.2012}. Radio indicates relativistic material \citep{Horesh.2013}.}
    \tablefoottext{h}{SN2012ap is an engine-driven type Ic-BL SN with evidence for a relativistic outflow observed in radio \citep{Chakraborti.2015, Milisa.2015}.}
    \tablefoottext{i}{SN2012aa has a luminosity second bump in the late time optical light curve that suggest it is intermediate between type Ic-BL SN and SLSN\citep{Roy.2016}. Spectroscopically it appears similar to a type Ic-BL SN such as SN2007gr and SN2006aj\citep{Roy.2016}.}
    \tablefoottext{j}{OGLE-2013-SN-134 was classified for this publication using SNID. It appears to show good fits to type Ic SNe and type Ic-BL SNe. However, this classification is based on a single spectrum.}
    \tablefoottext{k}{SN2016P is transitions from a type Ic-BL SN at early times \citep{Gangopadhyay.2020}), but later on becomes more like a type Ic SN \citep{Prentice.2019}.}
    \tablefoottext{l}{SN2016coi is a peculiar type Ic-BL SN in that it exhibits a helium feature early in its evolution, indicating that it is not fully stripped \citep{Terreran.2019, Prentice.2018}. This SN has similar radio and X-ray emission to engine-driven SNe such as SN2009bb and SN2012ap\citep{Terreran.2019}.}
    \tablefoottext{m}{SN2017dgk was classified for this publication using SNID. It appears to show good fits to type Ic SNe and type Ic-BL SNe. However, this classification is based on a single spectrum.}
    \tablefoottext{n}{SN2017ens undergoes a transition from a type Ic-BL SN to a type IIn SN, and is impacted by CSM interaction \citep{Chen.2018}. As such, we only obtained velocities from the first $\sim$1 month of spectra, and only one Fe II measurement was made. There is also some evidence that this SN is a SLSN \citep{Margutti.2023}.}
    \tablefoottext{o}{SN2017dio undergoes a transition from a type Ic-BL SN to a type IIn SN at around 18 days, and is impacted by CSM interaction \citep{Kuncarayakti.2018}. As such, we only obtained velocities from the first $\sim$15 days of spectra.}
    \tablefoottext{p}{SN2018gep is similar to fast-blue optical transients in terms of its rise time and its initial very blue continuum \citep{Pritchard.2021}.}
    \tablefoottext{q}{SN2020abdw was classified for this publication using SNID. Although there was a similar template type Ia SN, the better fits visually are for type Ic/Ic-BL SNe, with most being type Ic-BL SNe. Notably it matches PTF12gzk which is a type Ic/Ic-BL SN.}
}}
\end{longtable}

\end{tiny}
\end{landscape}
\section{Measuring the velocities of spectral features using splines}\label{sec:linemethod}
   An absorption feature is characterised by its width, depth (strength) and the wavelength at which it reaches its minimum flux. Absorption features in supernovae are generated above the photosphere, when some of the photons from the photosphere are absorbed by electrons bound within the atoms of the ejecta, forming a spectral line. The lines observed in supernovae do not appear at only one wavelength value, instead they show a broadened profile across a range of wavelengths. This occurs due to the bulk motion of the particles of the gas relative to the photons. This leads to several narrow lines side by side in the spectrum, producing a broadened line. If the rest-wavelength of a spectral line can be identified, then the expansion velocity of the gas can be measured using the Doppler shift formula for wavelength, 

   \begin{equation}
     v = \frac{1-\beta}{1+\beta}\times c,
     \label{eq:doppler}
   \end{equation}

   where $c$ is the speed of light and $\beta$ is given by

   \begin{equation}
      \beta = \frac{\lambda^2_{obs}}{\lambda^2_{true}},
   \end{equation}

   where $\lambda_{obs}$ is the wavelength of the absorption feature measured from the spectrum, and $\lambda_{true}$ is the wavelength of the same line emitted from a gas at rest with respect to the observer.

   \cite{Silverman.2012} proposed a method of determining the Doppler-shifted wavelength of an absorption feature in the spectrum of a Ia supernova. The basic tenet of this method is to fit a smooth function to the spectrum in the region of an absorption feature. The wavelength is determined from the point at which the fit reaches its minimum. This method has also been used for SESN in the past \citep{Liu.2016}, and a similar idea was used to measure the Si II 6355\,\AA\,feature for GRB130702A-SN2013dx \citep{DElia.2015}, a type Ic-BL SN with an associated GRB. 

   For this analysis, a method similar to those in the literature was implemented in Python. The spectrum is first smoothed using a Savitzky-Golay filter as described in Sect. \ref{sec:smoothing}. The user can adjust the smoothing window length, and a comparison of the result is displayed graphically.

   In order to estimate the error on the velocity, a Monte-Carlo analysis was conducted. Once the smoothed spectrum is generated, it is used to generate an uncertainty array for the analysis, in a similar manner to \cite{Liu.2016}. First, the residuals between the input and smoothed spectra are calculated. Then, a noise array is generated, where the noise at each is drawn from a Gaussian distribution with zero mean and standard deviation computed from the residual array. Samples from the noise array are added to the smoothed spectrum to produce 1000 mock spectra.

   Prior to fitting, the user selects the region of interest in the spectrum, in the case of type Ic-BL SN spectra, the features are: Fe II at 5169\,\AA and Si II at 6355\,\AA\. The red and blue boundaries of the features are selected manually on the spectrum by clicking the plot. The ultimate location of the boundaries is determined from the peaks in the smoothed mock spectra which are nearest to the user-selected points. A delta value can also be specified, which defines the region (in resolution elements) in which the code searches for a local maximum relative to the user's input. The value of $\delta$ is kept low, typically 30-50 resolution elements and is chosen so as to ensure the selection of the most appropriate local maximum for the feature in question (see Fig. \ref{fig:splinefit}). This approach facilitates setting an approximate starting boundary, allowing the wavelength to vary based on the exact maxima in the Monte-Carlo spectra. In any case, the method is relatively insensitive to the choice of the bounding region, a conclusion which was also made by \cite{Silverman.2012}.
   
   Once the boundaries are determined for a given spectral feature, a cubic spline is fit to each smoothed mock spectrum (see Fig. \ref{fig:splinefit}), producing a distribution of velocities. The median velocity and the 16th and 84th percentiles are taken as the estimate for velocity and its error respectively. The cubic spline is agnostic to the shape of the complex absorption and emission processes that may shape a spectral feature. For example, the absorption in a P-Cygni profile would not produce the same feature as pure absorption would. The wavelength of the spline fit at its minimum is converted to a Doppler velocity using Eq. \ref{eq:doppler}. Figure \ref{fig:splinefit} shows a detailed example of this method applied to one of the SNe analysed in this paper.

\clearpage
\section{Additional velocity evolution figures}\label{sec:additional_velocity_evolution_figures}

   \begin{figure*}[!h]
      \centering{}
      \includegraphics[width=0.9\linewidth]{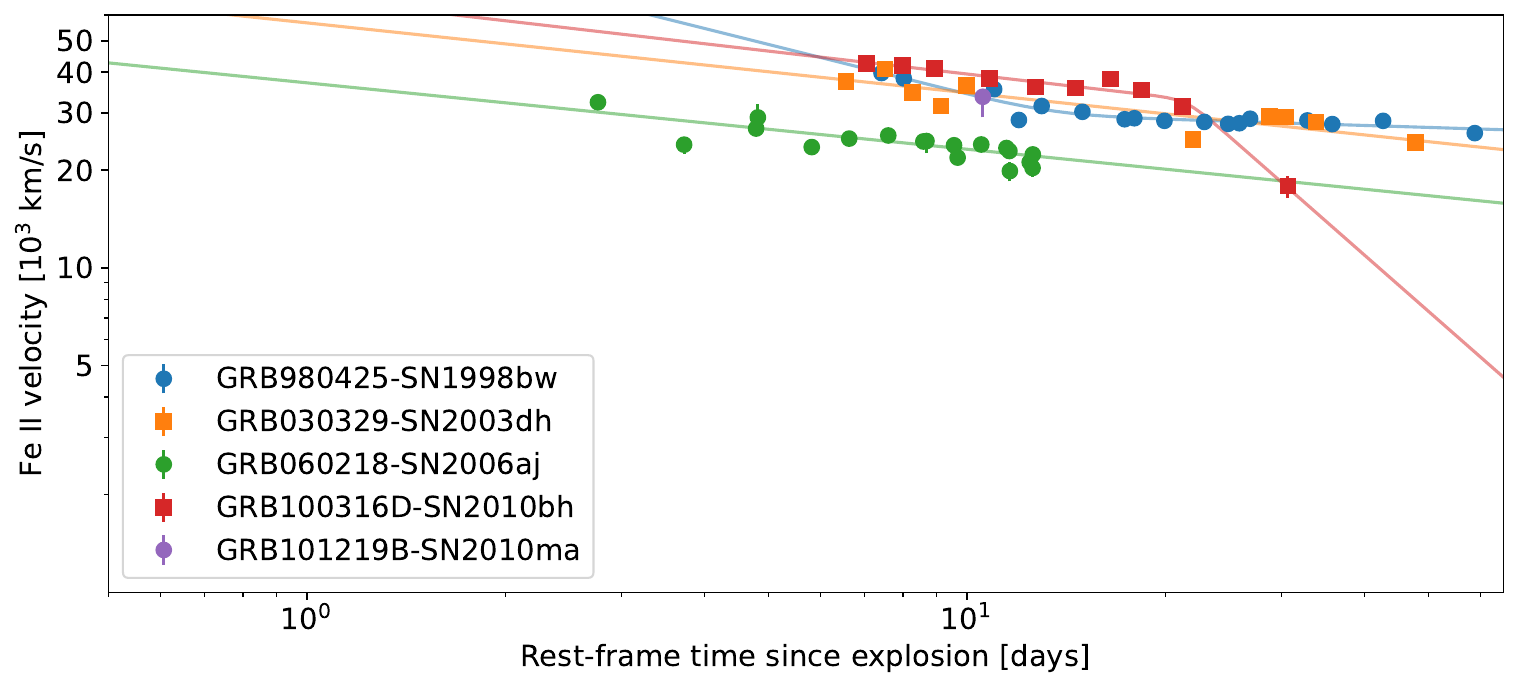}
      \includegraphics[width=0.9\linewidth]{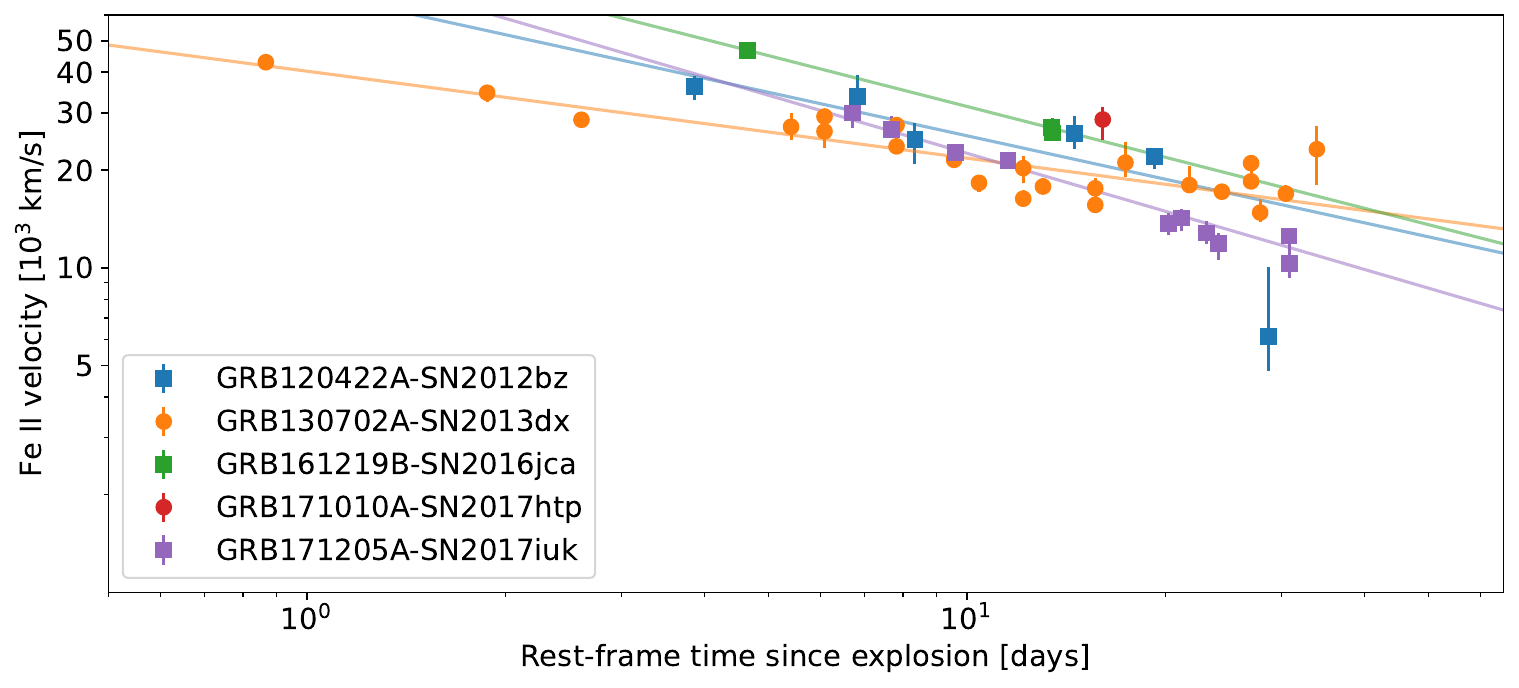}
      \includegraphics[width=0.9\linewidth]{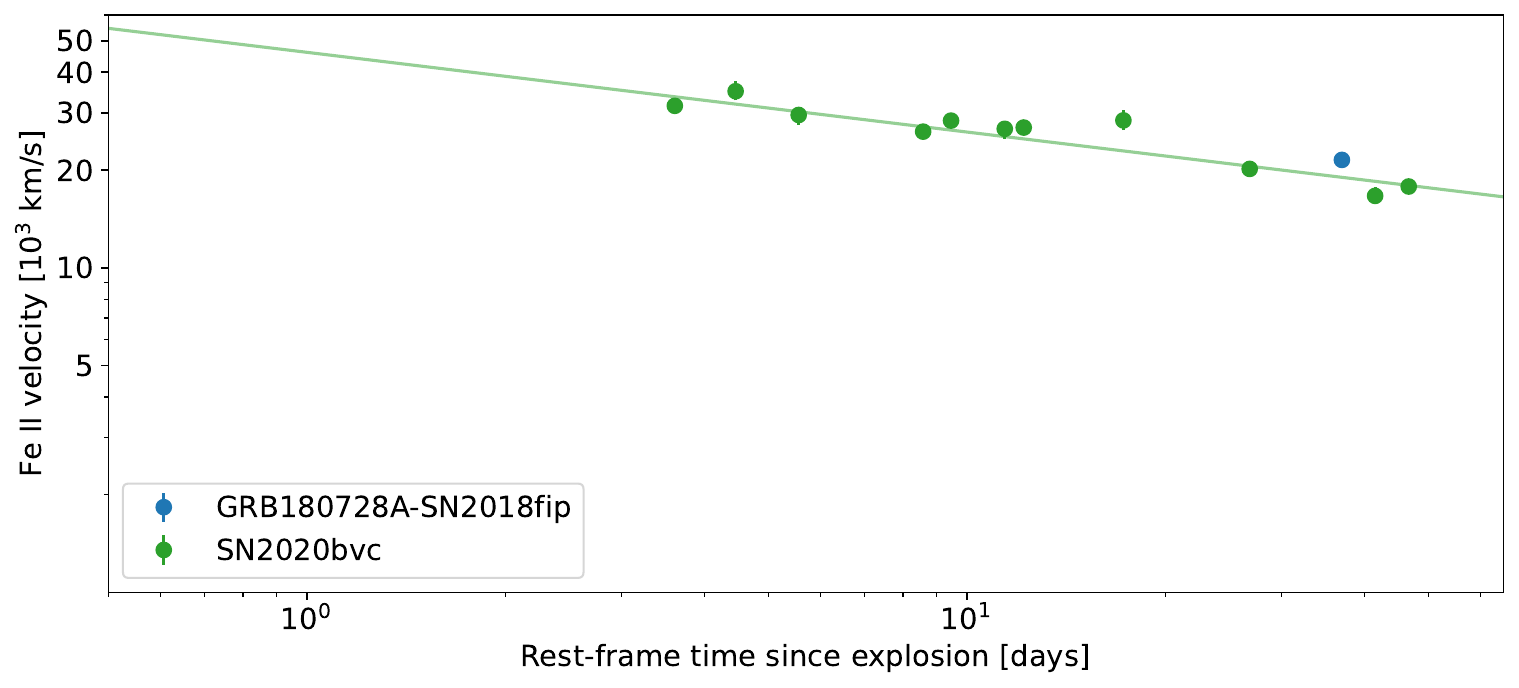}
      \caption{Fe II velocities of a subset of GRB-SNe from the \textit{Gold}, \textit{Silver} and \textit{Bronze} samples. Power-law and broken power-law fits are shown as solid lines.}
   \end{figure*}

   \begin{figure*}[!h]
      \centering
      \includegraphics[width=0.9\linewidth]{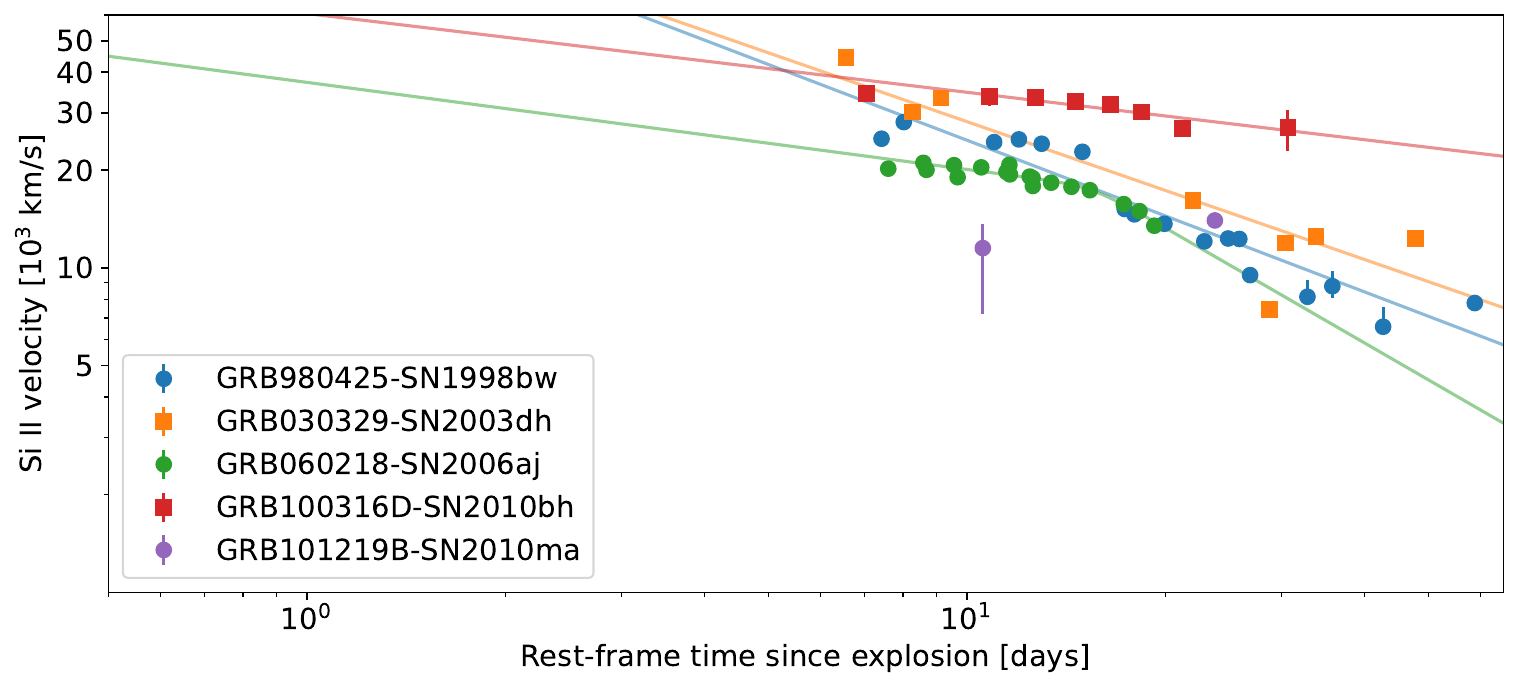}
      \includegraphics[width=0.9\linewidth]{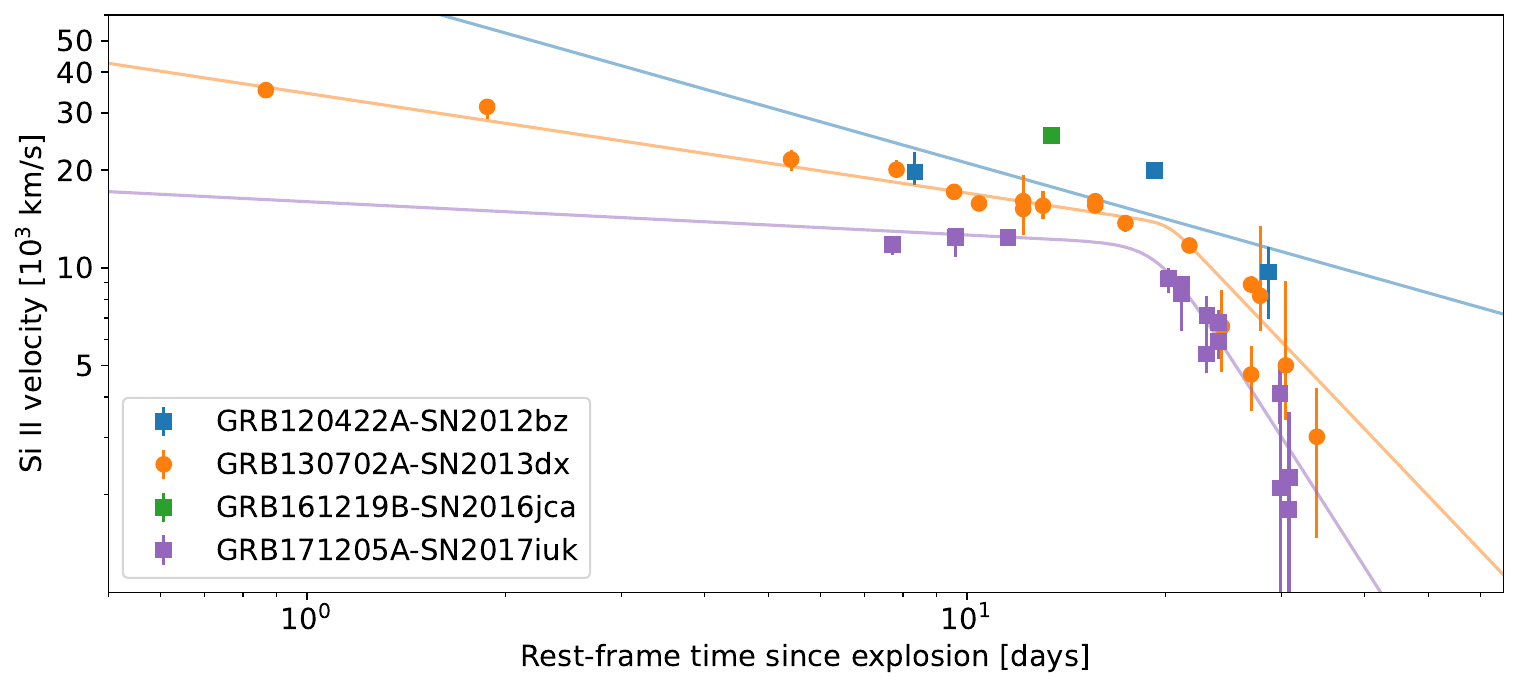}
      \includegraphics[width=0.9\linewidth]{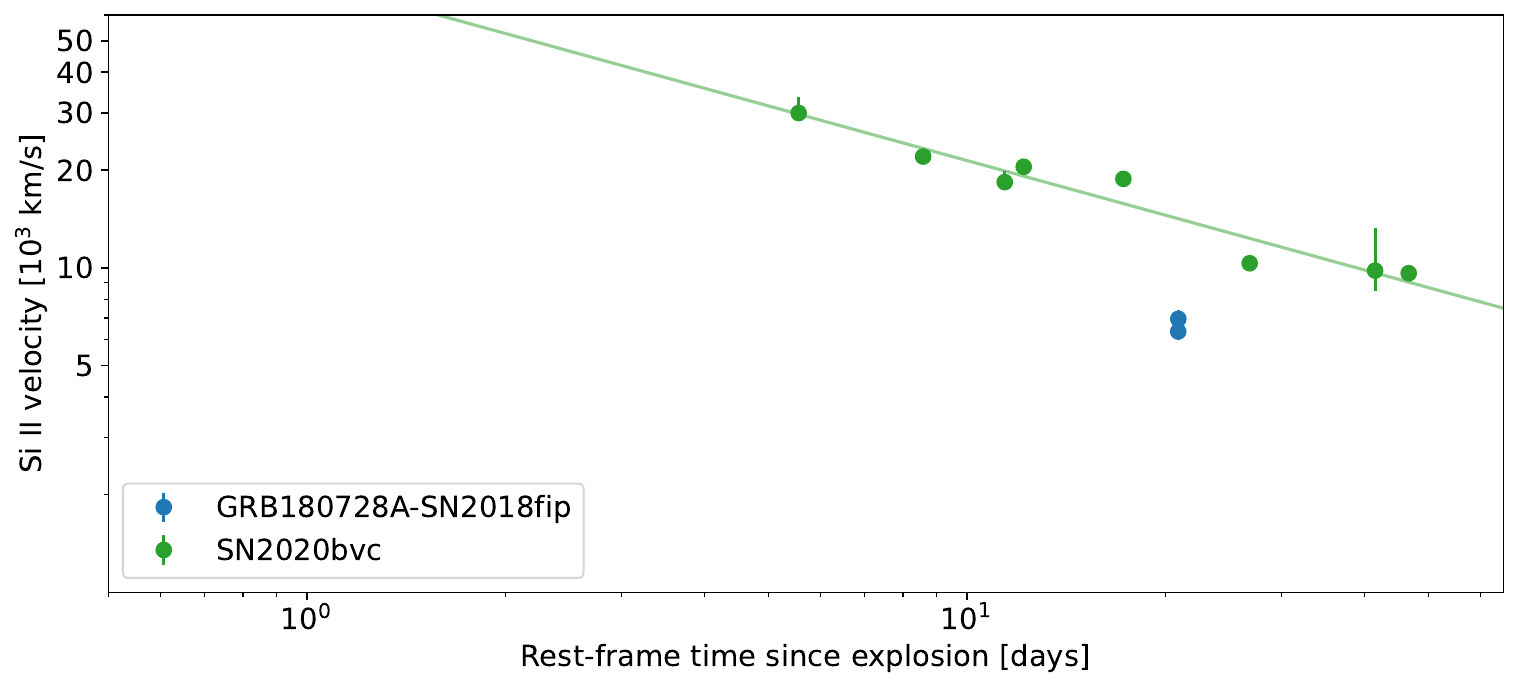}
      \caption{Si II velocities of a subset of GRB-SNe from the \textit{Gold}, \textit{Silver} and \textit{Bronze} samples. Power-law and broken power-law fits are shown as solid lines.}
   \end{figure*}

   \begin{figure*}[!h]
      \centering
      \includegraphics[width=0.9\linewidth]{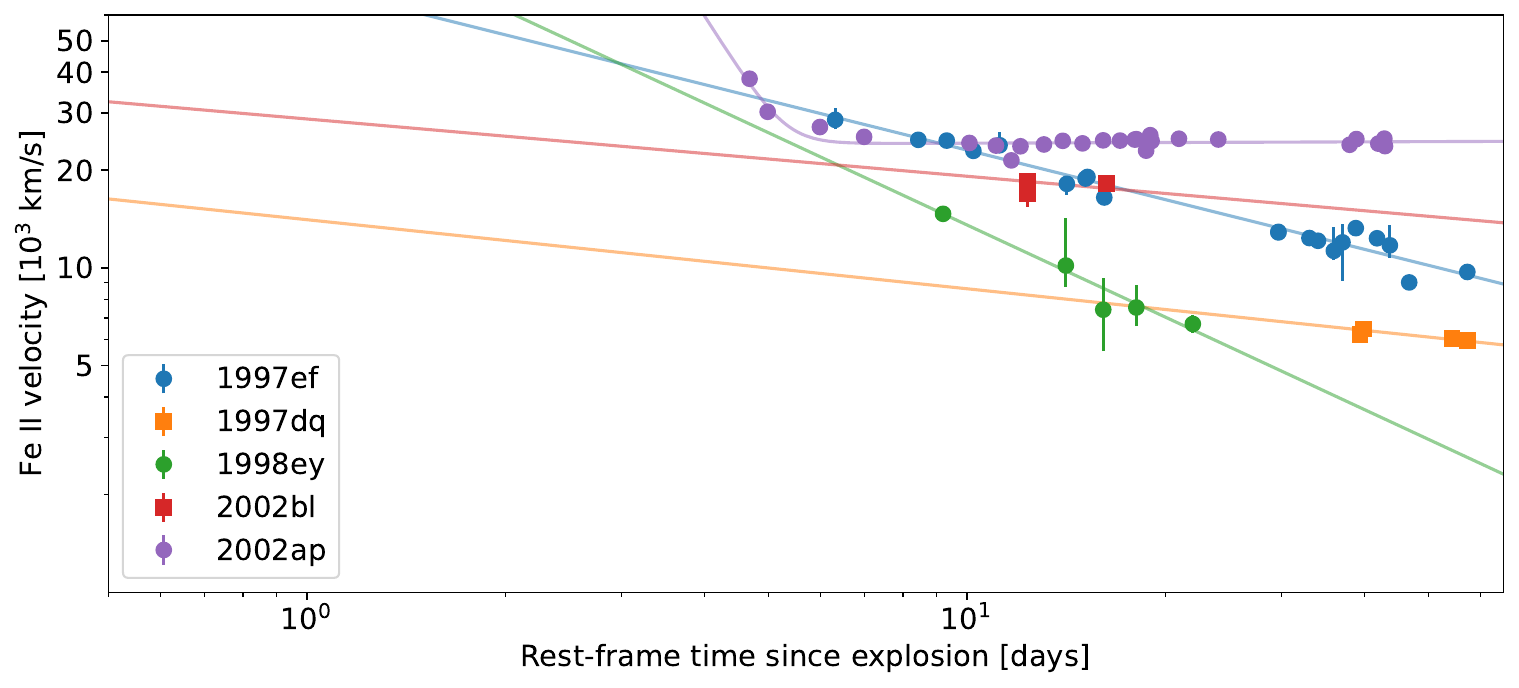}
      \includegraphics[width=0.9\linewidth]{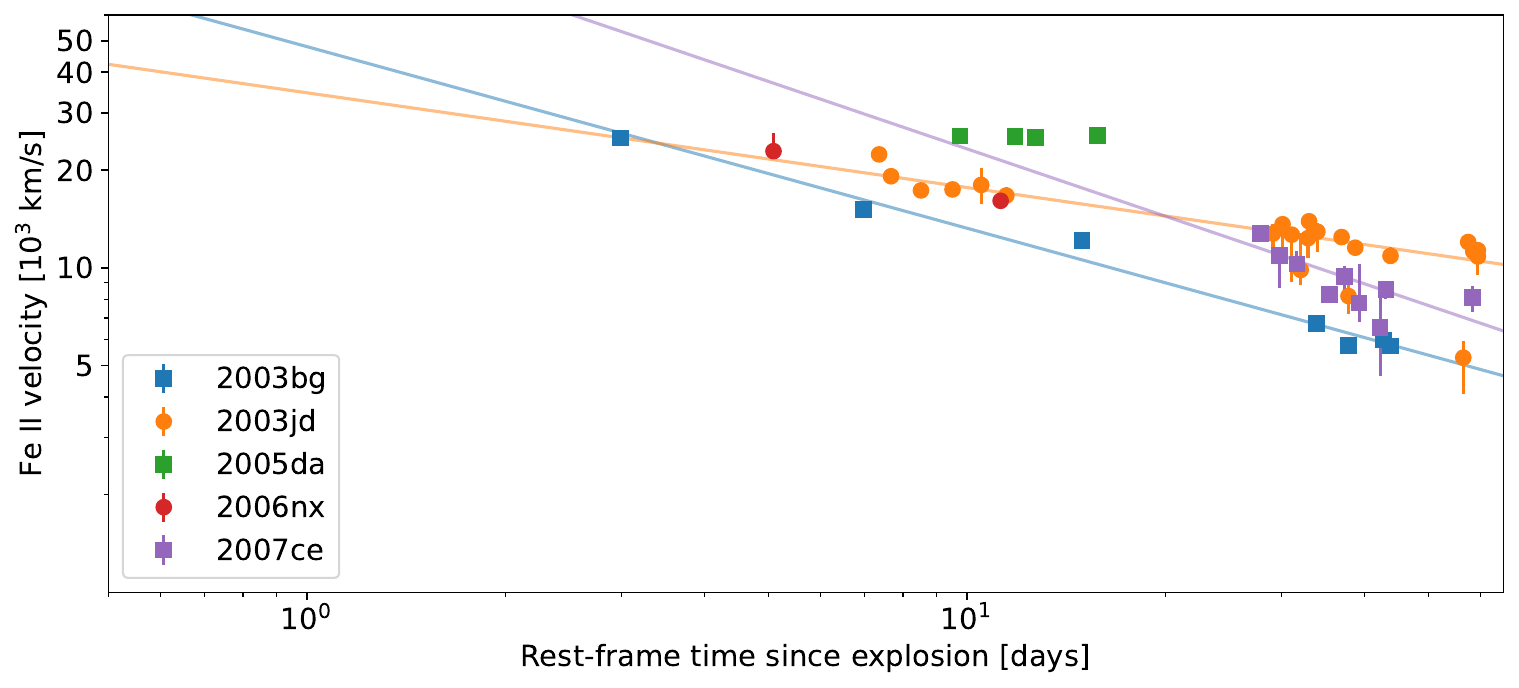}
      \includegraphics[width=0.9\linewidth]{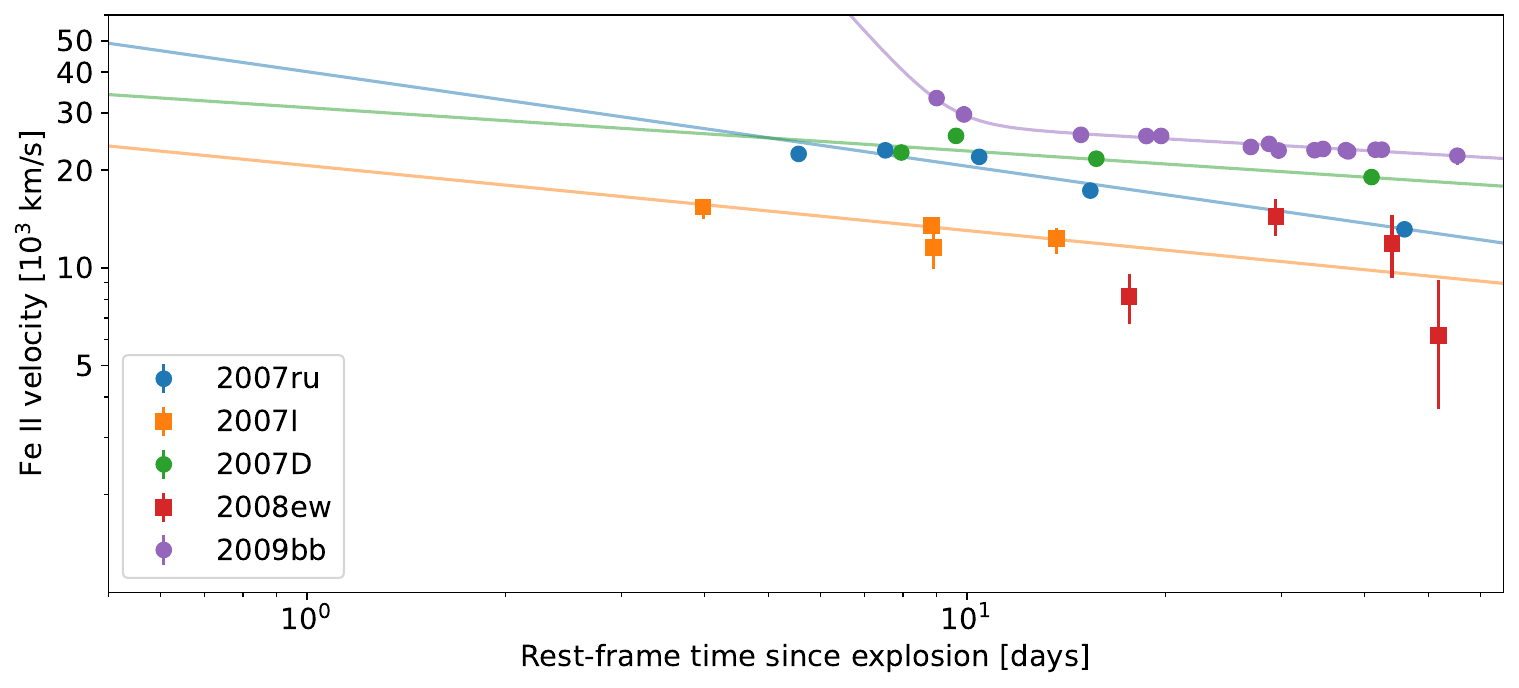}
      \caption{Fe II velocities of a subset of ordinary type Ic-BL SNe from the \textit{Gold}, \textit{Silver} and \textit{Bronze} samples. Power-law and broken power-law fits are shown as solid lines.}
   \end{figure*}

   \begin{figure*}[!h]
      \centering
      \includegraphics[width=0.9\linewidth]{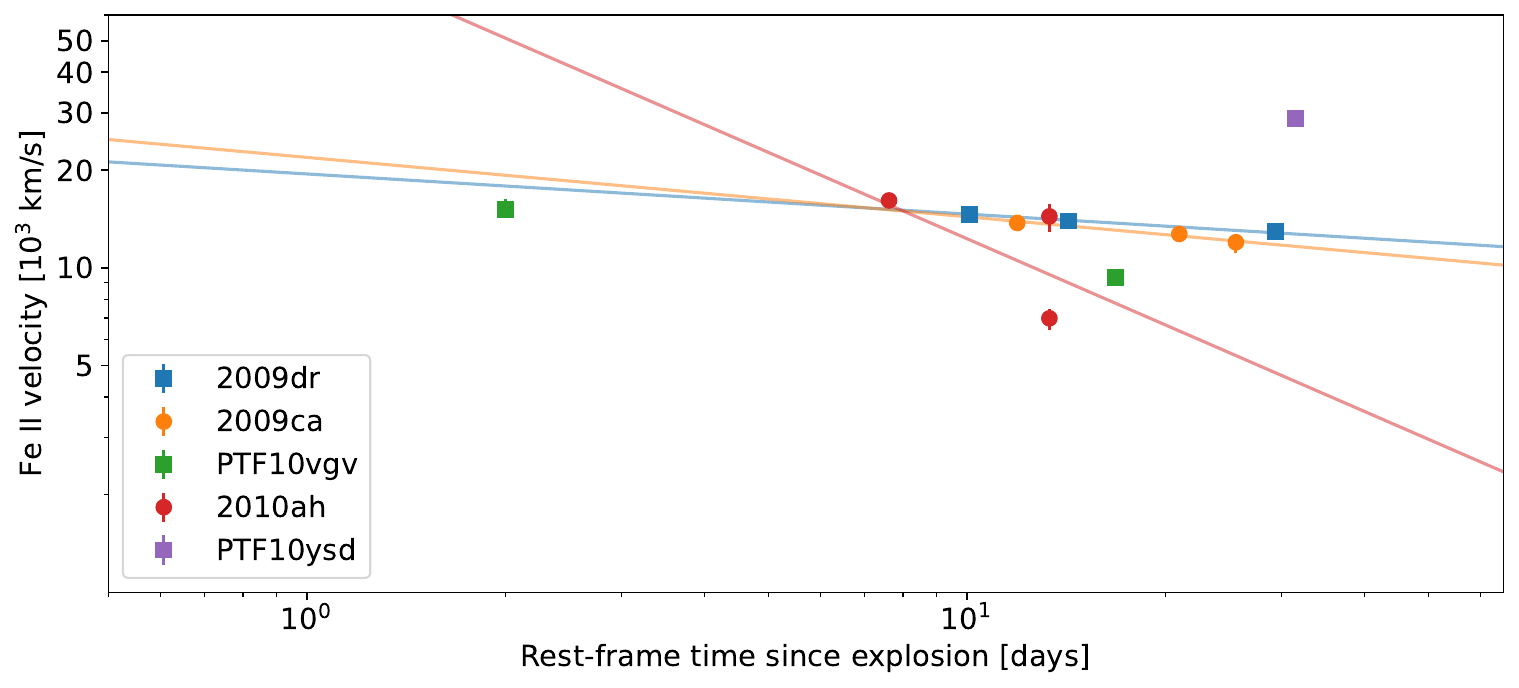}
      \includegraphics[width=0.9\linewidth]{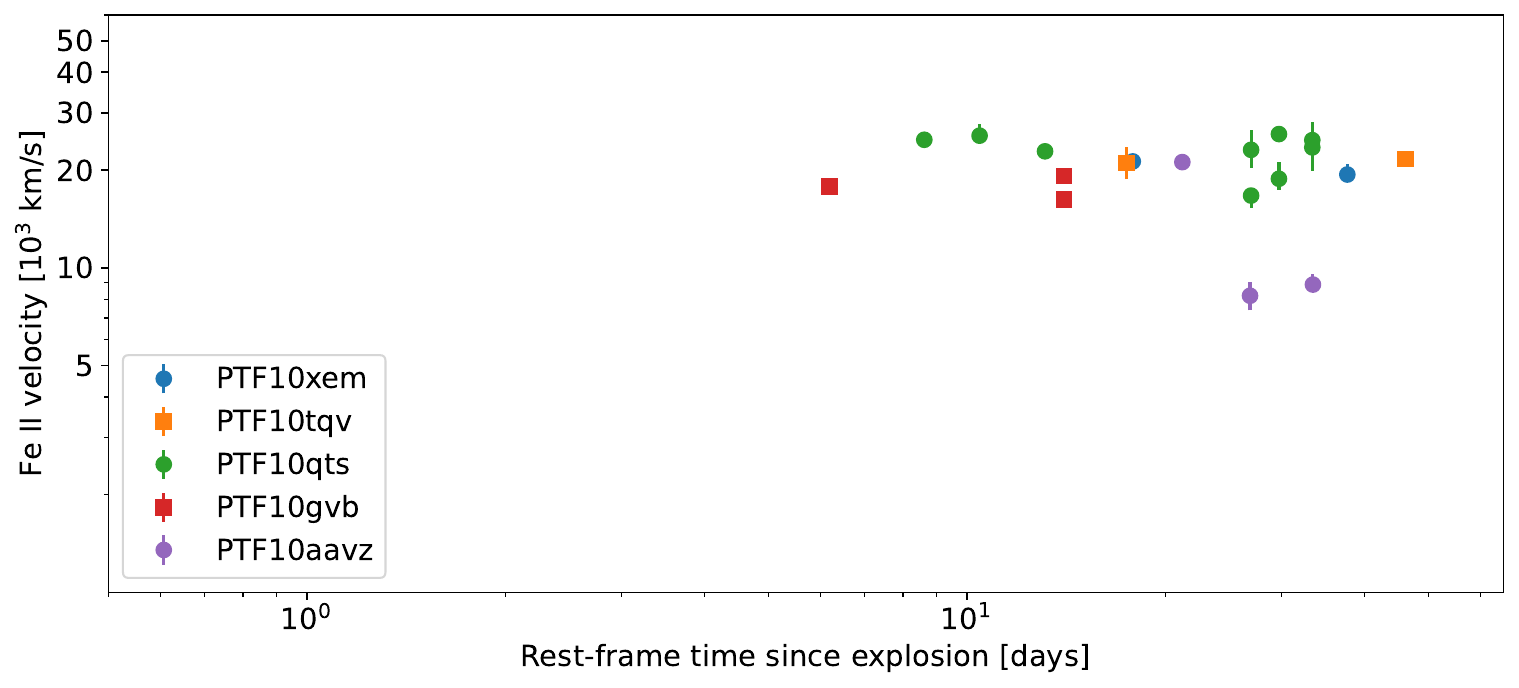}
      \includegraphics[width=0.9\linewidth]{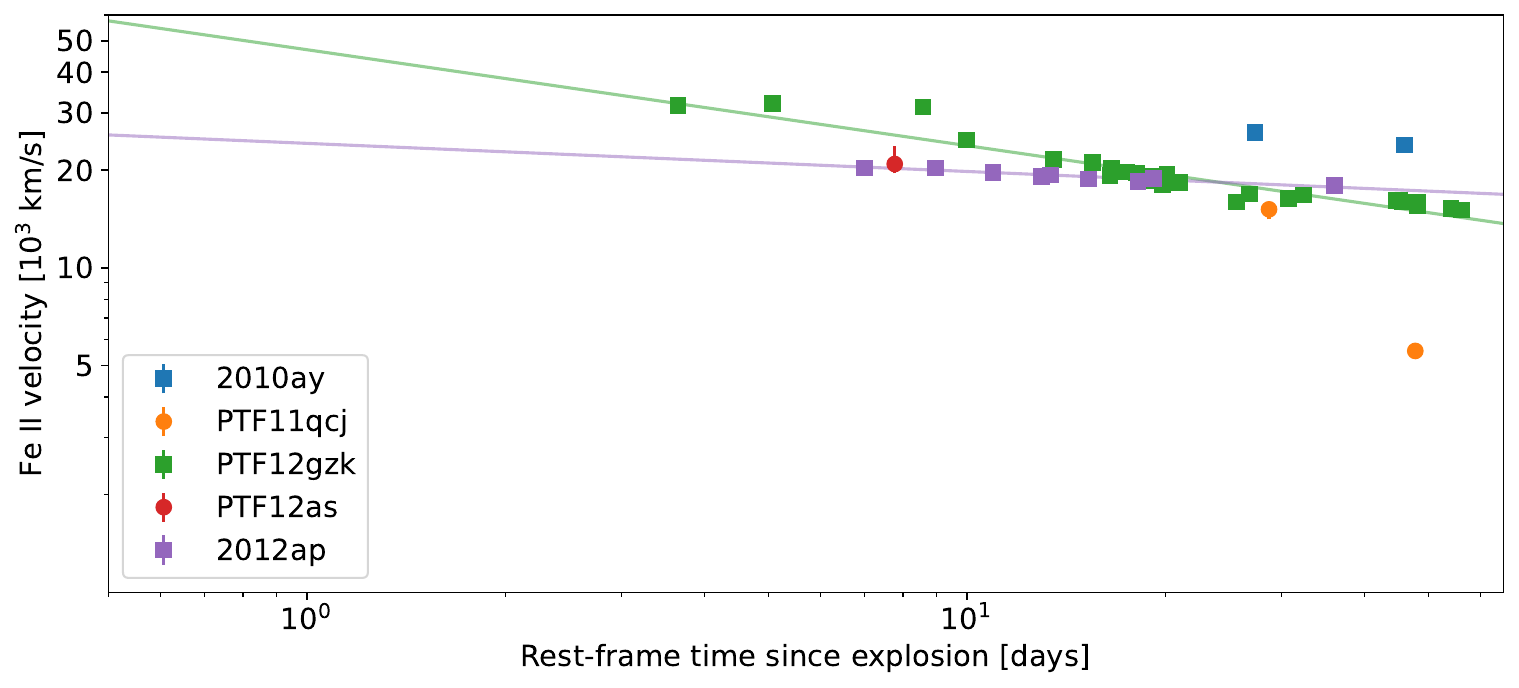}
      \caption{Fe II velocities of a subset of ordinary type Ic-BL SNe from the \textit{Gold}, \textit{Silver} and \textit{Bronze} samples. Power-law and broken power-law fits are shown as solid lines.}
   \end{figure*}

   \begin{figure*}[!h]
      \centering
      \includegraphics[width=0.9\linewidth]{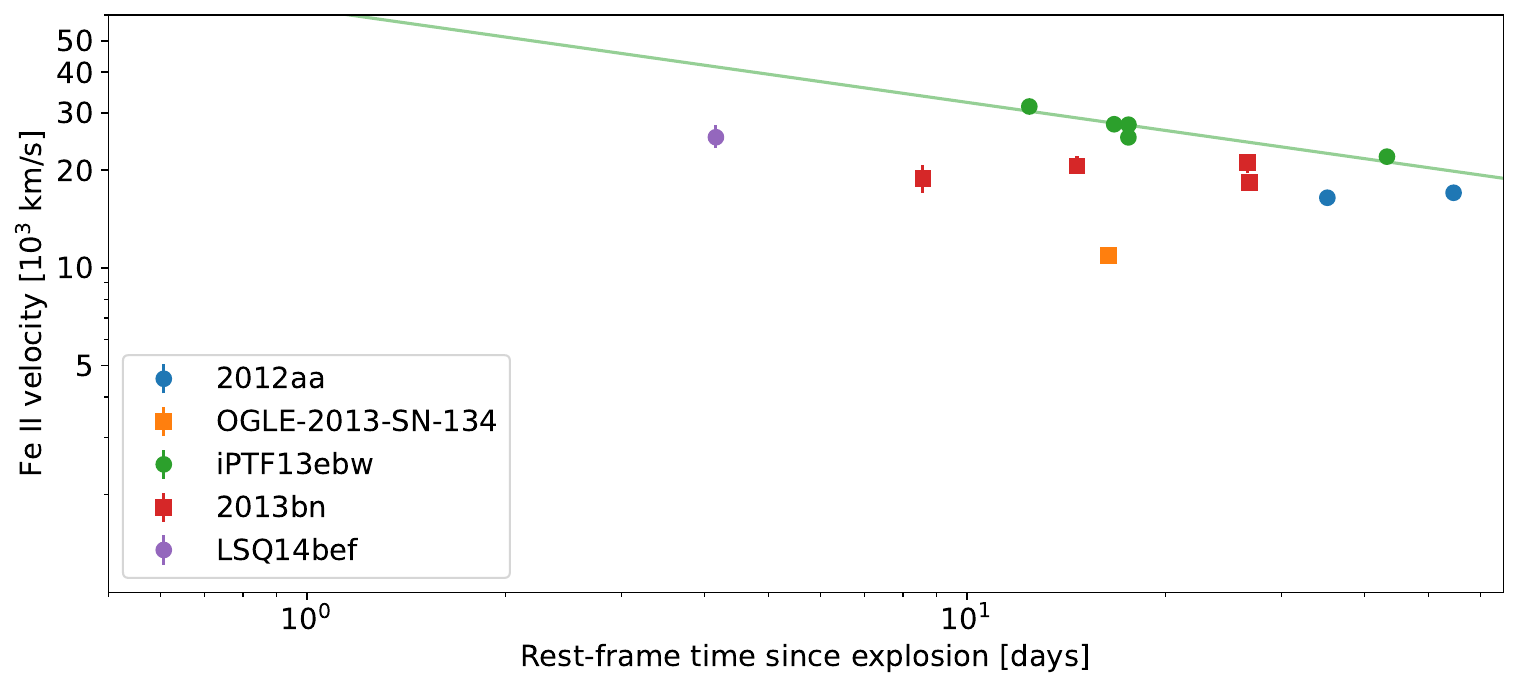}
      \includegraphics[width=0.9\linewidth]{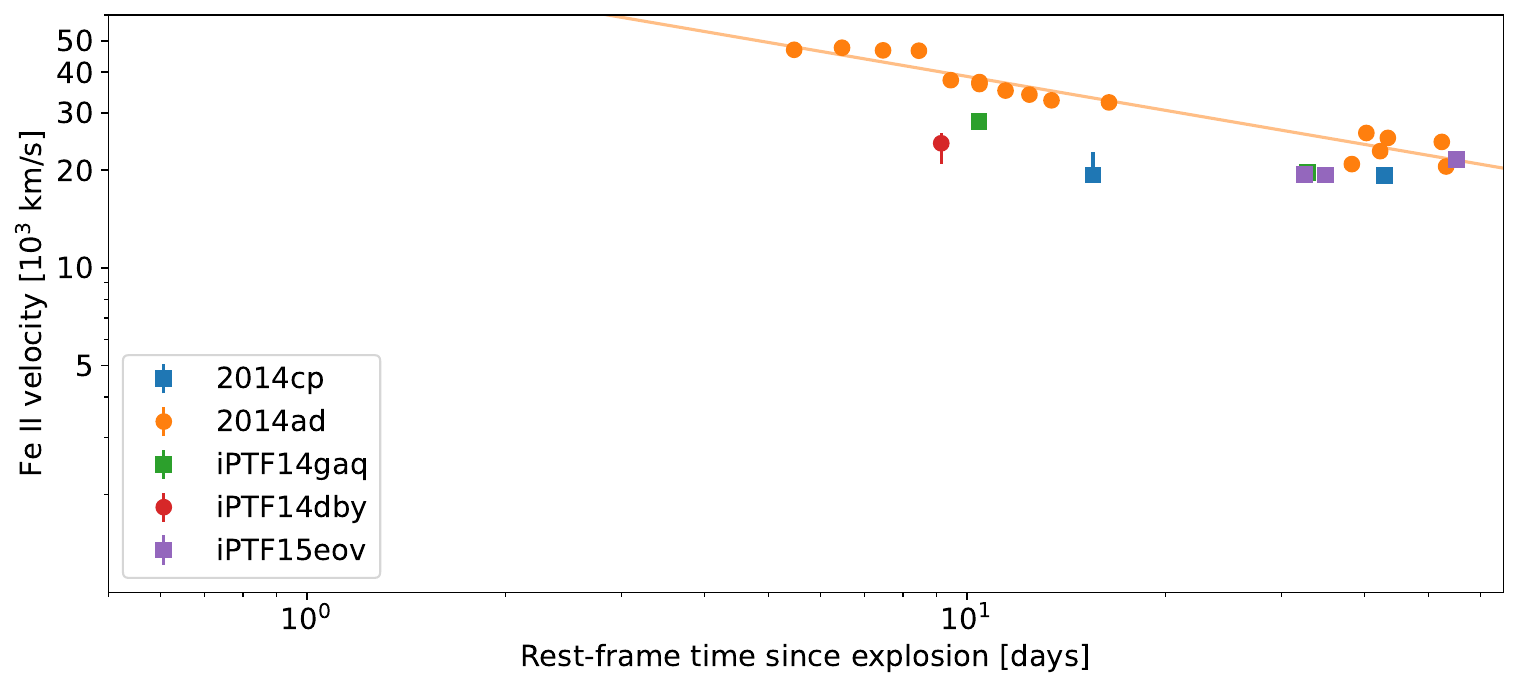}
      \includegraphics[width=0.9\linewidth]{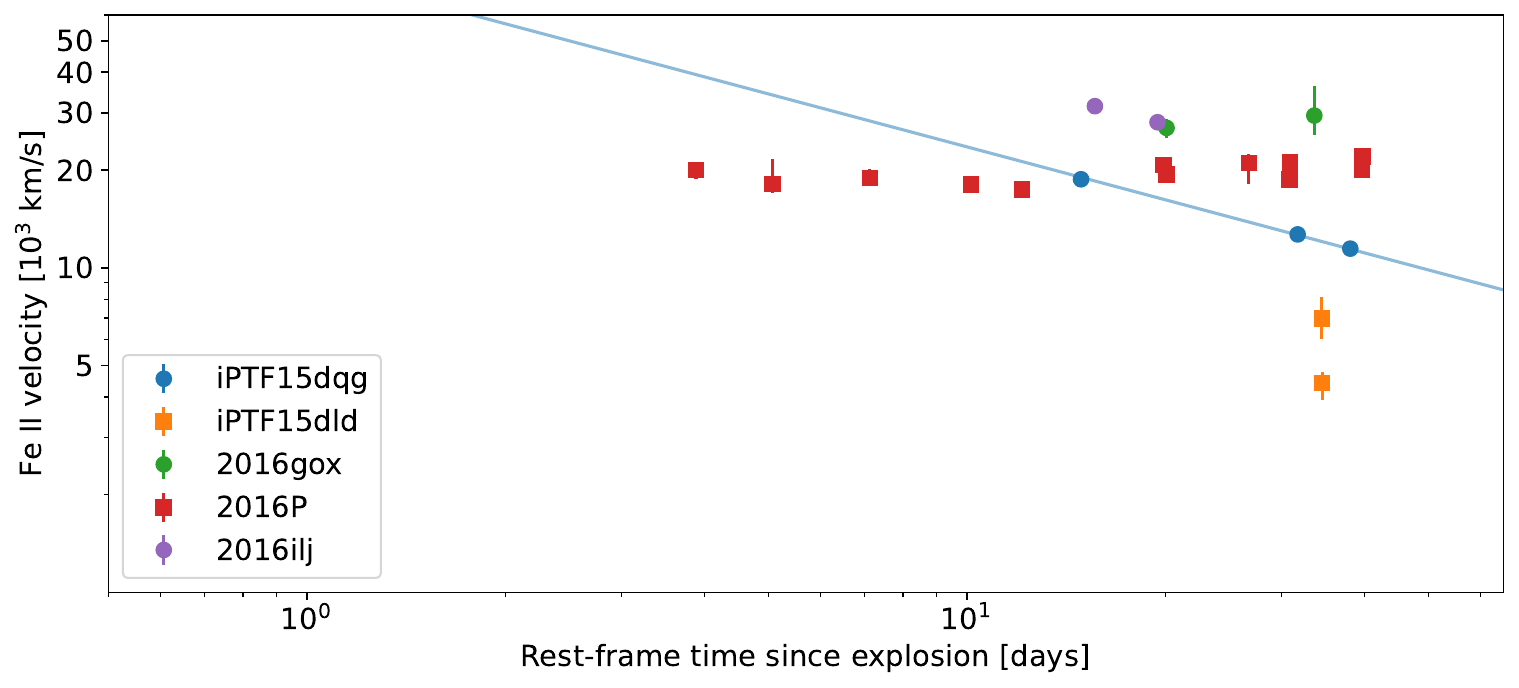}
      \caption{Fe II velocities of a subset of ordinary type Ic-BL SNe from the \textit{Gold}, \textit{Silver} and \textit{Bronze} samples. Power-law and broken power-law fits are shown as solid lines.}
   \end{figure*}

   \begin{figure*}[!h]
      \centering
      \includegraphics[width=0.9\linewidth]{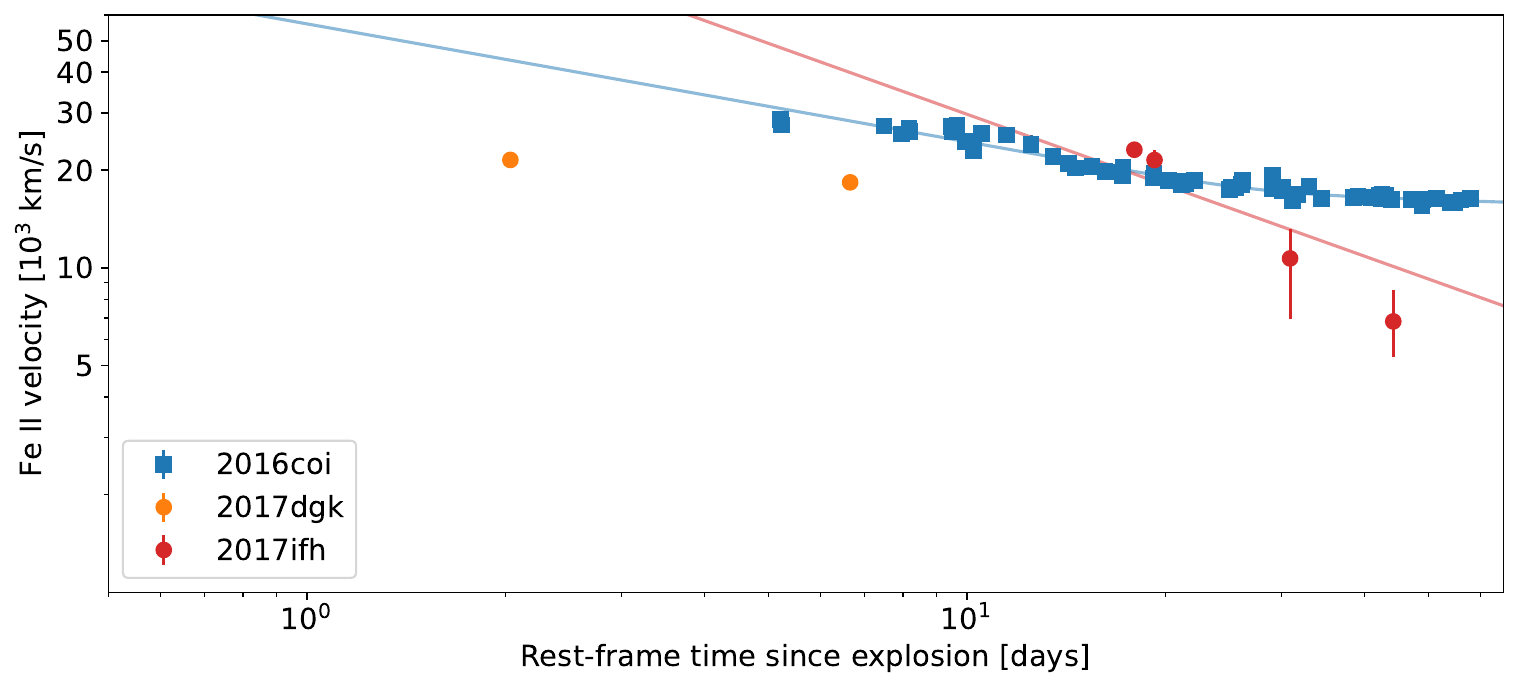}
      \includegraphics[width=0.9\linewidth]{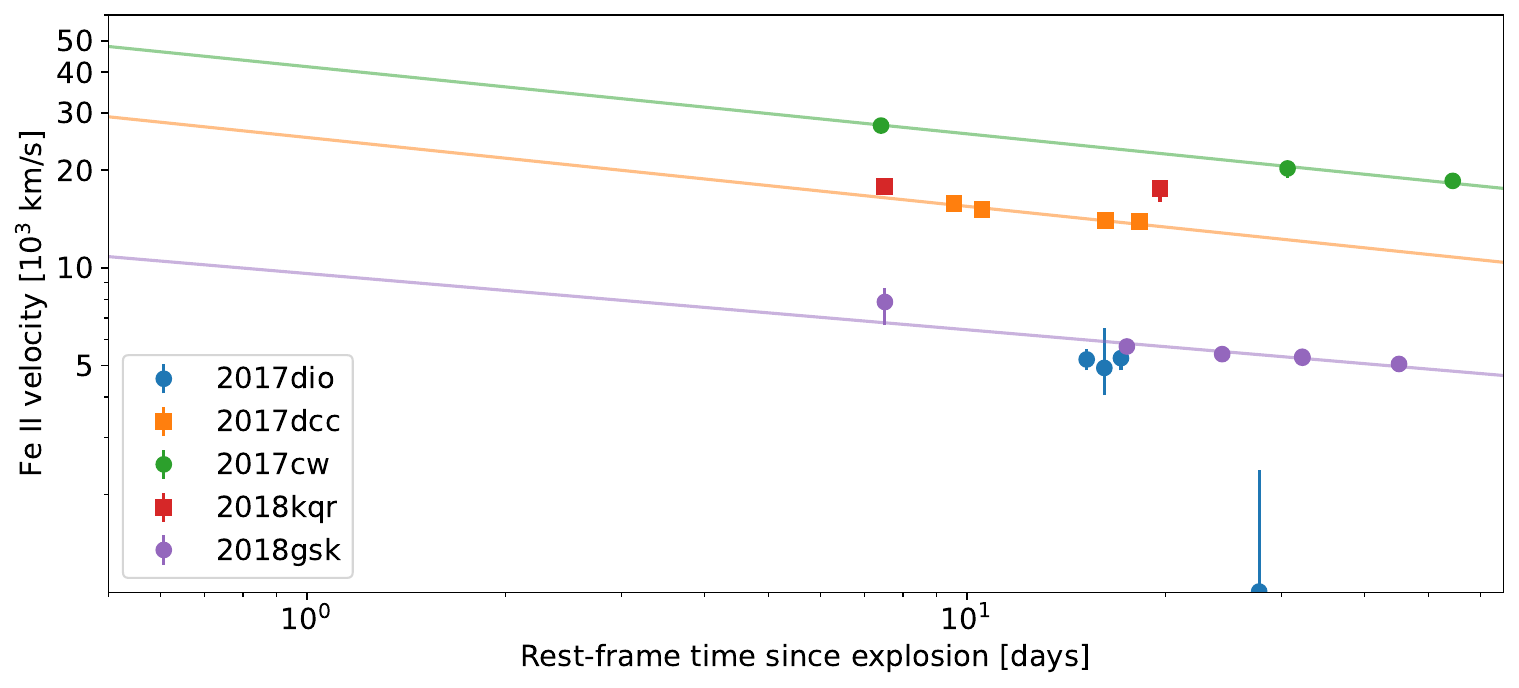}
      \includegraphics[width=0.9\linewidth]{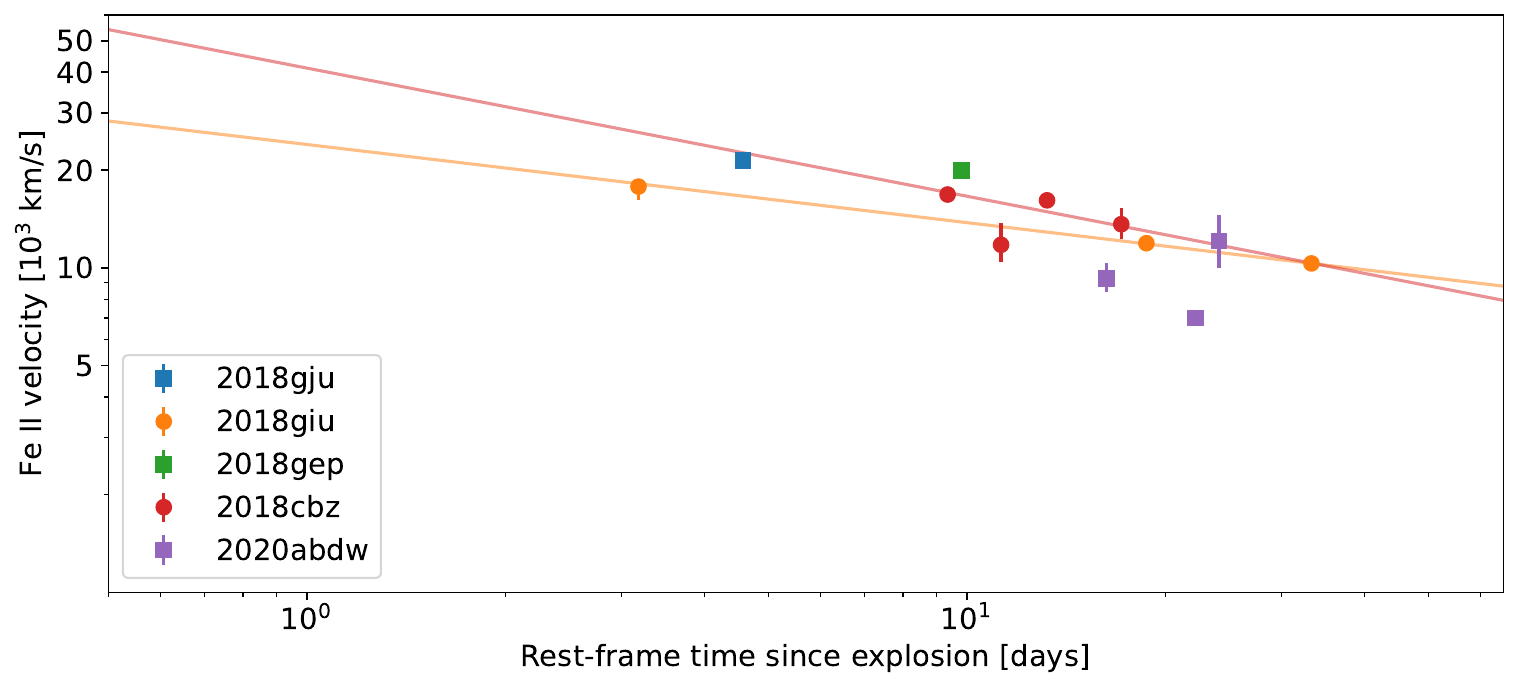}
      \caption{Fe II velocities of a subset of ordinary type Ic-BL SNe from the \textit{Gold}, \textit{Silver} and \textit{Bronze} samples. Power-law and broken power-law fits are shown as solid lines.}
   \end{figure*}

   \begin{figure*}[!h]
      \centering
      \includegraphics[width=0.9\linewidth]{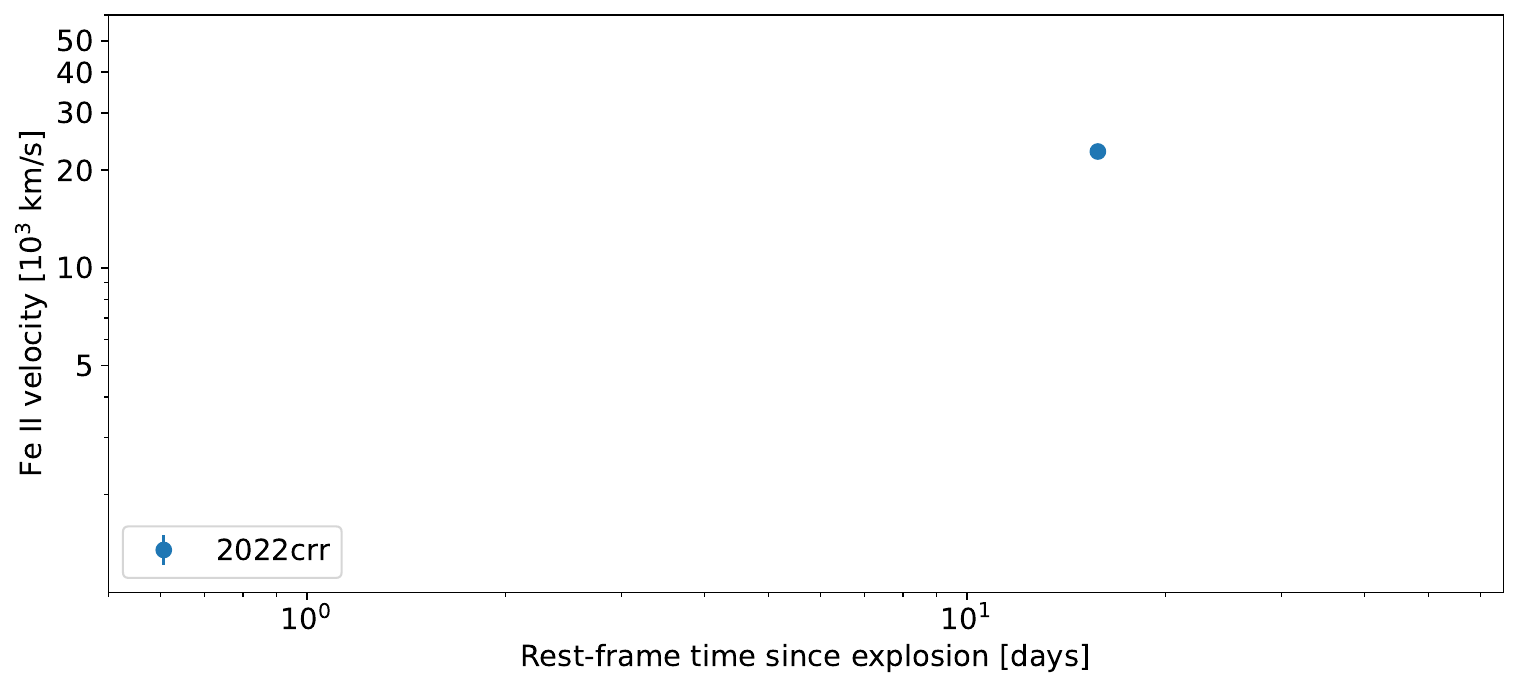}
      \caption{Fe II velocities of a subset of ordinary type Ic-BL SNe from the \textit{Gold}, \textit{Silver} and \textit{Bronze} samples. Power-law and broken power-law fits are shown as solid lines.}
   \end{figure*}

   \begin{figure*}[!h]
      \centering
      \includegraphics[width=0.9\linewidth]{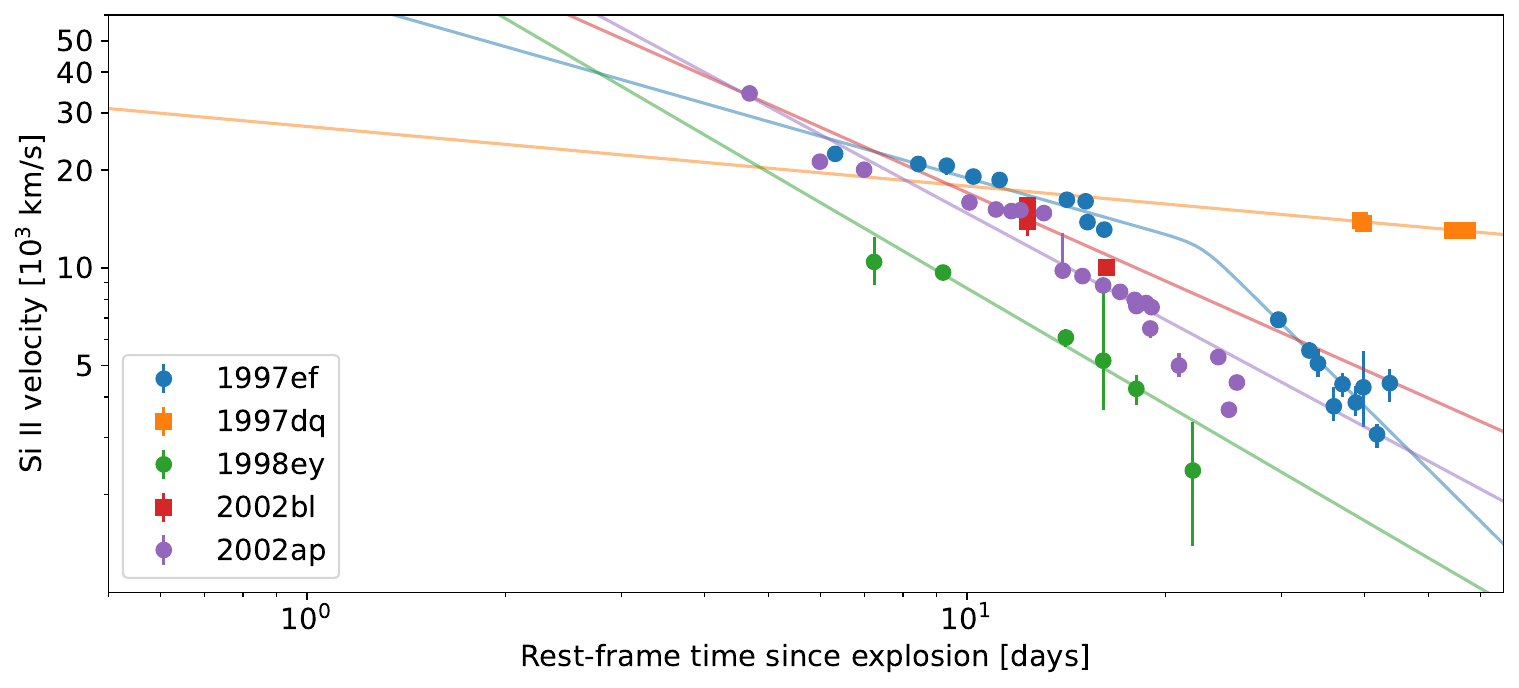}
      \caption{Si II velocities of a subset of ordinary type Ic-BL SNe from the \textit{Gold}, \textit{Silver} and \textit{Bronze} samples. Power-law and broken power-law fits are shown as solid lines.}
   \end{figure*}

   \begin{figure*}[!h]
      \centering

      \includegraphics[width=0.9\linewidth]{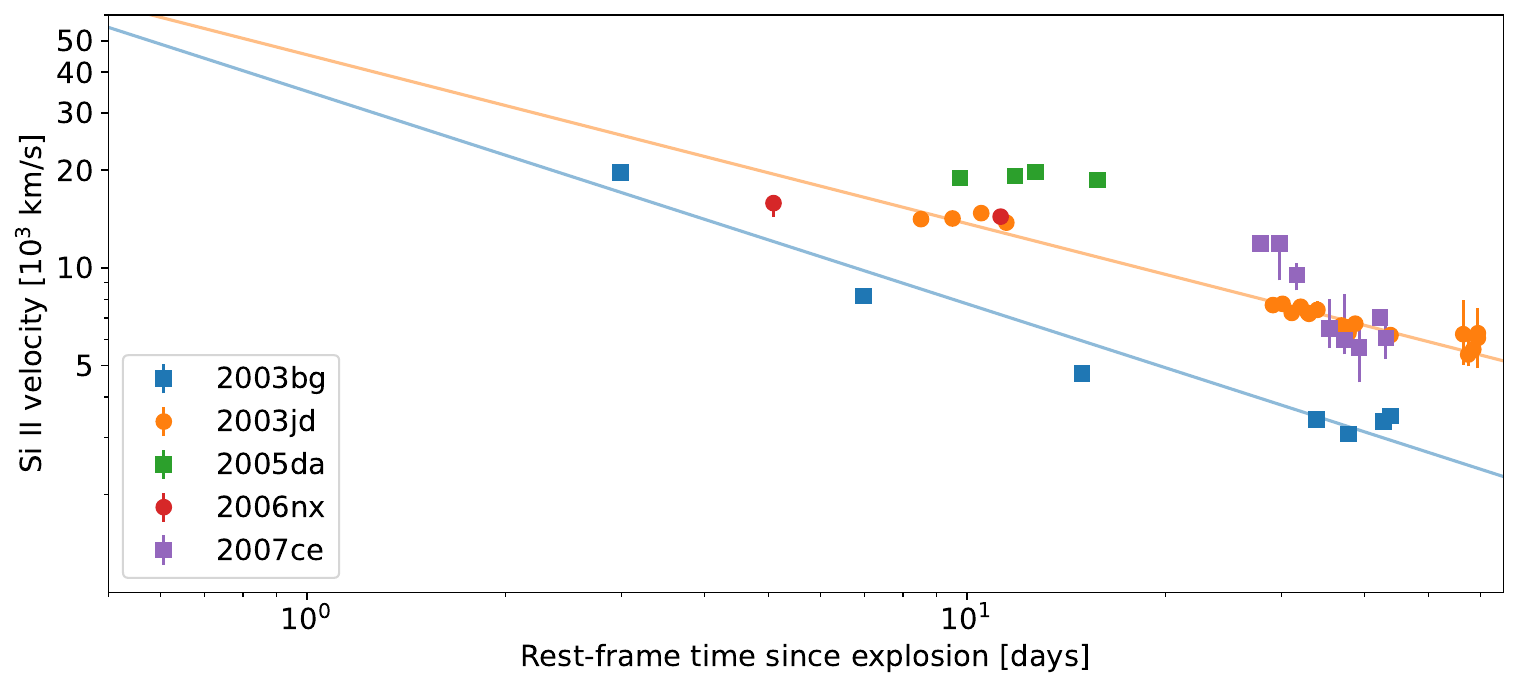}
      \includegraphics[width=0.9\linewidth]{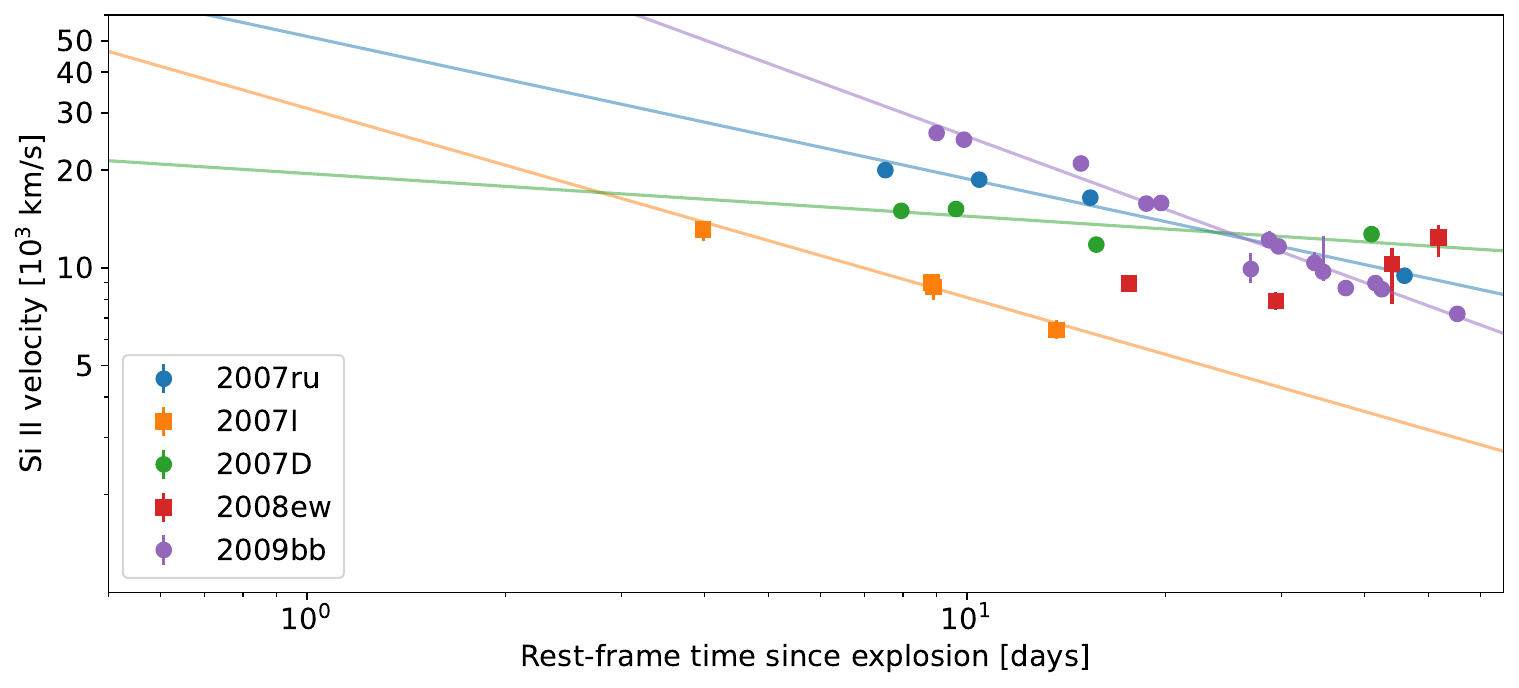}
      \includegraphics[width=0.9\linewidth]{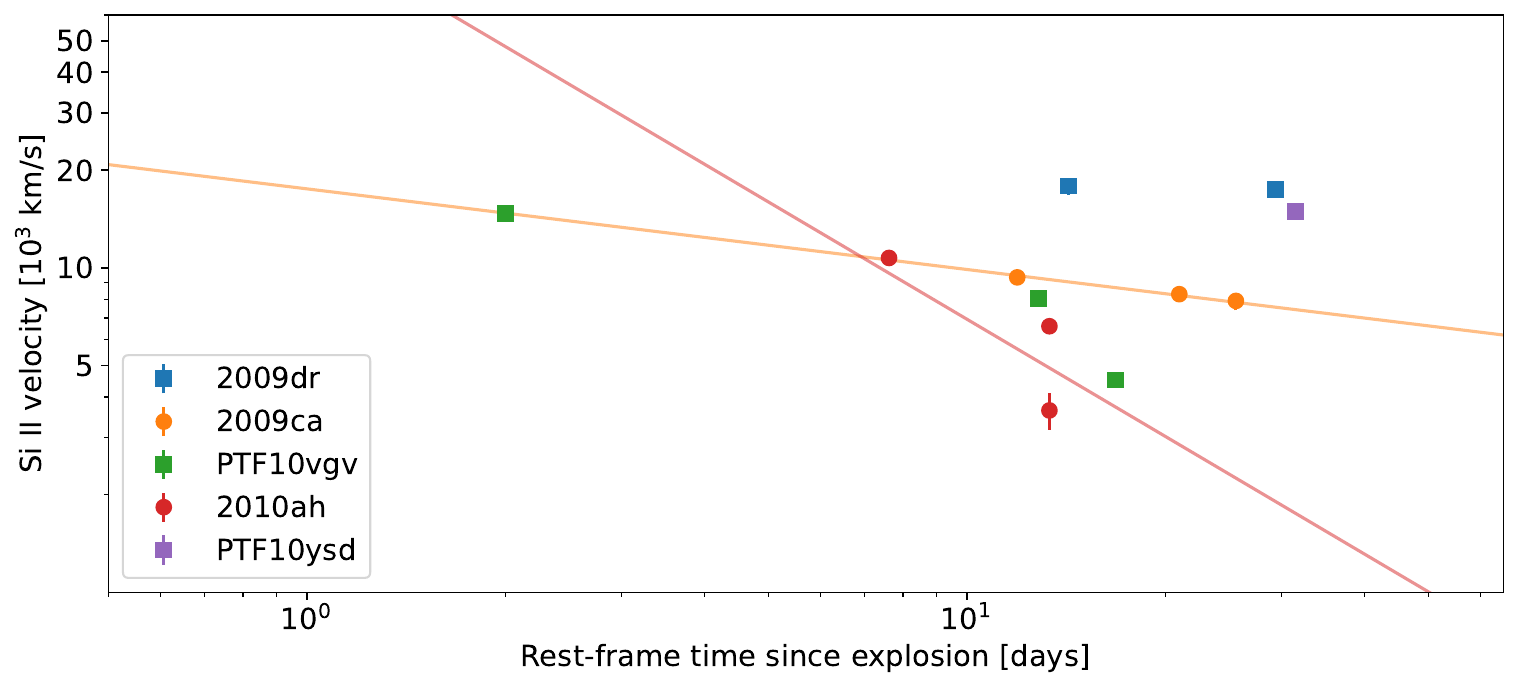}
      \caption{Si II velocities of a subset of ordinary type Ic-BL SNe from the \textit{Gold}, \textit{Silver} and \textit{Bronze} samples. Power-law and broken power-law fits are shown as solid lines.}
   \end{figure*}

   \begin{figure*}[!h]
      \centering

      \includegraphics[width=0.9\linewidth]{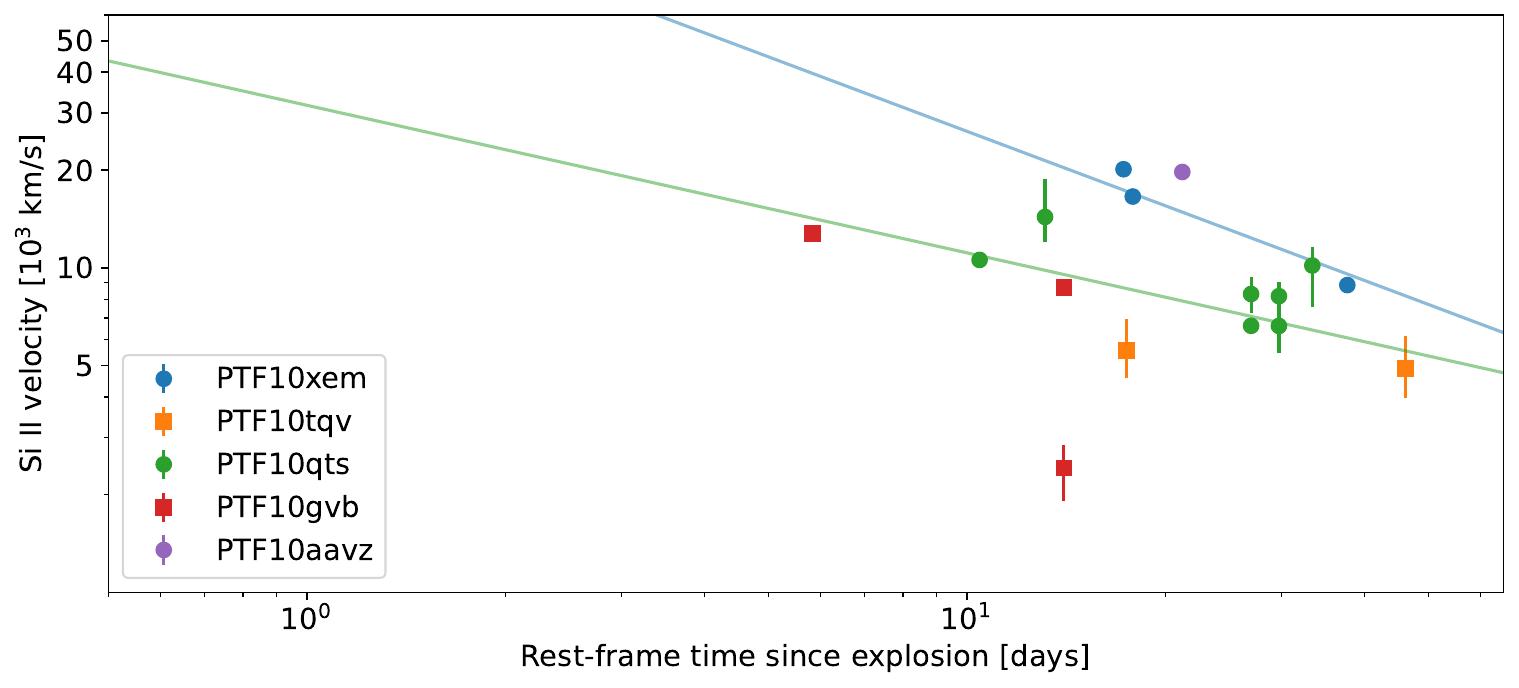}
      \includegraphics[width=0.9\linewidth]{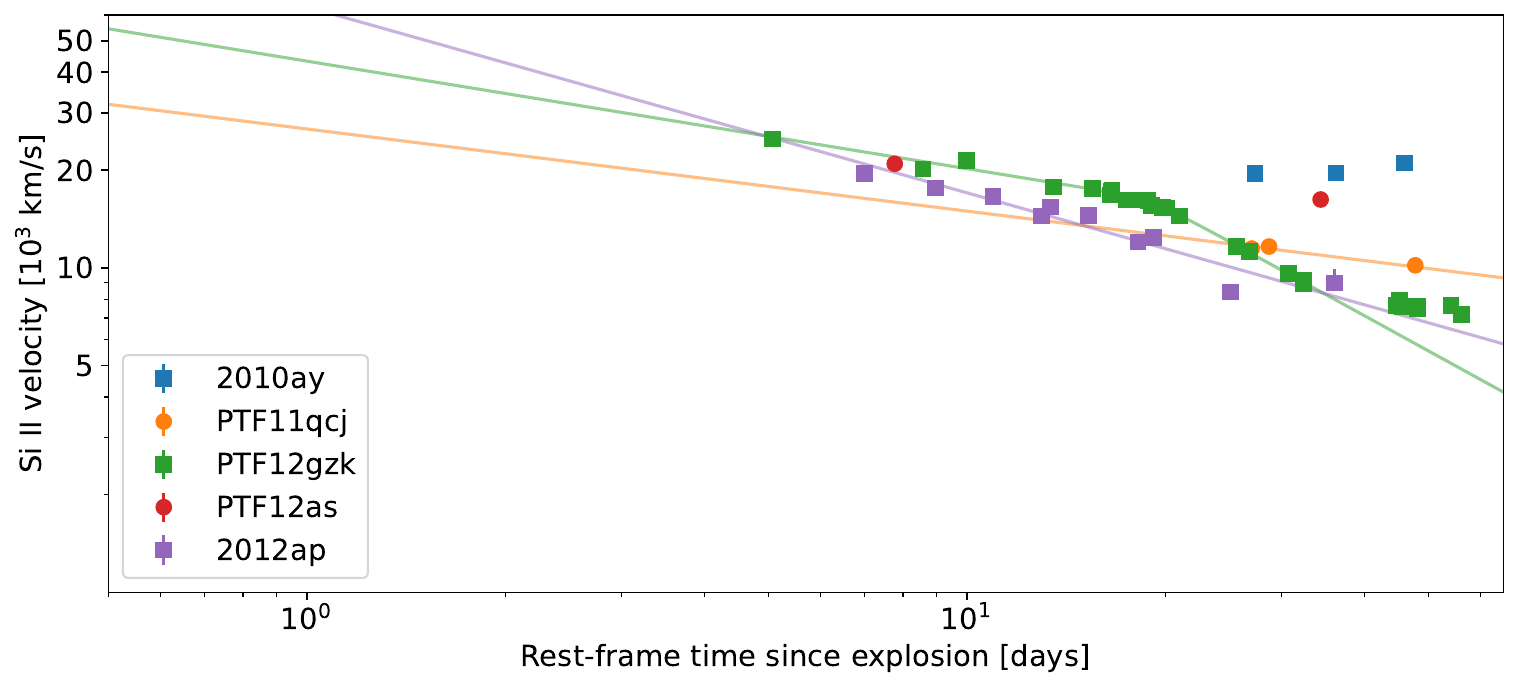}
      \includegraphics[width=0.9\linewidth]{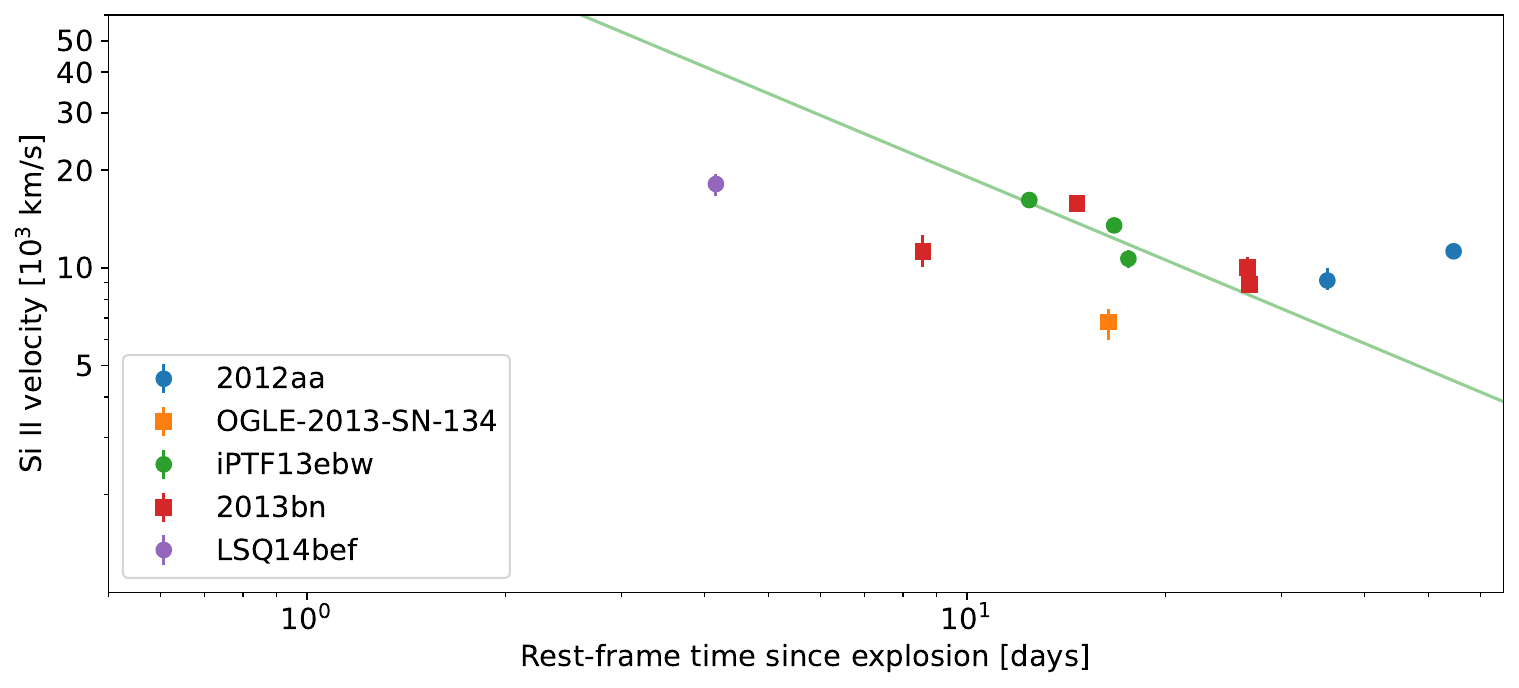}
      \caption{Si II velocities of a subset of ordinary type Ic-BL SNe from the \textit{Gold}, \textit{Silver} and \textit{Bronze} samples. Power-law and broken power-law fits are shown as solid lines.}
   \end{figure*}

   \begin{figure*}[!h]
      \centering
      \includegraphics[width=0.9\linewidth]{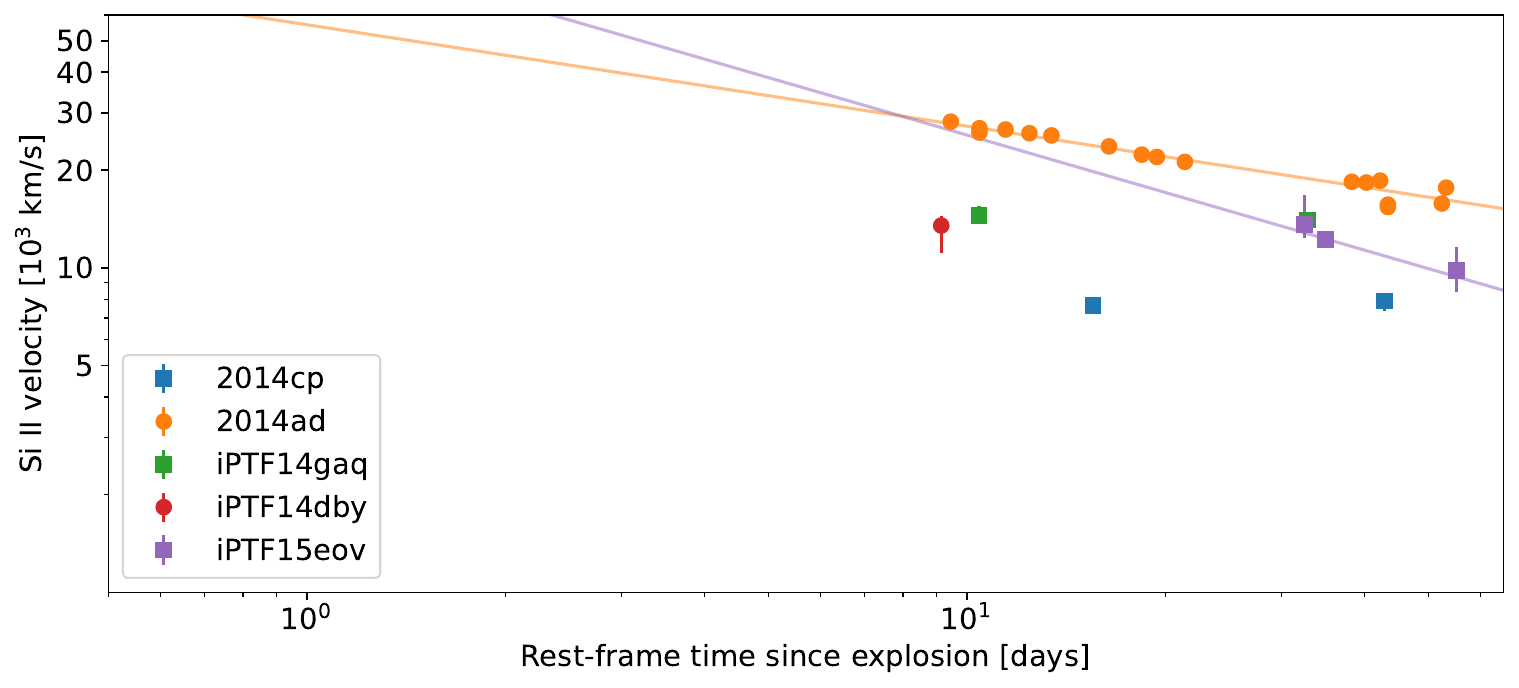}
      \includegraphics[width=0.9\linewidth]{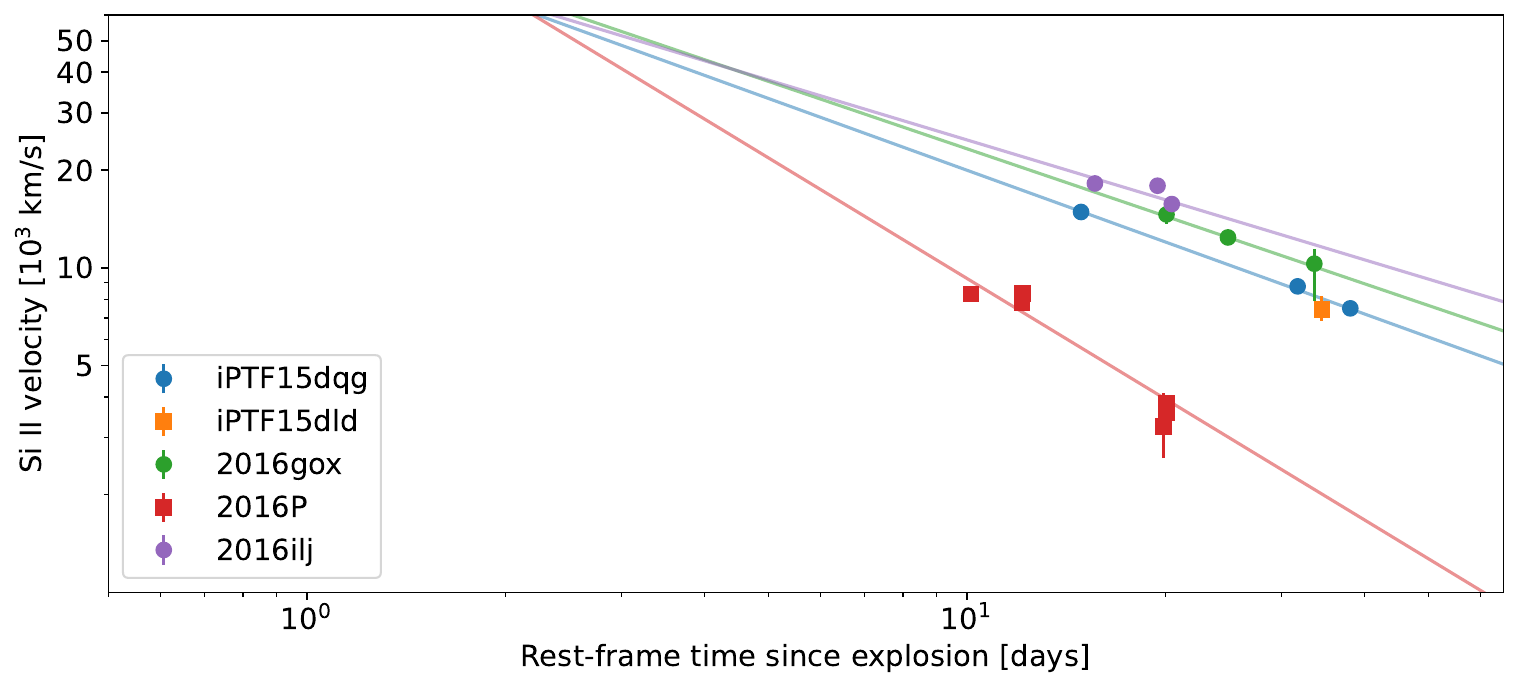}
      \includegraphics[width=0.9\linewidth]{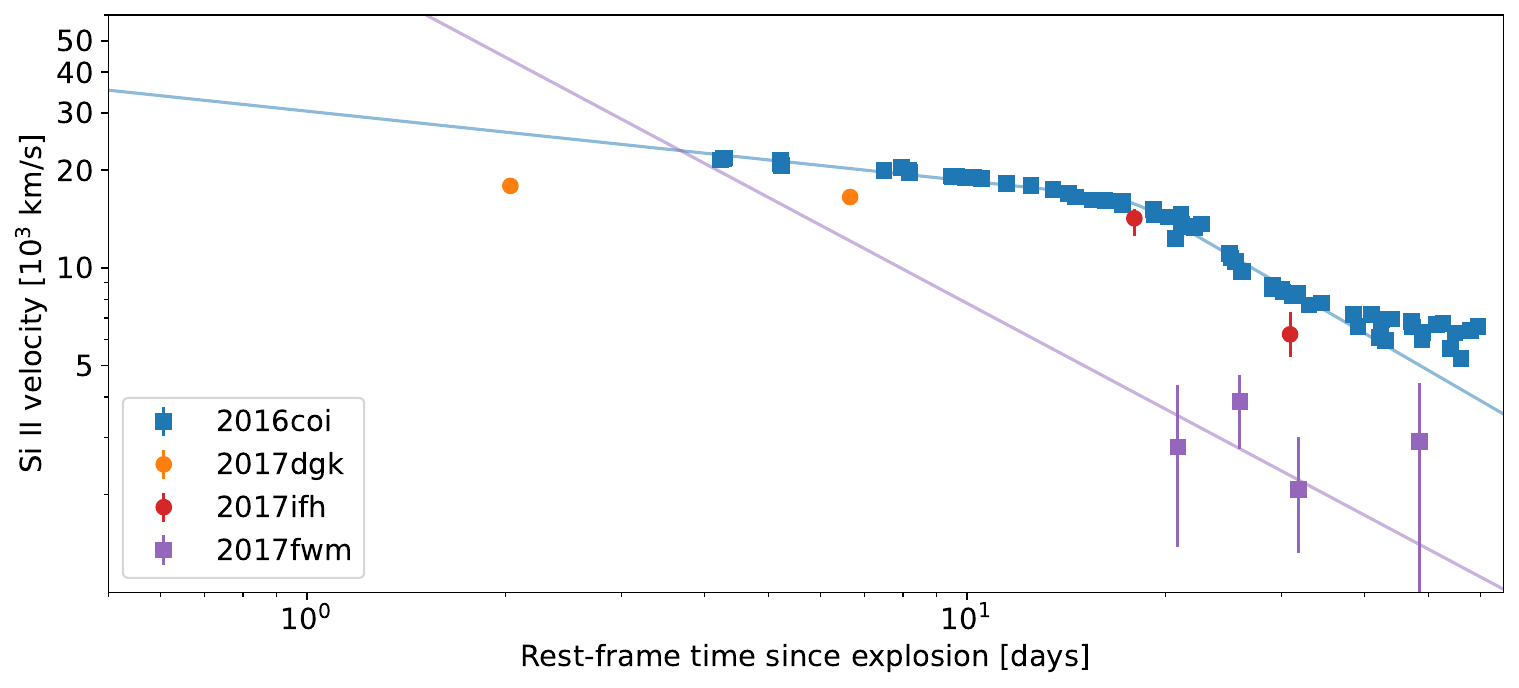}
      \caption{Si II velocities of a subset of ordinary type Ic-BL SNe from the \textit{Gold}, \textit{Silver} and \textit{Bronze} samples. Power-law and broken power-law fits are shown as solid lines.}
   \end{figure*}

   \begin{figure*}[!h]
      \centering
      \includegraphics[width=0.9\linewidth]{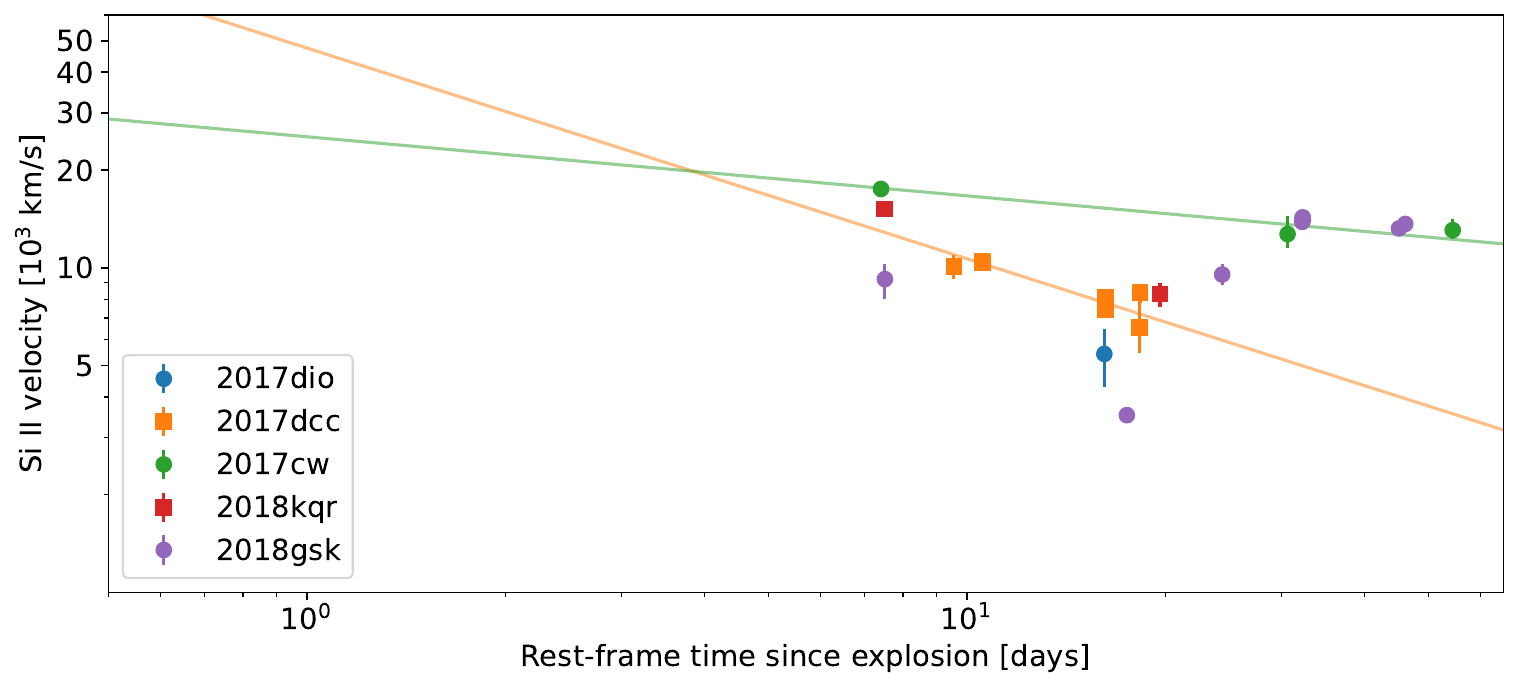}
      \includegraphics[width=0.9\linewidth]{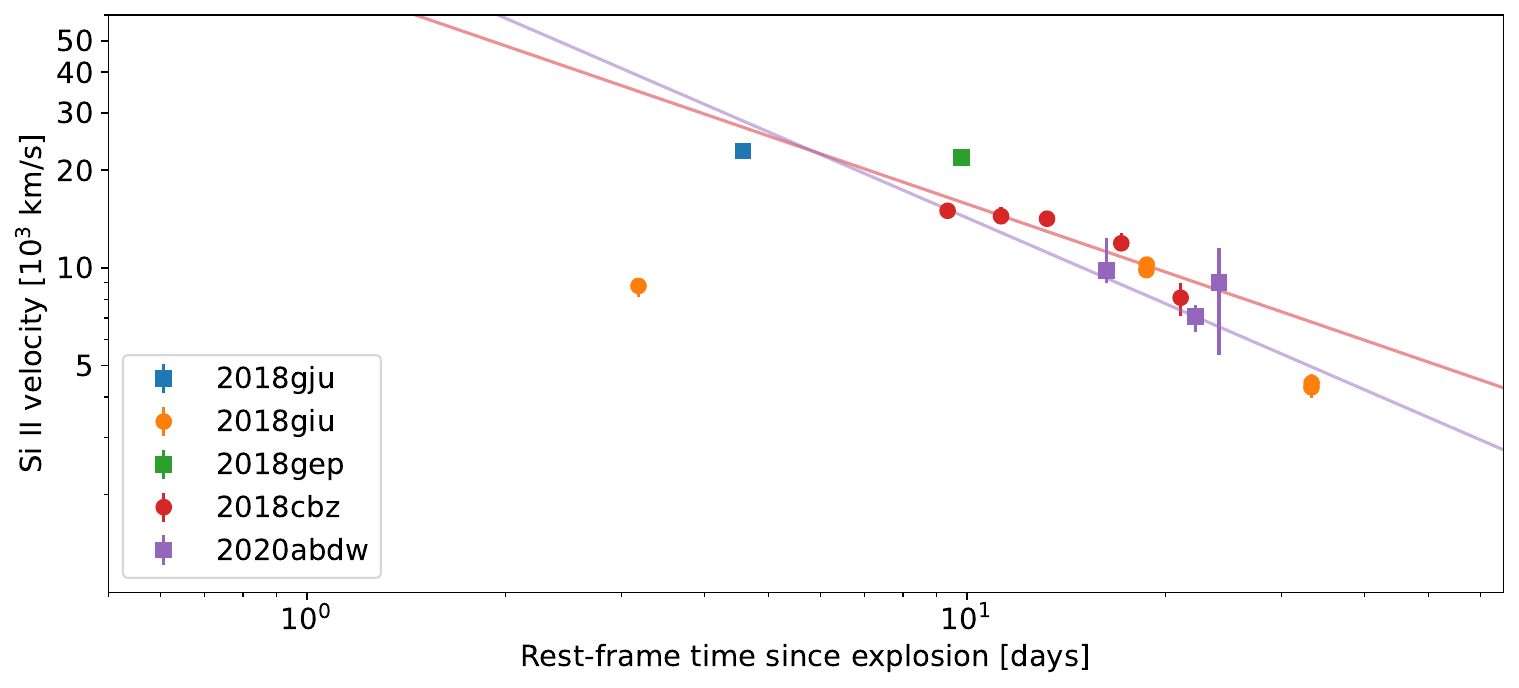}
      \includegraphics[width=0.9\linewidth]{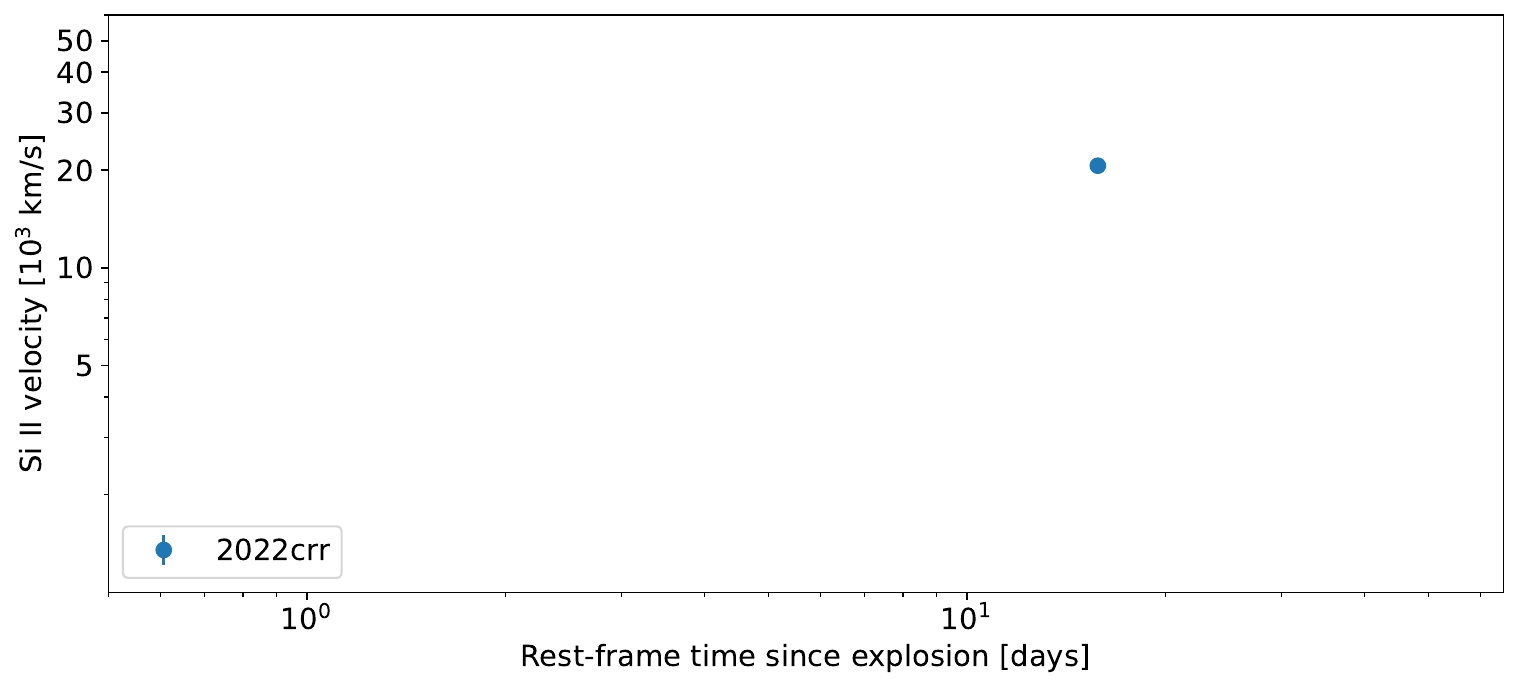}
      \caption{Si II velocities of a subset of ordinary type Ic-BL SNe from the \textit{Gold}, \textit{Silver} and \textit{Bronze} samples. Power-law and broken power-law fits are shown as solid lines.}
   \end{figure*}

   \FloatBarrier
\section{How robust are broken power-law fits to uncertainties in explosion time?}\label{sec:testrobustness}

   The broken power-law fits presented in this paper were motivated by the observation of apparent breaks in the slope of the velocity evolution for both Si II and Fe II, and their adoption as the best-fit to the observations is supported by the results of an F-test when compared to a power-law fit. However, it is possible that the model parameters, and potentially the best-fitting model, may change if the supernova explosion epoch, $t_0$, is uncertain. This is not the case for GRB-SNe, as their $t_0$ is very well constrained by the trigger time of gamma-ray observing satellites derived from the prompt gamma-ray emission. However, for the majority of the ordinary type Ic-BL supernovae in this sample, there is some uncertainty in $t_0$, and the level of uncertainty is not uniform across the sample. It is therefore important to understand the impact that uncertainty in $t_0$ may have on the marginalised parameters of the models.

   \begin{table}[h]
   \caption[Model parameters used to generate a simulated dataset]{Model parameters chosen to generate a simulated dataset used in verifying the robustness of the broken power-law fits. These values are the average best-fit values for the velocity evolution of the Si II feature in ordinary Ic-BL supernovae.}             
   \label{tab:simdataparams}
   \centering
   \begin{tabular}{l c l}
   \toprule
   \textbf{Parameter} & \textbf{Value} & \textbf{Unit}\\  
   \midrule
   $A$ & 15000 & km/s \\
   $\alpha_1$ & -0.35 & \\
   $\alpha_2$ & -1.4 & \\
   $t_b$ & 20 & days\\
   $s$ & 20 & days\\
   \bottomrule
   \end{tabular}
   \end{table}

   \begin{figure}[h]
      \centering
      \includegraphics[width=0.5\linewidth]{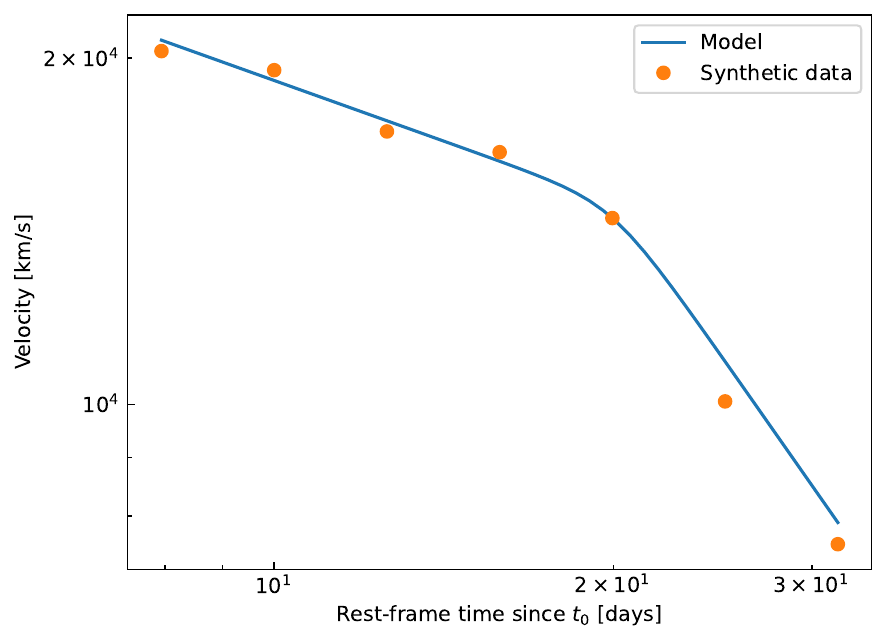}
      \caption[Simulated broken power-law and synthetic data]{Simulated broken power-law and synthetic data used to test robustness of broken-power law fits. The circular points have been sampled from the broken power-law, represented by the solid line. The parameters of the broken power-law can be found in Table \ref{tab:simdataparams}. A 500 km/s Gaussian scatter has been added to the sample data points to make the data more realistic.}\label{fig:syntheticdata}
   \end{figure}

   \begin{table*}[h]
   \caption[Best-fit parameters for power-law and broken power-law fits for different time-shifts]{Best-fit parameters for power-law (PL) and broken power-law (BPL) fits for different time-shifts. Values shown are the median and 16th and 84th percentile uncertainties. The BPL results for a time-shift of 0 days agree within uncertainties with the parameters of the generating function shown in Table \ref{tab:simdataparams}.}             
   \label{tab:t0testparams}
   \centering
   \begin{tabular}{cccccccccl}
   \toprule
    \multicolumn{1}{c}{}  & \multicolumn{2}{c}{\textbf{PL fit}} & \multicolumn{4}{c}{\textbf{BPL fit}} & \multicolumn{3}{c}{\textbf{Best fit stats.}} \\
    \cmidrule(lr){2-3}
   \cmidrule(lr){4-7}
   \cmidrule(lr){8-10}
   \textbf{Time-shift} & \textbf{$a$} & \textbf{$b$} & \textbf{$t_b$} & \textbf{$\alpha_1$} & \textbf{$\alpha_2$} & \textbf{$A$} & \textbf{F-stat.} & \textbf{$p$-value} & \textbf{Best-fit} \\  
   \textbf{[days]} & \textbf{[km/s]} &  & \textbf{[days]} &  &  & \textbf{[km/s]} & &  &  \\ 
   \midrule
   -6 &  ${30000_{-7000}^{+10000}}$   &  ${-0.40_{-0.10}^{+0.10}}$ &  ${12.7_{-0.8}^{+0.9}}$ &  ${-0.13_{-0.03}^{+0.03}}$ &  ${-1.1_{-0.1}^{+0.1}}$ &  ${16100_{-700}^{+700}}$   &  74.824   &  0.0028   &  BPL  \vspace{0.1cm} \\
   -5 &  ${40000_{-10000}^{+10000}}$  &  ${-0.40_{-0.10}^{+0.10}}$ &  ${13.8_{-0.8}^{+1.0}}$ &  ${-0.17_{-0.04}^{+0.04}}$ &  ${-1.1_{-0.1}^{+0.1}}$ &  ${15900_{-800}^{+700}}$   &  76.968   &  0.0026   &  BPL  \vspace{0.1cm} \\
   -4 &  ${50000_{-10000}^{+20000}}$  &  ${-0.50_{-0.10}^{+0.10}}$ &  ${14.8_{-0.8}^{+1.0}}$ &  ${-0.20_{-0.04}^{+0.04}}$ &  ${-1.2_{-0.1}^{+0.1}}$ &  ${15800_{-700}^{+700}}$   &  75.161   &  0.0027   &  BPL  \vspace{0.1cm} \\
   -3 &  ${60000_{-20000}^{+30000}}$  &  ${-0.60_{-0.10}^{+0.10}}$ &  ${15.8_{-0.9}^{+1.0}}$ &  ${-0.23_{-0.05}^{+0.05}}$ &  ${-1.3_{-0.1}^{+0.1}}$ &  ${15800_{-800}^{+700}}$   &  68.606   &  0.0031   &  BPL  \vspace{0.1cm} \\
   -2 &  ${70000_{-20000}^{+40000}}$  &  ${-0.60_{-0.10}^{+0.10}}$ &  ${17.0_{-1.0}^{+1.0}}$ &  ${-0.26_{-0.05}^{+0.05}}$ &  ${-1.3_{-0.1}^{+0.1}}$ &  ${15700_{-800}^{+800}}$   &  64.125   &  0.0035   &  BPL  \vspace{0.1cm} \\
   -1 &  ${90000_{-30000}^{+40000}}$  &  ${-0.70_{-0.10}^{+0.10}}$ &  ${18.0_{-1.0}^{+1.0}}$ &  ${-0.28_{-0.06}^{+0.06}}$ &  ${-1.4_{-0.1}^{+0.1}}$ &  ${15600_{-900}^{+800}}$   &  56.881   &  0.0041   &  BPL  \vspace{0.1cm} \\
   0  &  ${110000_{-30000}^{+40000}}$ &  ${-0.70_{-0.10}^{+0.10}}$ &  ${19.0_{-1.0}^{+1.0}}$ &  ${-0.31_{-0.07}^{+0.07}}$ &  ${-1.4_{-0.1}^{+0.1}}$ &  ${15600_{-900}^{+900}}$   &  52.735   &  0.0046   &  BPL  \vspace{0.1cm} \\
   1  &  ${120000_{-40000}^{+40000}}$ &  ${-0.80_{-0.10}^{+0.10}}$ &  ${20.0_{-1.0}^{+1.0}}$ &  ${-0.34_{-0.07}^{+0.08}}$ &  ${-1.5_{-0.1}^{+0.1}}$ &  ${15600_{-900}^{+900}}$   &  44.667   &  0.0059   &  BPL  \vspace{0.1cm} \\
   2  &  ${140000_{-40000}^{+40000}}$ &  ${-0.78_{-0.09}^{+0.10}}$ &  ${21.0_{-1.0}^{+1.0}}$ &  ${-0.36_{-0.08}^{+0.09}}$ &  ${-1.5_{-0.1}^{+0.2}}$ &  ${16000_{-900}^{+1000}}$  &  36.955   &  0.0077   &  BPL  \vspace{0.1cm} \\
   3  &  ${150000_{-40000}^{+40000}}$ &  ${-0.79_{-0.07}^{+0.10}}$ &  ${22.0_{-1.0}^{+1.0}}$ &  ${-0.39_{-0.09}^{+0.09}}$ &  ${-1.6_{-0.2}^{+0.2}}$ &  ${16000_{-1000}^{+1000}}$ &  31.043   &  0.0099   &  BPL  \vspace{0.1cm} \\
   4  &  ${150000_{-40000}^{+30000}}$ &  ${-0.79_{-0.06}^{+0.10}}$ &  ${23.0_{-1.0}^{+1.0}}$ &  ${-0.40_{-0.10}^{+0.10}}$ &  ${-1.7_{-0.2}^{+0.2}}$ &  ${16000_{-1000}^{+1000}}$ &  26.984   &  0.0121   &  BPL  \vspace{0.1cm} \\
   5  &  ${160000_{-40000}^{+30000}}$ &  ${-0.79_{-0.05}^{+0.10}}$ &  ${24.0_{-1.0}^{+2.0}}$ &  ${-0.40_{-0.10}^{+0.10}}$ &  ${-1.7_{-0.2}^{+0.2}}$ &  ${16000_{-1000}^{+1000}}$ &  25.628   &  0.013 &  BPL  \vspace{0.1cm} \\
   6  &  ${160000_{-40000}^{+30000}}$ &  ${-0.78_{-0.05}^{+0.10}}$ &  ${25.0_{-2.0}^{+1.0}}$ &  ${-0.50_{-0.10}^{+0.10}}$ &  ${-1.8_{-0.2}^{+0.2}}$ &  ${16000_{-1000}^{+1000}}$ &  25.731   &  0.0129   &  BPL  \vspace{0.1cm} \\
   7  &  ${170000_{-40000}^{+30000}}$ &  ${-0.78_{-0.05}^{+0.10}}$ &  ${26.0_{-2.0}^{+2.0}}$ &  ${-0.50_{-0.10}^{+0.10}}$ &  ${-1.8_{-0.2}^{+0.2}}$ &  ${16000_{-1000}^{+1000}}$ &  27.953   &  0.0115   &  BPL  \vspace{0.1cm} \\
   8  &  ${160000_{-50000}^{+30000}}$ &  ${-0.76_{-0.05}^{+0.10}}$ &  ${27.0_{-2.0}^{+2.0}}$ &  ${-0.50_{-0.10}^{+0.10}}$ &  ${-1.9_{-0.2}^{+0.2}}$ &  ${16000_{-1000}^{+1000}}$ &  30.732   &  0.01  &  BPL  \vspace{0.1cm} \\
   9  &  ${160000_{-50000}^{+30000}}$ &  ${-0.75_{-0.05}^{+0.10}}$ &  ${28.0_{-2.0}^{+2.0}}$ &  ${-0.50_{-0.10}^{+0.20}}$ &  ${-1.9_{-0.2}^{+0.2}}$ &  ${16000_{-1000}^{+1000}}$ &  34.483   &  0.0085   &  BPL  \vspace{0.1cm} \\
   10 &  ${160000_{-50000}^{+30000}}$ &  ${-0.74_{-0.05}^{+0.10}}$ &  ${29.0_{-2.0}^{+1.0}}$ &  ${-0.60_{-0.10}^{+0.20}}$ &  ${-2.0_{-0.2}^{+0.2}}$ &  ${16000_{-1000}^{+1000}}$ &  37.033   &  0.0077   &  BPL  \vspace{0.1cm} \\
   \bottomrule
   \end{tabular}
   \end{table*}

   \begin{figure*}[h]
      \centering
      \includegraphics[width=0.8\linewidth]{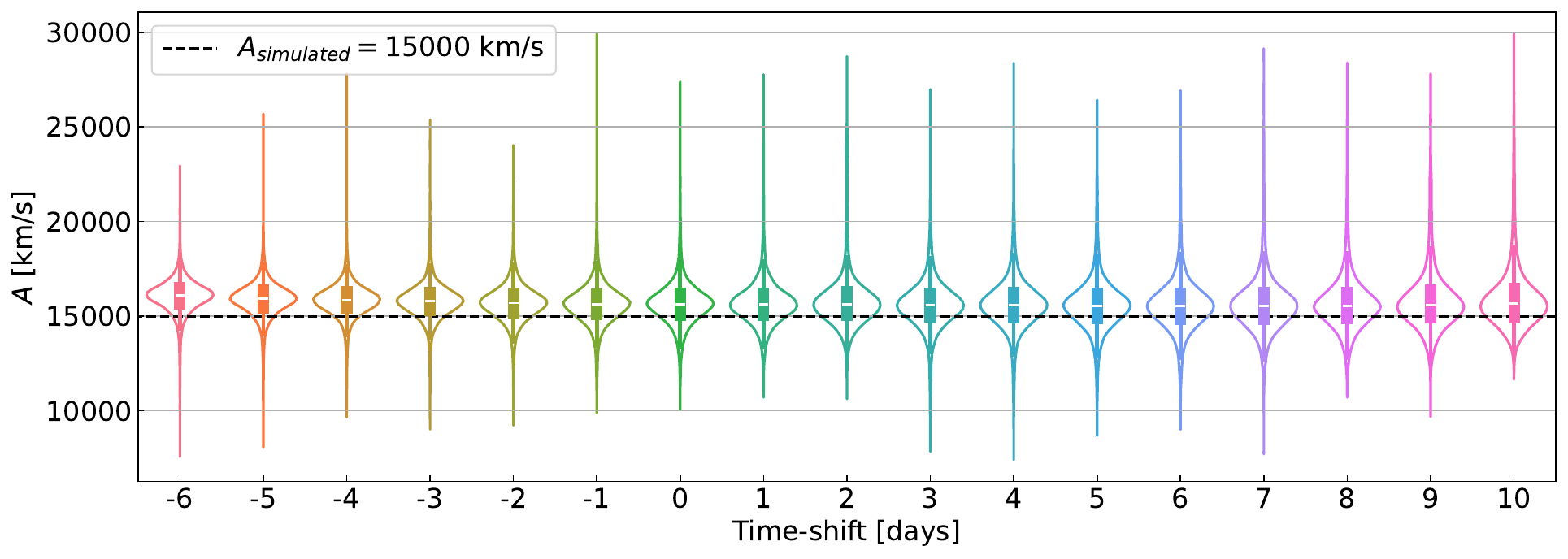}
      \includegraphics[width=0.8\linewidth]{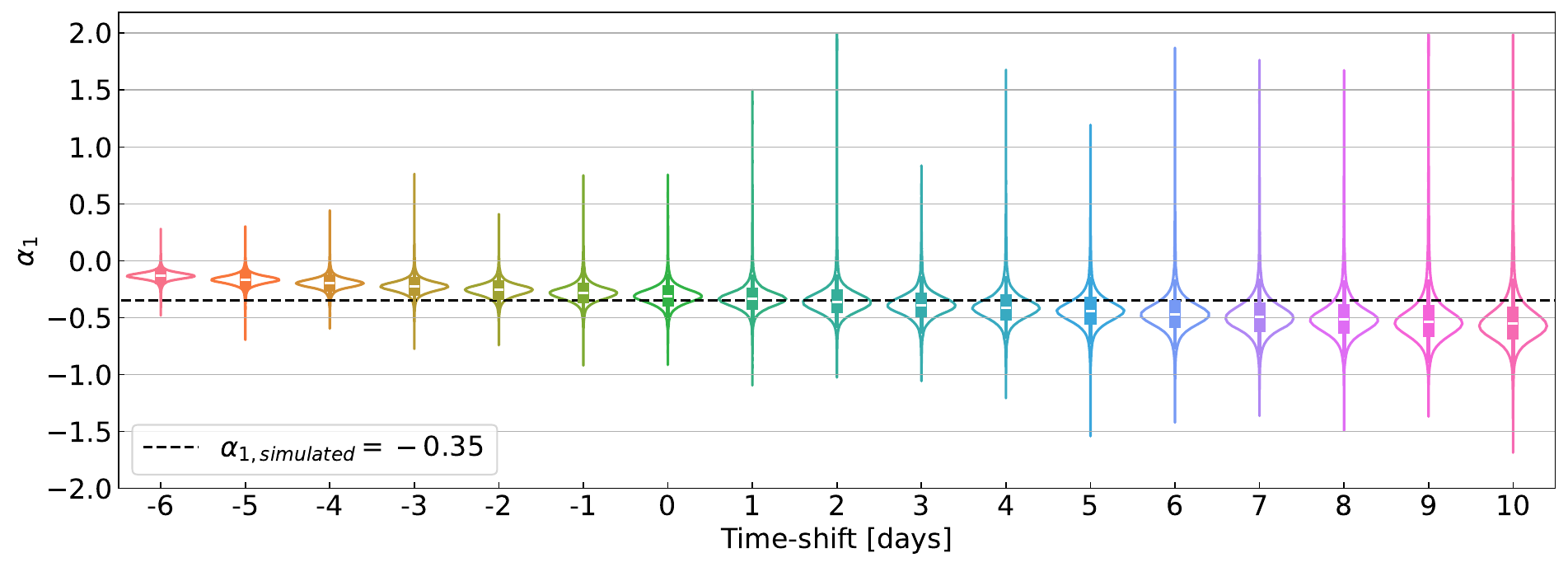}
      \includegraphics[width=0.8\linewidth]{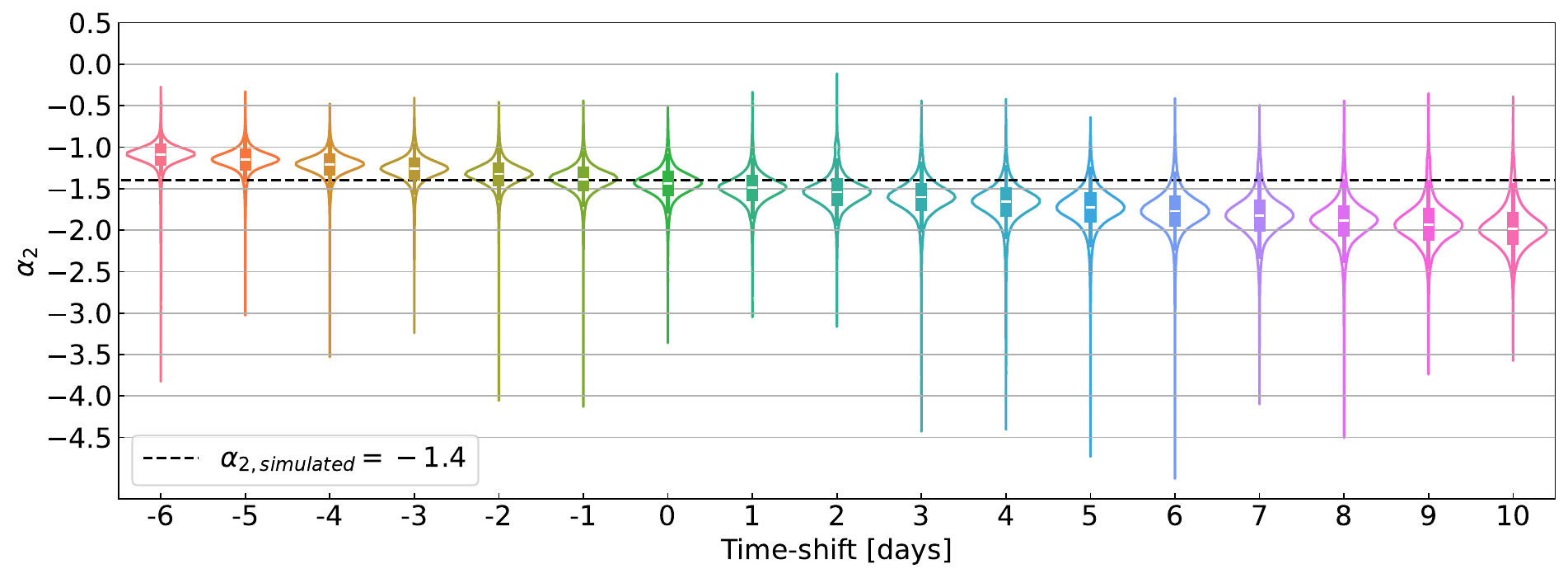}
      \includegraphics[width=0.8\linewidth]{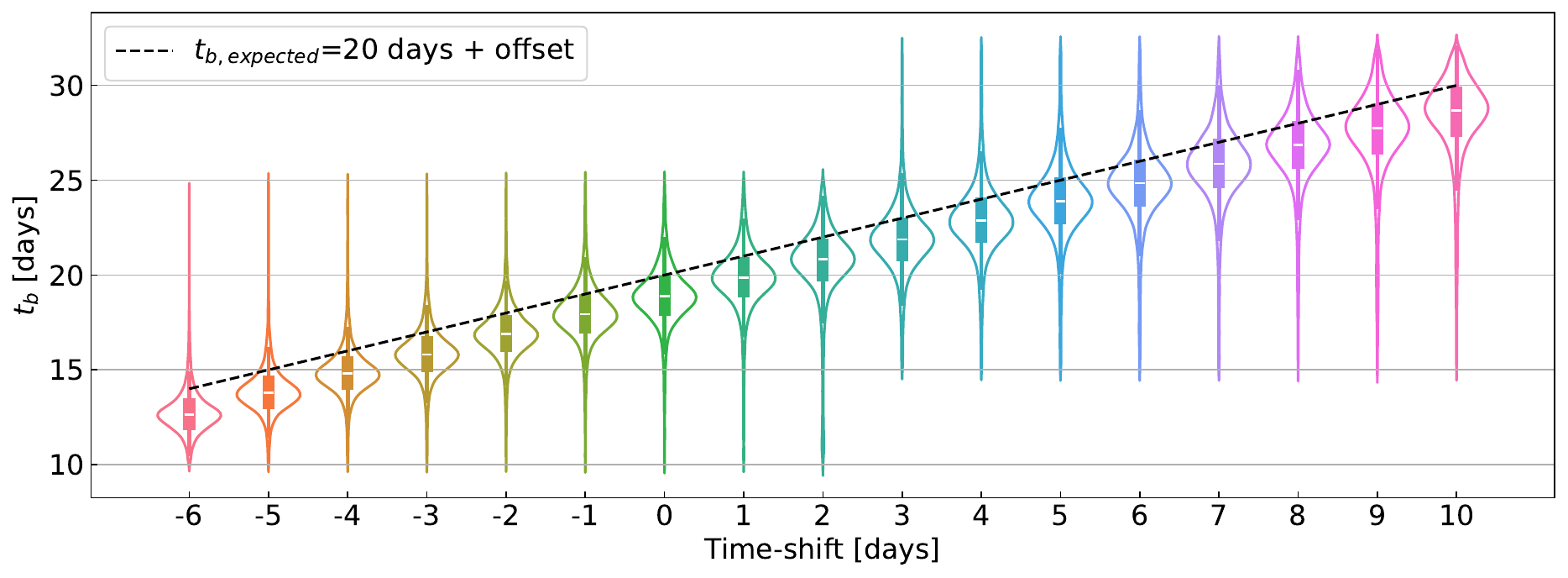}
      \caption[Violin plots showing the distributions of broken power-law parameters for a given shift in $t_0$]{Violin plots showing distributions of broken power-law parameters for a given shift in $t_0$. The dotted lines show the simulated values of these parameters for a given $t_0$ shift. All values are subject to variation due to these $t_0$ shifts. \textit{Top:} Distributions for the Si II velocity at the break, $A$. \textit{Second from top:} Distributions for the pre-break power-law index, $\alpha_1$. \textit{Second from bottom:} Distributions for the post-break power-law index, $\alpha_2$. \textit{Bottom:} Distributions for the break time, $t_b$.}\label{fig:t0testviolins}
   \end{figure*}

   To estimate the impact of this source of uncertainty, a simulated dataset was created using the broken power-law given in equation \ref{eq:bpl}. Table \ref{tab:simdataparams} shows the model parameters used to generate the simulated data. These values are the average best-fit values for the velocity evolution of the Si II feature in ordinary type Ic-BL supernovae (all of these supernovae are in the \textit{Gold} sample). Figure \ref{fig:syntheticdata} shows the broken power-law model and a sample of 7 data points taken from this curve between $t_0$+8 and $t_0$+32 days with regular spacing. A 500 km/s Gaussian scatter was added to make the data more representative of real observations.

   From this simulated dataset, multiple datasets were generated in which the data were shifted along the time axis by between -6 and +10 days (in increments of 1 day). This simulates the potential uncertainty on the $t_0$ value, with a shift of -1 day representing a true $t_0$ that occurs 1 day after the $t_0$ of the non-shifted data-set. These datasets were then fit using the same MCMC fitting procedure used to compute velocity evolution parameters in previous analyses. Both power-law and broken power-law models were fit to these datasets, and an F-test was then used to determine which model was the best fit. Based on a 2-$\sigma$ confidence level, the $p$-values of the F-tests indicate that this range of uncertainty in $t_0$ does not alter the best-fit from a broken power-law to a power-law. This result suggests that the observation of broken power-laws is robust. The parameters arising from this analysis are shown in Table \ref{tab:t0testparams}.

   The parameters presented in Table \ref{tab:t0testparams} show that although the observation of broken power-laws is robust, some of the parameters of the best-fit broken power-law are subject to significant variance under changes in $t_0$. Figure \ref{fig:t0testviolins} shows how the best-fit parameters and their distributions are altered as $t_0$ is shifted. This figure shows that $A$ is consistently over-estimated and $t_b$ consistently under-estimated compared to their values used when generating the model dataset, no matter the $t_0$ shift applied. This may have been caused by the introduction of Gaussian noise to the simulated dataset. Table \ref{tab:t0testparams} confirms that the values for a time-shift of 0 days are consistent within their uncertainties with those of the parameters used to generate this data.

   The value of $A$ is robust to changes in $t_0$, and the value of $t_b$ changes as expected when the data are shifted in time. However, the values of the power-law indices are not robust to shifting the $t_0$. Both $\alpha_1$ and $\alpha_2$ are reduced (relative to the 0-day shift value) when data are shifted forward in time, i.e. the power-laws become steeper. The opposite is true when the data are shifted backwards in time. This effect is more pronounced for $\alpha_2$. However, there is no evidence that $\alpha_1$ and $\alpha_2$ converge towards each other, which would lead to a power-law becoming the best-fit. This is in line with the results of the F-tests.

   It can be inferred that if data generated from a power-law was used in such analysis, a similar change in $b$ (the power-law index) would result. This has implications for all conclusions drawn from the power-law indices used in this paper. It is for this reason that statistical tests used to draw conclusions from quantities dependent on the power-law index are performed using only the \textit{Gold} sample, which has the smallest uncertainty on $t_0$. The GRB-SNe are not impacted by this effect, because they have highly constrained $t_0$ values.

   \FloatBarrier

\newpage
\section{Spectral sequence figures}\label{sec:specsequences}

   \begin{figure*}[!h]
      \centering
      \includegraphics[width=0.49\linewidth]{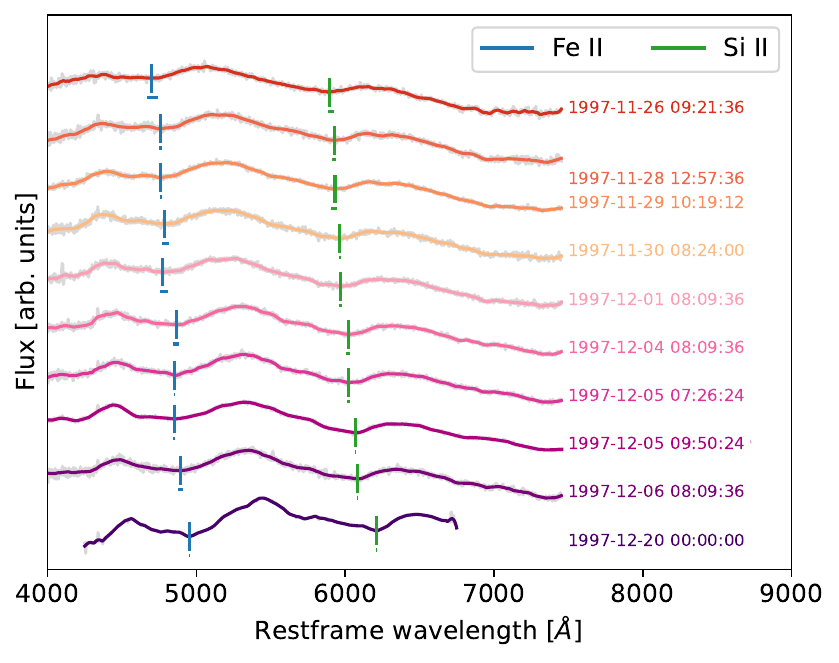}
      \includegraphics[width=0.49\linewidth]{figures/1997ef_spectra_0_colorful}
      \caption{Spectral sequences of the ordinary type Ic-BL supernova SN1997ef. The observed spectra (grey) and smoothed spectra (multiple colours) are plotted as solid lines. Also shown are the determined minimum wavelengths of the Fe II (blue) and Si II (green) features, and the corresponding uncertainty on the minimum wavelength.}
   \end{figure*}

   \begin{figure*}[!h]
      \centering
      \includegraphics[width=0.49\linewidth]{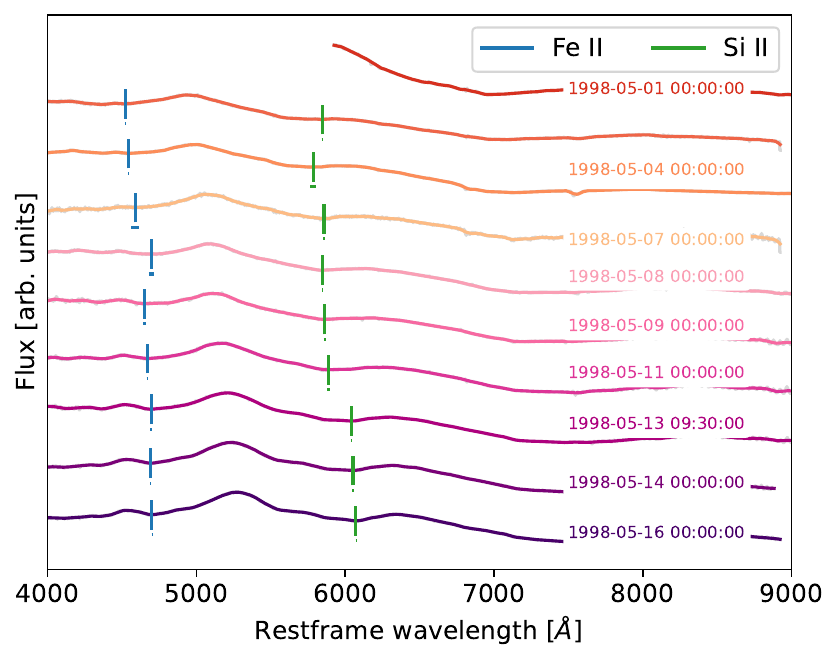}
      \includegraphics[width=0.49\linewidth]{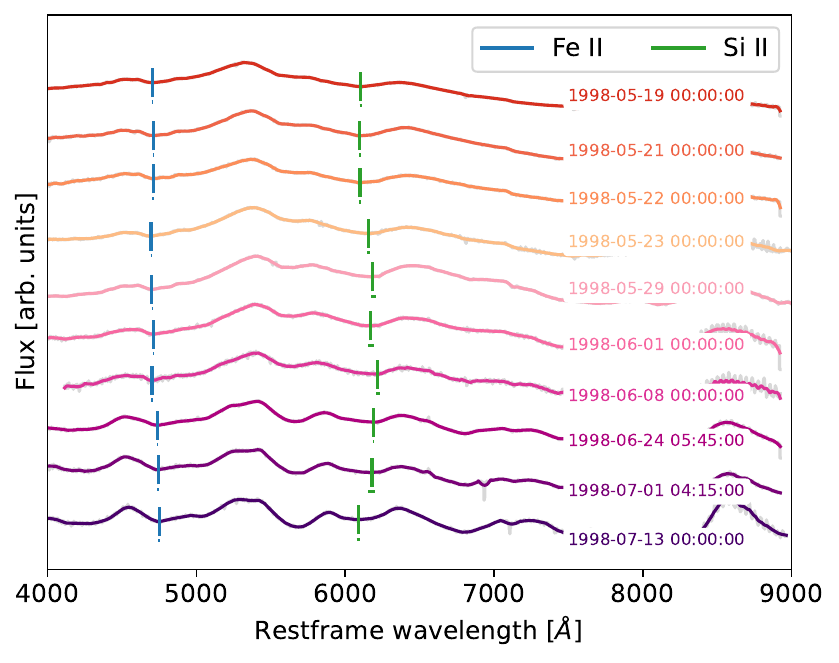}
      \caption{Spectral sequences of the GRB-SN GRB980425-SN1998bw. The observed spectra (grey) and smoothed spectra (multiple colours) are plotted as solid lines. Also shown are the determined minimum wavelengths of the Fe II (blue) and Si II (green) features, and the corresponding uncertainty on the minimum wavelength.}
   \end{figure*}

   \begin{figure*}[!h]
      \centering
      \includegraphics[width=0.49\linewidth]{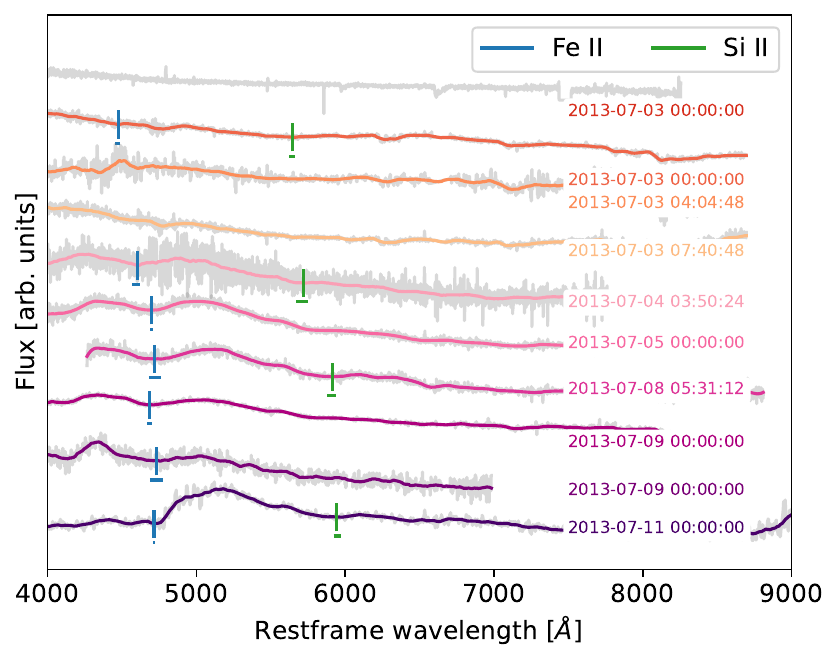}
      \includegraphics[width=0.49\linewidth]{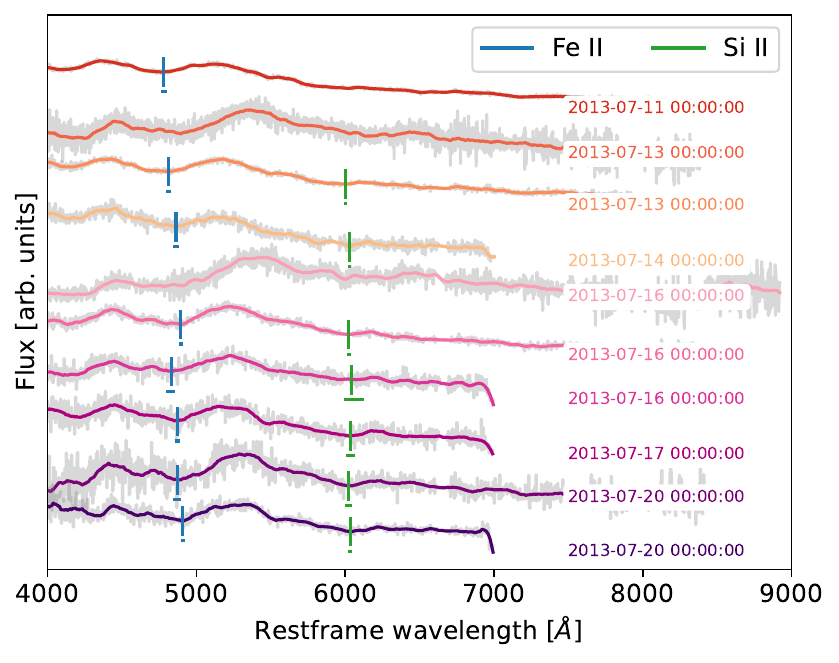}
      \caption{Spectral sequences of the GRB-SN GRB130702A-SN2013dx. The observed spectra (grey) and smoothed spectra (multiple colours) are plotted as solid lines. Also shown are the determined minimum wavelengths of the Fe II (blue) and Si II (green) features, and the corresponding uncertainty on the minimum wavelength.}
   \end{figure*}

\FloatBarrier

\begin{landscape}
\section{Tables}
\begin{tiny}
\begin{longtable}{lcccccc}
\caption{Measured velocities of Fe II and Si II for the type Ic-BL supernovae analysed here. Each line in the table represents one observation. The name of the event, SG filter width, redshift ($z$), observation date of the spectrum and rest-frame time relative to the explosion date ($\Delta T_{rest}$) are all given. The velocities (and uncertainties) of the Fe II ($v_\mathrm{Fe\,II}$) and Si II ($v_\mathrm{Si\,II}$) features are reported for each observation. In some cases, no velocity is reported for a feature, for example because the wavelength range of the feature in question was not covered by the spectrum.} \label{tab:silverresultsIcBL} \\
\toprule
\textbf{Event name} & \textbf{SG filter width} & \textbf{$z$} & \textbf{Obs. date [UT]} & \textbf{$\Delta T_{rest}$ [days]} & \textbf{$v_\mathrm{Fe\,II}$ [km/s]} & \textbf{$v_\mathrm{Si\,II}$ [km/s]} \\
\midrule
\endfirsthead
\caption[]{Measured velocities of Fe II and Si II for the type Ic-BL supernovae analysed here, continued.}  \\
\toprule
\textbf{Event name} & \textbf{SG filter width} & \textbf{$z$} & \textbf{Obs. date [UT]} & \textbf{$\Delta T_{rest}$ [days]} & \textbf{$v_\mathrm{Fe\,II}$ [km/s]} & \textbf{$v_\mathrm{Si\,II}$ [km/s]} \\
\midrule
\endhead
\midrule
\multicolumn{7}{r}{Continued on next page} \\
\midrule
\endfoot
\bottomrule
\endlastfoot
1997ef & 65 & 0.011805 & 1997-12-24 10:33:36.00 & 34.04 & $12106^{-503}_{+421}$ & $5073^{-555}_{+453}$ \\
1997ef & 80 & 0.011805 & 1997-12-29 07:12:00.00 & 38.84 & $13240^{-368}_{+603}$ & $3845^{-467}_{+358}$ \\
1997ef & 63 & 0.011805 & 1997-11-26 09:21:36.00 & 6.32 & $28533^{-2597}_{+1876}$ & $22417^{-1347}_{+450}$ \\
1997ef & 63 & 0.011805 & 1997-12-04 08:09:36.00 & 14.17 & $18127^{-1059}_{+1352}$ & $16192^{-467}_{+708}$ \\
1997ef & 65 & 0.011805 & 1997-12-23 10:19:12.00 & 33.04 & $12341^{-515}_{+463}$ & $5559^{-337}_{+237}$ \\
1997ef & 121 & 0.011805 & 1997-12-30 08:24:00.00 & 39.88 &  & $4283^{-1249}_{+1039}$ \\
1997ef & 67 & 0.011805 & 1998-01-20 05:31:12.00 & 60.52 & $9031^{-537}_{+548}$ &  \\
1997ef & 65 & 0.011805 & 1998-01-01 07:40:48.00 & 41.83 & $12322^{-303}_{+309}$ & $3069^{-235}_{+278}$ \\
\end{longtable}

\newpage
\begin{longtable}{lcccccc}
\caption{Measured velocities of Fe II and Si II for the GRB-SNe analysed here. Each line in the table represents one observation. The name of the event, SG filter width, redshift ($z$), observation date of the spectrum and rest-frame time relative to the explosion date ($\Delta T_{rest}$) are all given. The velocities (and uncertainties) of the Fe II ($v_\mathrm{Fe\,II}$) and Si II ($v_\mathrm{Si\,II}$) features are reported for each observation. In some cases, no velocity is reported for a feature, for example because the wavelength range of the feature in question was not covered by the spectrum.} \label{tab:silverresultsGRB} \\
\toprule
\textbf{Event name} & \textbf{SG filter width} & \textbf{$z$} & \textbf{Obs. date [UT]} & \textbf{$\Delta T_{rest}$ [days]} & \textbf{$v_\mathrm{Fe\,II}$ [km/s]} & \textbf{$v_\mathrm{Si\,II}$ [km/s]} \\
\midrule
\endfirsthead
\caption[]{Measured velocities of Fe II and Si II for the GRB-SNe analysed for this work, continued.} \\
\toprule
\textbf{Event name} & \textbf{SG filter width} & \textbf{$z$} & \textbf{Obs. date [UT]} & \textbf{$\Delta T_{rest}$ [days]} & \textbf{$v_\mathrm{Fe\,II}$ [km/s]} & \textbf{$v_\mathrm{Si\,II}$ [km/s]} \\
\midrule
\endhead
\midrule
\multicolumn{7}{r}{Continued on next page} \\
\midrule
\endfoot
\bottomrule
\endlastfoot
GRB980425-SN1998bw & 51 & 0.008499 & 1998-09-12 00:00:00.00 & 137.92 & $23934^{-146}_{+130}$ & $14966^{-146}_{+170}$ \\
GRB980425-SN1998bw & 47 & 0.008499 & 1998-05-03 09:30:00.00 & 7.42 & $39762^{-324}_{+287}$ & $24954^{-229}_{+250}$ \\
GRB980425-SN1998bw & 40 & 0.008499 & 1998-06-08 00:00:00.00 & 42.73 & $28326^{-320}_{+377}$ & $6580^{-989}_{+340}$ \\
GRB980425-SN1998bw & 21 & 0.008499 & 1998-05-29 00:00:00.00 & 32.81 & $28411^{-70}_{+94}$ & $8139^{-1037}_{+478}$ \\
GRB980425-SN1998bw & 47 & 0.008499 & 1998-05-07 00:00:00.00 & 11.00 & $35464^{-1766}_{+1749}$ & $24363^{-253}_{+229}$ \\
GRB980425-SN1998bw & 71 & 0.008499 & 1998-06-24 05:45:00.00 & 58.83 & $25970^{-157}_{+163}$ & $7785^{-244}_{+225}$ \\
GRB980425-SN1998bw & 57 & 0.008499 & 1998-05-08 00:00:00.00 & 11.99 & $28501^{-1187}_{+1103}$ & $24838^{-163}_{+197}$ \\
GRB980425-SN1998bw & 31 & 0.008499 & 1998-05-04 00:00:00.00 & 8.02 & $38268^{-186}_{+166}$ & $28085^{-910}_{+1005}$ \\
\end{longtable}

\newpage
\begin{longtable}{llcccccccll}
\caption[Fit results for Fe II velocity evolution in Ic-BLs and GRB-SNe]{Fit results for Fe II velocity evolution in Ic-BLs and GRB-SNe. The name and type of the event, best fit type (power-law (PL) or broken power-law (BPL)) and sample that the event belongs to (\textit{Gold, Silver, Bronze, Exclude)} are all provided. For power-law fits, the Fe II velocity at $t_0$+1 day ($a$) and the slope of the evolution ($b$) are listed. In cases where a broken power-law fit was used, the SN velocity at the break time ($A$); the slopes of the power-law segments ($\alpha_1$ and $\alpha_2$); rest-frame time of the break ($t_b$); and $s$, a parameter which controls the smoothness of the break; are given. In some cases a broken power-law fit was not attempted, and so there are no parameters to report.}\label{tab:Feevoresults} \\
\toprule
\textbf{Event name} & \textbf{Type} & $a$ \textbf{[$\mathrm{10^3}$ kms/s]} & $b$ & \textbf{A [$\mathrm{10^3}$ km/s]} & \textbf{$\alpha_1$} & \textbf{$\alpha_2$} & \textbf{$t_b$ [days]} & $s$ & \textbf{Best fit} & \textbf{Sample} \\
\midrule
\endfirsthead
\caption[]{Fit results for Fe II velocity evolution in Ic-BLs and GRB-SNe, continued.} \\
\toprule
\textbf{Event name} & \textbf{Type} & $a$ \textbf{[$\mathrm{10^3}$ kms/s]} & $b$ & \textbf{A [$\mathrm{10^3}$ km/s]} & \textbf{$\alpha_1$} & \textbf{$\alpha_2$} & \textbf{$t_b$ [days]} & $s$ & \textbf{Best fit} & \textbf{Sample}  \\
\midrule
\endhead
\midrule
\multicolumn{11}{r}{Continued on next page} \\
\midrule
\endfoot
\bottomrule
\endlastfoot
GRB980425-SN1998bw & GRB-SN & ${52}^{+4}_{-4}$ & ${-0.18}^{+0.03}_{-0.03}$ & ${29.4}^{+0.7}_{-0.8}$ & ${-0.51}^{+0.11}_{-0.07}$ & ${-0.06}^{+0.02}_{-0.02}$ & ${14}^{+2}_{-2}$ & -20 & BPL & Gold \\
GRB030329-SN2003dh & GRB-SN & ${57}^{+6}_{-7}$ & ${-0.21}^{+0.04}_{-0.04}$ & ${34}^{+7}_{-5}$ & ${-0.3}^{+1.3}_{-0.1}$ & ${-0.2}^{+1.2}_{-0.1}$ & ${8}^{+2}_{-10}$ & -20 & PL & Silver \\
GRB060218-SN2006aj & GRB-SN & ${37}^{+2}_{-3}$ & ${-0.20}^{+0.03}_{-0.03}$ & ${21}^{+1}_{-1}$ & ${-0.27}^{+3.05}_{-0.09}$ & ${-0.26}^{+3.14}_{-0.07}$ & ${18}^{+4}_{-5}$ & 20 & PL & Gold \\
GRB100316D-SN2010bh & GRB-SN & ${100}^{+20}_{-30}$ & ${-0.41}^{+0.10}_{-0.10}$ & ${33}^{+1}_{-2}$ & ${-0.23}^{+0.04}_{-0.04}$ & ${-1.8}^{+0.6}_{-0.4}$ & ${22}^{+2}_{-2}$ & 20 & BPL & Bronze \\
GRB120422A-SN2012bz & GRB-SN & ${70}^{+20}_{-40}$ & ${-0.4}^{+0.2}_{-0.1}$ & ${22}^{+3}_{-4}$ & ${-0.5}^{+3.3}_{-0.3}$ & ${-0.8}^{+3.1}_{-0.5}$ & ${20}^{+2}_{-3}$ & 20 & PL & Silver \\
GRB130702A-SN2013dx & GRB-SN & ${40}^{+3}_{-3}$ & ${-0.27}^{+0.03}_{-0.03}$ & ${17}^{+1}_{-1}$ & ${-0.25}^{+0.08}_{-0.59}$ & ${-0.23}^{+0.09}_{-0.59}$ & ${21}^{+5}_{-3}$ & -20 & PL & Gold \\
GRB161219B-SN2016jca & GRB-SN & ${100}^{+10}_{-20}$ & ${-0.52}^{+0.07}_{-0.06}$ &  &  &  &  &  & PL & Silver \\
GRB171205A-SN2017iuk & GRB-SN & ${90}^{+10}_{-10}$ & ${-0.59}^{+0.05}_{-0.05}$ & ${28}^{+9}_{-4}$ & ${-0.55}^{+0.09}_{-1.13}$ & ${-0.54}^{+0.10}_{-1.30}$ & ${7}^{+1}_{-7}$ & 20 & PL & Gold \\
\end{longtable}

\newpage
\begin{longtable}{llcccccccll}
\caption[Fit results for Si II velocity evolution in Ic-BLs and GRB-SNe]{Fit results for Si II velocity evolution in Ic-BLs and GRB-SNe. The name and type of the event, best fit type (power-law (PL) or broken power-law (BPL)) and sample that the event belongs to (\textit{Gold, Silver, Bronze, Exclude)} are all provided. For power-law fits, the Si II velocity at $t_0$+1 day ($a$) and the slope of the evolution ($b$) are listed. In cases where a broken power-law fit was used, the SN velocity at the break time ($A$); the slopes of the power-law segments ($\alpha_1$ and $\alpha_2$); rest-frame time of the break ($t_b$); and $s$, a parameter which controls the smoothness of the break; are given. In some cases a broken power-law fit was not attempted, and so there are no parameters to report.} \label{tab:Sievoresults} \\
\toprule
\textbf{Event name} & \textbf{Type} & $a$ \textbf{[$\mathrm{10^3}$ kms/s]} & $b$ & \textbf{A [$\mathrm{10^3}$ km/s]} & \textbf{$\alpha_1$} & \textbf{$\alpha_2$} & \textbf{$t_b$ [days]} & $s$ & \textbf{Best fit} & \textbf{Sample} \\
\midrule
\endfirsthead
\caption[]{Fit results for the evolution of Si II for Ic-BLs with and without GRBs.} \\
\toprule
\textbf{Event name} & \textbf{Type} & \textbf{a [$\mathrm{10^3}$ kms/s]} & \textbf{b} & \textbf{A [$\mathrm{10^3}$ km/s]} & \textbf{$\alpha_1$} & \textbf{$\alpha_2$} & \textbf{$t_b$ [days]} & \textbf{s} & \textbf{Best fit} & \textbf{Sample} \\
\midrule
\endhead
\midrule
\multicolumn{11}{r}{Continued on next page} \\
\midrule
\endfoot
\bottomrule
\endlastfoot
GRB980425-SN1998bw & GRB-SN & ${150}^{+20}_{-30}$ & ${-0.78}^{+0.06}_{-0.06}$ & ${29}^{+4}_{-6}$ & ${-0.7}^{+0.2}_{-1.6}$ & ${-0.7}^{+0.1}_{-1.5}$ & ${9}^{+2}_{-2}$ & 20 & PL & Gold \\
GRB030329-SN2003dh & GRB-SN & ${140}^{+30}_{-30}$ & ${-0.70}^{+0.07}_{-0.09}$ & ${40}^{+10}_{-10}$ & ${-0.6}^{+0.2}_{-1.0}$ & ${-0.6}^{+0.2}_{-1.1}$ & ${8}^{+2}_{-8}$ & 20 & PL & Gold \\
GRB060218-SN2006aj & GRB-SN & ${55}^{+7}_{-8}$ & ${-0.44}^{+0.05}_{-0.05}$ & ${17.9}^{+0.6}_{-1.1}$ & ${-0.27}^{+0.06}_{-0.08}$ & ${-1.2}^{+0.4}_{-0.3}$ & ${16}^{+2}_{-1}$ & 20 & BPL & Gold \\
GRB100316D-SN2010bh & GRB-SN & ${60}^{+10}_{-10}$ & ${-0.24}^{+0.07}_{-0.07}$ & ${34}^{+3}_{-4}$ & ${-0.2}^{+0.4}_{-0.4}$ & ${-0.3}^{+0.3}_{-0.4}$ & ${13}^{+5}_{-5}$ & 20 & PL & Silver \\
GRB120422A-SN2012bz & GRB-SN & ${80}^{+50}_{-70}$ & ${-0.6}^{+0.3}_{-0.3}$ &  &  &  &  &  & PL & Bronze \\
GRB130702A-SN2013dx & GRB-SN & ${46}^{+8}_{-13}$ & ${-0.48}^{+0.10}_{-0.07}$ & ${13.7}^{+0.6}_{-0.6}$ & ${-0.31}^{+0.02}_{-0.02}$ & ${-2.1}^{+0.6}_{-0.4}$ & ${20}^{+1}_{-1}$ & 20 & BPL & Gold \\
GRB171205A-SN2017iuk & GRB-SN & ${100}^{+40}_{-50}$ & ${-0.9}^{+0.1}_{-0.2}$ & ${12}^{+2}_{-2}$ & ${-0.1}^{+0.2}_{-0.3}$ & ${-3.2}^{+0.6}_{-0.5}$ & ${20}^{+2}_{-2}$ & 5 & BPL & Gold \\
SN2020bvc & GRB-SN & ${80}^{+20}_{-20}$ & ${-0.56}^{+0.08}_{-0.07}$ & ${24}^{+7}_{-6}$ & ${-0.5}^{+0.2}_{-0.6}$ & ${-0.5}^{+0.2}_{-0.5}$ & ${9}^{+3}_{-9}$ & 20 & PL & Gold \\
\end{longtable}

\end{tiny}
\end{landscape}
\end{appendix}

\end{document}